# Structure and Composition of Pluto's atmosphere from the New Horizons Solar Ultraviolet Occultation

Leslie A. Young[1*], Joshua A. Kammer[2], Andrew J. Steffl[1], G. Randall Gladstone[2], Michael E. Summers[3], Darrell F. Strobel[4], David P. Hinson[5], S. Alan Stern[1], Harold A. Weaver[6], Catherine B. Olkin[1], Kimberly Ennico[7], David J. McComas[2,8], Andrew F. Cheng[6], Peter Gao[9], Panayotis Lavvas[10], Ivan R. Linscott[11], Michael L. Wong[9], Yuk L. Yung[9], Nathanial Cunningham[1], Michael Davis[2], Joel Wm. Parker[1], Eric Schindhelm[1,12], Oswald H.W. Siegmund[13], John Stone[2], Kurt Retherford[2], Maarten Versteeg[2]

[1] Southwest Research Institute, Boulder CO, [2] Southwest Research Institute, San Antonio TX, [3] George Mason University, Fairfax VA, [4] The Johns Hopkins University, Baltimore MD, [5] SETI Institute, Mountain View CA, [6] Johns Hopkins University Applied Physics Laboratory, Columbia MD, [7] NASA Ames Research Center, Moffett Field CA, [8] Princeton University, Princeton NJ, [9] California Institute of Technology, Pasadena CA, [10] Université de Reims Champagne-Ardenne, 51687 Reims, France, [11] Stanford University, Stanford CA, [12] Ball Aerospace, Boulder CO, [13] Sensor Sciences, Pleasant Hill, CA 94523, USA,

[*] To whom correspondence should be sent. layoung@boulder.swri.edu

## Abstract

The Alice instrument on NASA's New Horizons spacecraft observed an ultraviolet solar occultation by Pluto's atmosphere on 2015 July 14. The transmission vs. altitude was sensitive to the presence of $N_2$, $CH_4$, $C_2H_2$, $C_2H_4$, $C_2H_6$, and haze. We derived line-of-sight abundances and local number densities for the 5 molecular species, and line-of-sight optical depth and extinction coefficients for the haze. We found the following major conclusions: (1) We confirmed temperatures in Pluto's upper atmosphere that were colder than expected before the New Horizons flyby, with upper atmospheric temperatures near 65-68 K. The inferred enhanced Jeans escape rates were (3-7) x $10^{22}$ $N_2$ $s^{-1}$ and (4-8) x $10^{25}$ $CH_4$ $s^{-1}$ at the exobase (at a radius of ~ 2900 km, or an altitude of ~1710 km). (2) We measured $CH_4$ abundances from 80 to 1200 km above the surface. A joint analysis of the Alice $CH_4$ and Alice and REX $N_2$ measurements implied a very stable lower atmosphere with a small eddy diffusion coefficient, most likely between 550 and 4000 $cm^2$ $s^{-1}$. Such a small eddy diffusion coefficient placed the homopause within 12 km of the surface, giving Pluto a small planetary boundary layer. The inferred $CH_4$ surface mixing ratio was ~ 0.28-0.35%. (3) The abundance profiles of the "$C_2H_x$ hydrocarbons" ($C_2H_2$, $C_2H_4$, $C_2H_6$) were not simply exponential with altitude. We detected local maxima in line-of-sight abundance near 410 km altitude for $C_2H_4$, near 320 km for $C_2H_2$, and an inflection point or the suggestion of a local maximum at 260 km for $C_2H_6$. We also detected local minima near 200 km altitude for $C_2H_4$, near 170 km for $C_2H_2$, and an inflection point or minimum near 170-200 km for $C_2H_6$. These compared favorably with models for hydrocarbon production near 300-400 km and haze condensation near 200 km, especially for $C_2H_2$ and $C_2H_4$ (Wong et al. 2017). (4) We found haze that had an extinction coefficient approximately proportional to $N_2$ density.

# 1. Introduction

We report here on the ultraviolet solar occultation by Pluto's atmosphere observed with the Alice spectrograph on NASA's New Horizons spacecraft. Ultraviolet occultations have proven invaluable for measuring the structure and composition of the other two $N_2$-rich atmospheres in the outer solar system, Titan (Smith et al. 1982, Herbert et al. 1987, Koskinen et al. 2011, Kammer et al. 2013, Capalbo et al. 2015) and Triton (Broadfoot et al. 1989, Herbert & Sandel 1991, Stevens et al. 1992, Krasnopolsky et al. 1992). By observing how the absorption by molecular species and extinction by haze particles vary with altitude as the Sun passes behind an atmosphere, it is possible to measure their vertical density profiles, and infer the pressure and temperature from the density of the majority species. Because pressure, temperature and composition are central to nearly every aspect of atmospheric science, the Pluto ultraviolet (UV) solar occultation was ranked as a Group 1 (required) observation for the New Horizons mission (Young et al. 2008). The UV solar occultation drove aspects of both the design of the Alice Ultraviolet Imaging Spectrograph (Stern et al. 2008) and the mission design of the New Horizons flyby past Pluto (Guo & Farquhar 2008). We built the Alice instrument to observe the occulted solar flux from 52 to 187 nm, covering absorption by the $N_2$ continuum on the short end and extinction by haze on the long end. We designed the spacecraft trajectory to pass through the Sun and Earth shadows of both Pluto and Charon, nearly diametrically for Pluto.

The UV solar occultation occurred from approximately 2015 July 14 12:15 to 13:32 UTC (spacecraft time). Roughly one terrestrial day later, at approximately 2015 July 15 12:38 UTC (ground receipt time), we received confirmation that the observations were successful and that the spacecraft successfully flew through the Pluto's solar shadow. Downlink data volume constraints meant that this first "contingency download" of the UV solar occultation contained only the Alice housekeeping data, which included the total number of photons detected across all wavelengths each second; these data were discussed by Stern et al. (2015). The full downlink of the Pluto solar occultation (0.67 Gigabits) was completed on 2015 Oct 2. Initial analysis of Pluto's ultraviolet solar occultation (Gladstone et al. 2016) presented line-of-sight column abundances of $N_2$, $CH_4$, $C_2H_2$, $C_2H_4$, and $C_2H_6$, and the densities of $N_2$, $CH_4$, $C_2H_2$, and $C_2H_4$.

This paper extends the analysis of Gladstone et al. (2016) in the following ways: (i) it uses an improved reduction of the raw observations, and includes more details about the observation and reduction process, (ii) it presents error analysis, including correlations between the measurements of various species, (iii) it includes analysis of extinction by haze at the long-wavelength end of the Alice range, (iv) it improves or extends the density retrievals of $N_2$, $CH_4$, $C_2H_2$, $C_2H_4$, $C_2H_6$ and haze, and (v) it includes a joint analysis with new results from the New Horizons radio occultation (Hinson et al. 2017).

# 2. Observations and Reduction

We recap here the salient features of the Alice ultraviolet spectrograph on the New Horizons spacecraft and its observation of Pluto's atmosphere during the solar occultation. Alice (which is a name, not an acronym) is described in more detail in Stern et al. (2008), and a previous Alice stellar occultation by Jupiter is described in Greathouse et al. (2010). Alice is an imaging spectrograph that has a bandpass from 52 to 187 nm, with a photocathode gap from 118 to 125 nm designed to decrease the count rate near Ly-α. Alice has two data collection modes (pixel list and histogram), two adjoined slit elements (a wider 2° x 2° "box" and a narrower 4° x 0.1° "slot"), and two apertures (the lower-throughput "solar occultation channel" or SOCC and the higher-throughput "airglow aperture"). For the Pluto solar occultation, we used the pixel list data collection mode for higher time cadence. We also placed the Sun in the "box" of the SOCC to avoid slit losses, to avoid oversaturation, and to observe the UV solar occultation simultaneously with the radio earth occultation (Fig 1).

The SOCC is roughly co-aligned with the field of view of REX (Radio EXperiment, Tyler et al. 2008), to allow for simultaneous observations of the solar and the uplink radio occultations. During the Pluto solar occultation, we centered the REX field of view on Earth (Fig. 1), which placed the Sun within a few tenths of a degree from the center of the 2° x 2° "box." Thrusters were fired to keep REX centered on the Earth within 0.0143° (deadband half-width). Thus, the Sun moved only slightly within the 2° x 2° "box" during the solar occultation observation.

The Sun's diameter as seen from New Horizons in July 2015 was 0.016°, which was much smaller than the size of the "box" and slightly smaller than the Alice pixel size (Alice pixels subtend 0.019° in the spectral axis, and 0.308° in the spatial axis). Pluto, by contrast, was large compared to the box (Fig. 1), varying from 3.0° at 2015 July 14 12:41 UT, when Pluto first entered the box, to 2.2° at 13:01 UT, when Pluto exited the box.

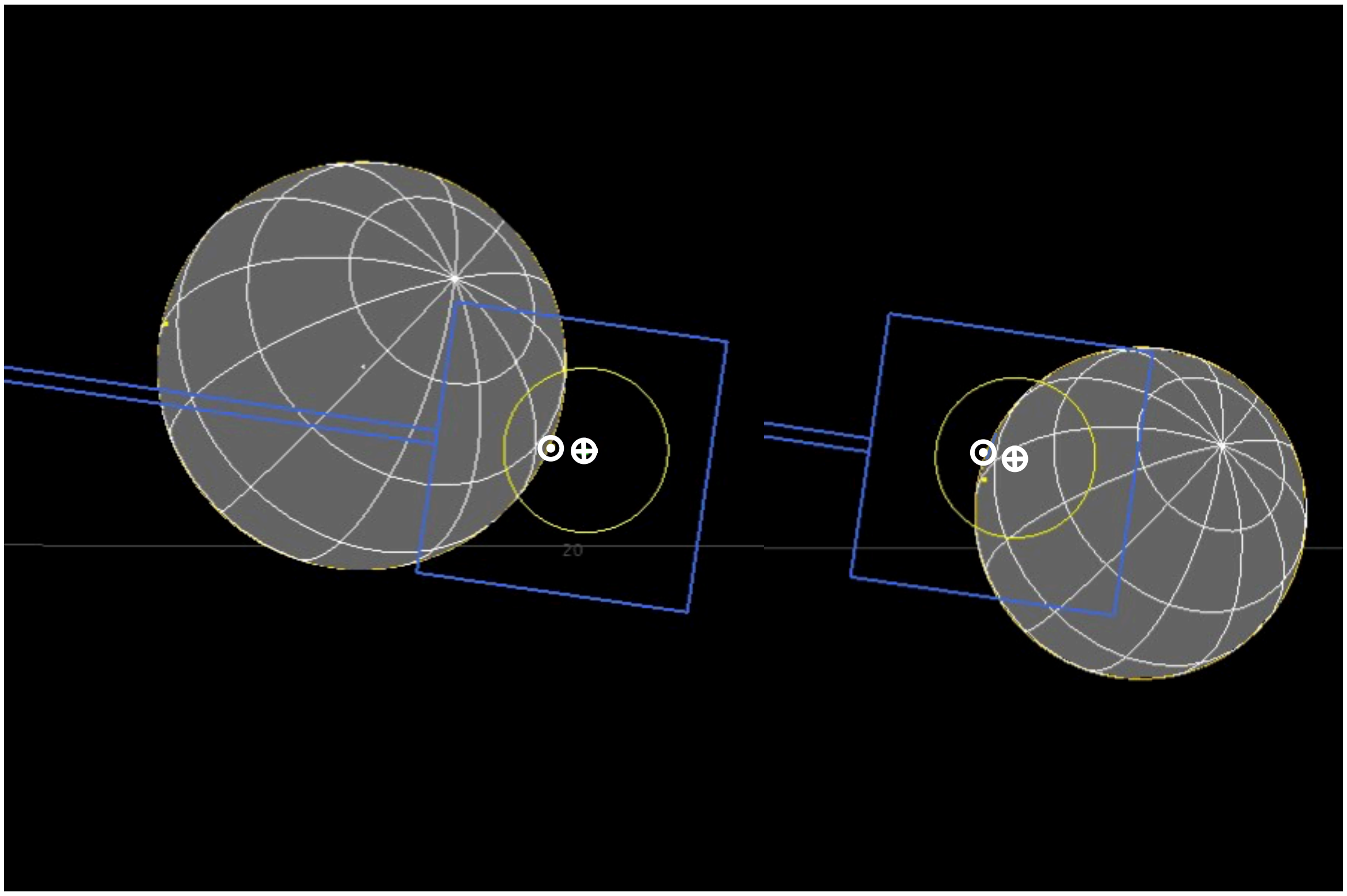

**FIGURE 1**. The Alice Solar Occultation Channel (SOCC), Pluto, and the Sun at the time of solar ingress at 2015 Jul 14 12:44 UT (left) and egress at 12:55 UT (right) as seen from New Horizons. The REX field of view (yellow circle, indicating the 1.2° diameter of the REX 3 dB beamwidth) was centered on the Earth (white Earth symbol). The Alice slit (the blue "box" and "slot") had the box portion centered on the Sun (white Sun symbol). The figure is oriented with celestial North up and East to the left. This figure also shows the scale of Pluto at ingress and egress, and a latitude-longitude grid (at 30° intervals) on Pluto. The southern (winter) pole was in view.

The locations probed by the solar occultation depended only on the relative positions of the Sun, Pluto, and New Horizons, and not on the pointing of the Alice field of view. Pluto passed across the Sun at a sky-plane velocity (that is, the component of the velocity perpendicular to the spacecraft-Sun line) of 3.586 km $s^{-1}$, so it took ~11 minutes for the solid body of Pluto to pass across the Sun. (This is slightly faster than the Earth's sky-plane velocity of 3.531 km $s^{-1}$ during the Earth occultation; Hinson et al. 2017.) Solar ingress occurred at longitude 195.3° E, latitude 15.5° S, while egress occurred at longitude 13.3° E, latitude 16.5° N (Gladstone et al. 2016). Thus, ingress probed the atmosphere just off of the southern tip of the left-hand side of the bright heart-shaped feature, informally named Sputnik Planitia, and egress probed the atmosphere near the transition between dark equatorial regions and the mid-latitude areas (Fig. 2).

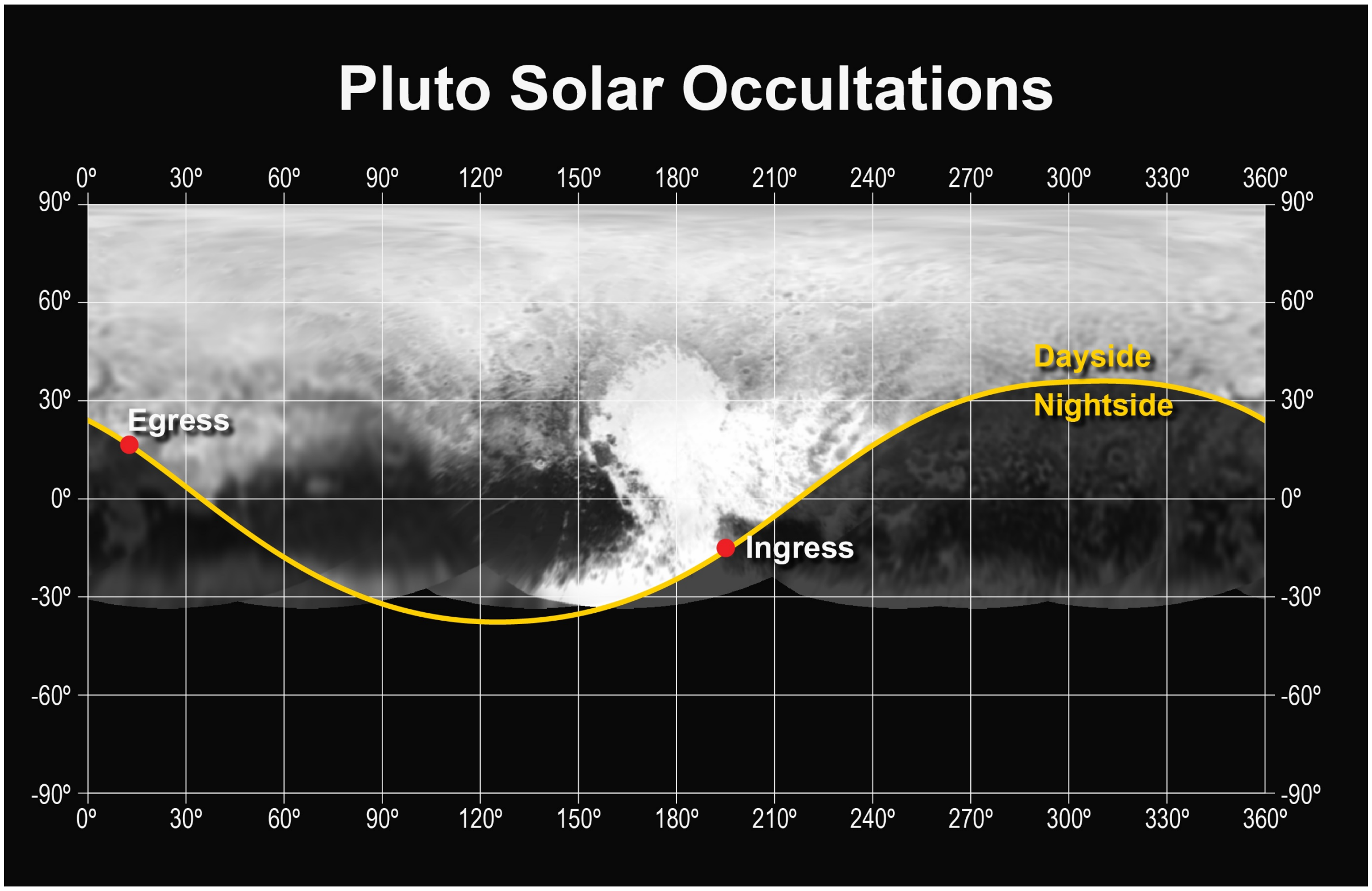

**FIGURE 2**. The locations of ingress (dusk) and egress (dawn) of the solar occultation by Pluto.

The observations were designed to observe the solar flux prior to ingress and after egress, to directly measure the unocculted solar flux. The timing had to account for possible differences between the nominal and actual trajectory (that is, the uncertainties in the final trajectory correction maneuver). In order to be robust in case of a spacecraft Command and Data Handling (C&DH) reset or a problem with the Alice instrument, the Pluto solar UV occultation was commanded as two separate observations, called PEAL_01_Pocc and PEAL_01_PoccEgress. These spanned 2015 Jul 14 12:15:27 to 12:50:35 and 12:53:12 to 13:32:51 UTC, respectively, with a planned, precautionary Alice power reset between the two observations.

Observations were taken in "pixel list" mode, in which each detected photon is tagged with its location on the detector. Effectively, this location was a measure of which of the 1024 spectral and 32 spatial pixels was stimulated by each detected photon. The pixel resolution was 0.177-0.183 nm pixel$^{-1}$, which Nyquist samples the instrumental spectral resolution (0.35 nm when operated in pixel list mode). Timing was determined by the insertion of special "time hack" values into the instrument's memory buffer every 4 ms. This 4 ms timing was much finer than that required for the analysis presented here, and we summed the counts into 1-second time bins. This resulted in a 1024 by 32 image of counts per second at each 1 second interval, called a count rate image. Fig. 3 shows the average of 100 one-second count rate images of the unocculted solar flux.

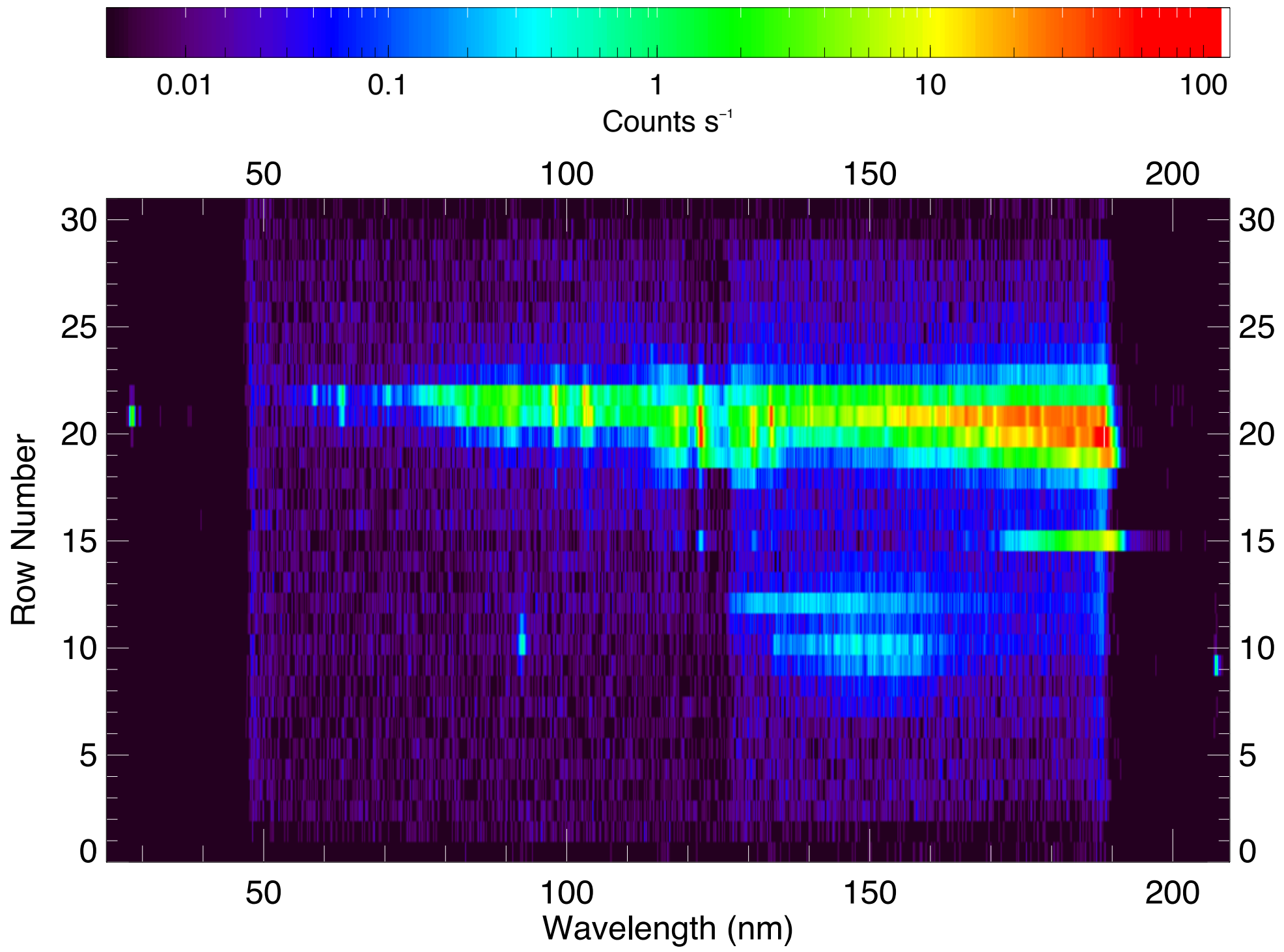

**FIGURE 3.** Average count-rate image of the unocculted Sun. The solar spectrum was contained primarily in rows 20 and 21. The detected background flux was lower near 121.6 nm, where the photocathode gap was designed to have lower sensitivity. We extracted the total solar spectrum by a simple sum of rows 19-22, inclusive. The features centered near [90 nm, row 10], [150 nm, row 12], and [180 nm, row 15] were all instrumental artifacts or ghosts, whose contribution was negligible (they are seen in the plot because the count rates are plotted in a logarithmic scale).

In order to extract 1-second count-rate spectra from the pixel list data, we performed the following analysis steps. First, we calculated dead time correction in the raw pixel list stream. The detector electronics took a finite amount of time to process each count. During this time, the detector was "dead" i.e., it was insensitive to any additional counts. We therefore weighted each detected photon by a factor of 1 / (1 - $\tau_d$ * *C*), where $\tau_d$ =18 μs is the dead time constant of the electronics, and *C* is the count rate, measured over a 4 ms interval.

Second, we summed the pixel stream to construct 2-D count-rate images at 1-second resolution (Fig. 3). During one second, the tangent altitude probed by the Sun moved ~3.586 km through Pluto's atmosphere (c.f. Fig. 6). The choice of 1-second binning was chosen as a balance between increasing signal-to-noise ratio per image and sub-sampling the 4 to 5 second (~16 km) smoothing caused by the Sun's finite size.

Third, we extracted a 1-D solar spectrum from each 2-D count rate image. The Sun varied in its deadband by 0.0143° (half-width). Since this was much smaller than the pixel size of 0.308 deg pixel$^{-1}$ in the spatial direction, the variation within the deadband did not change which detector row contained the counts from the Sun. We extracted the solar spectra by a simple sum of rows 19-22, inclusive (Fig. 3), which accounted for the width of the spatial point-spread function (Stern et al. 2008, their Fig 12) and the motion within the deadband. The contribution of Pluto's nightside to the UV signal (Fig. 1) was negligible compared to the direct solar flux.

Fourth, we used Alice "stim pixels" (Stern et al. 2008) to correct the wavelength scale for temperature effects in the Alice detector. The mapping between the physical location of an event on the detector and its pixel number in data space depended on the resistivity of the readout anode, which itself depended on temperature. Essentially, the detector electronics produced counts at two known physical locations on opposite ends of the detector. These counts were then mapped into data space, allowing for a linear correction to the apparent position of detected photons.

Fifth, each one-second count-rate spectrum was then corrected for the wavelength dependence on the location of the Sun within the 2° by 2° "box" portion of the slit. The Sun was offset slightly from the center line of the slit (Fig. 1), which introduced an overall wavelength shift of 0.396 nm. There was some variation in wavelength as the Sun's position moved in the ±0.0143° deadband, since a pixel subtended 0.019° in the dispersion direction. For unocculted spectra, we determined the wavelength shift by fitting a Gaussian line profile to five solar lines, including the Lyman-alpha (Ly-α) line at 121.6 nm. The shift was also calculated from the spacecraft attitude, and the two methods agreed to within 0.10 pixels; the shift calculated from the spacecraft altitude was used for spectra taken when the solar lines were obscured by Pluto's atmosphere. The resulting spectra were placed on a common wavelength grid using a sinc interpolation. We ignored Doppler shifts in this analysis. Pluto's heliocentric motion, which affected the interaction of the solar lines and absorption in Pluto's atmosphere, was at most 1.075 km s$^{-1}$ during this observation (0.00035 nm shift at 100 nm). This was much less than the width of the solar lines (Curdt et al., 2001). The rate at which the spacecraft receded from Pluto, which affected how the spectrum is recorded on the Alice spectrograph, ranged from 11.66 to 13.60 km/s over the POCC observation (0.0038 to 0.0045 nm shift at 100 nm). This led to a shift of only ~2% of a pixel.

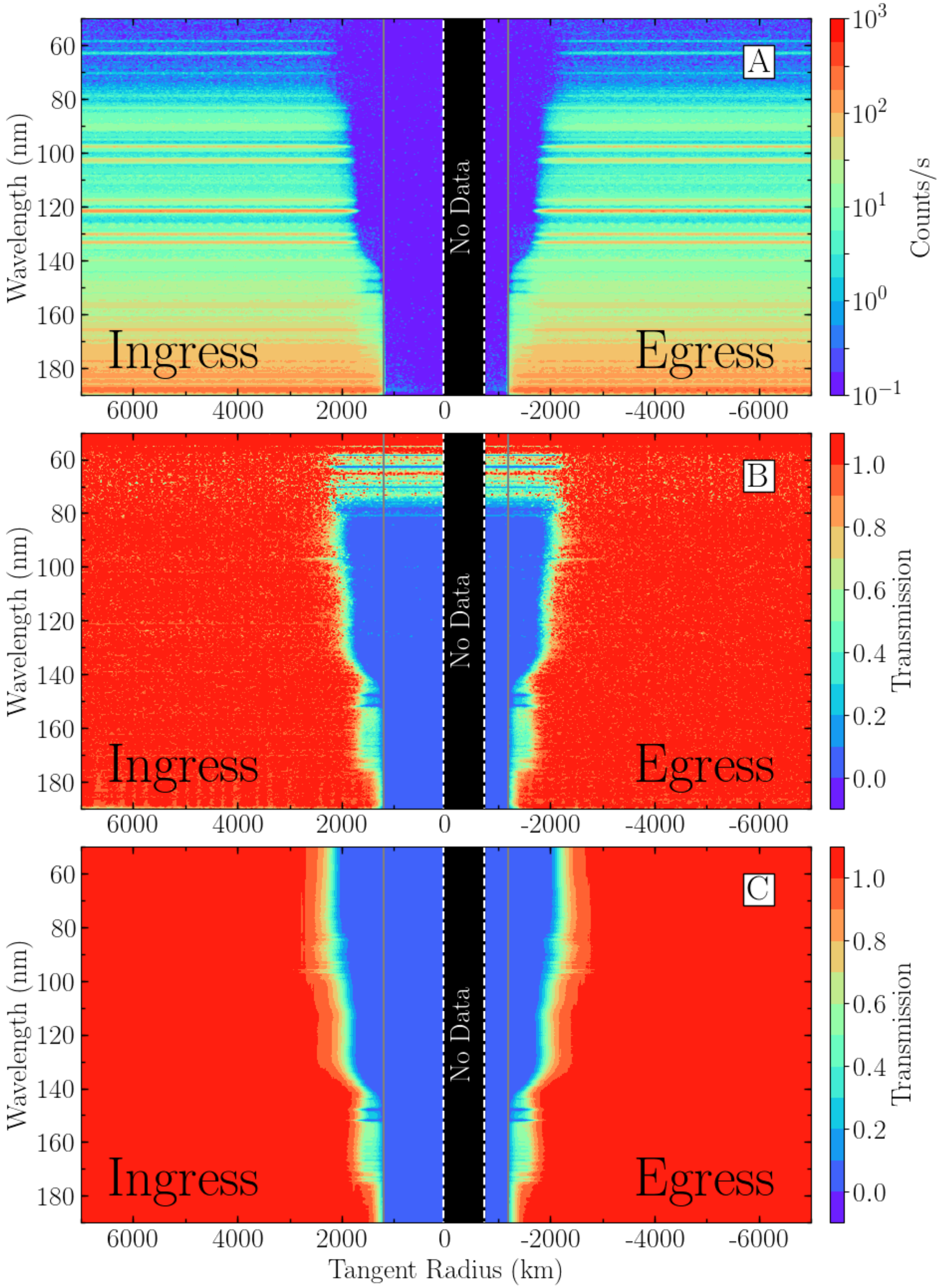

**FIGURE 4**. A. The entire occultation displayed as one-second count rate spectra as a function of the tangent radius and wavelength. To distinguish ingress and egress, we plot egress tangent radii as negatives. B. Observed transmission spectra, plotted linearly from -0.1 to 1.1 as a function of tangent radius (X axis) and wavelength (Y axis). C. Model spectra (See Sections 3 and 4). Pluto's radius is also indicated by vertical lines, at 1190 km. Transmission is plotted from -0.1 to 1.1 to show the scatter around the upper and lower baselines. The high transmission for most wavelengths less than 70 nm even during the Pluto solid body occultation is an indication of the difficulty in normalization when there is low solar flux. The black "no data" periods in this figure are during the commanded Alice power reset described in the text.

Sixth, we corrected for the gradual decrease in sensitivity of regions of the detector that saw the highest solar flux. This localized phenomenon, known as "gain sag," is a function of the total amount of charge extracted from the micro-channel plates, per unit area, over the lifetime of the detector (Stern et al. 2008). The magnitude of the sensitivity loss due to gain sag varied from zero at the short wavelength end of the detector, where the solar flux is low, up to several percent at the long wavelength end of the detector, where the solar flux is greatest. We derived a correction for the gain sag during the occultation observation, on a pixel by pixel basis, by fitting a line to the observed count rate when the sun was unocculted by Pluto or its atmosphere as a function of the total integrated number of counts during the observation, using both ingress and egress spectra.

Seventh, after correcting for gain sag, we subtracted the dark rate spectrum. This was calculated from a composite dark spectrum, shifted (with sinc interpolation) to match the wavelength of each count-rate spectrum. The resulting extracted one-second count-rate spectra vs. tangent radius are shown in Fig. 4A. The tangent radius was the closest distance between Pluto's center and the ray connecting the spacecraft and the center of the solar disk, and was calculated at each second using SPICE kernels supplied by the New Horizons Kinetix navigation team for science analysis: nh_pred_20141201_20190301_od122.bsp, NavSE_plu047_od122.bsp, and NavPE_de433_od122.bsp.

The eighth and final step in the reduction was to create a reference solar spectrum for each one-second spectrum. Because of the small entrance aperture of the SOCC, the repeller grid, a lattice-like pattern of wires designed to remove stray ions (Stern et al. 2008, their Fig 5), produced discrete shadows on the surface of the detector. Even a sub-pixel change in the Sun's position on the focal plane affected the observed shadow pattern on the detector, and caused up to a 40% variation in apparent throughput at certain regions of the spectrum. Failure to correct for this effect would leave periodic reductions in counts that mimic absorption features. To correct for this, we first used the spacecraft attitude information derived from the star trackers to derive the position of the Sun in the focal plane. For each one-second count rate spectrum, we then identified other spectra (in either the Pluto or Charon solar occultations) that (i) were obtained when the Sun was within 0.002° of the spectrum in question, and (ii) were unocculted. This selection gave a median of 60 reference spectra for each one-second sample. These were averaged to produce a reference solar spectrum for each one-second count-rate spectrum. In Fig. 4B, we show the result of dividing each one-second count rate spectrum by its reference solar spectrum at full one-second, 0.17 nm resolution.

Because Alice is a photon-counting device, the probability distribution in the count rate spectra was essentially a Poisson distribution. At large count rates, the variance on the one-second count rate spectra was close to the count rate itself (it differs by a multiplicative factor due to dead time and gain sag corrections, which are close to one, and includes a small error due to uncertainty in the subtracted dark rate). At very small count rates, things were more problematic, including that a Poisson distribution has asymmetric confidence limits (Kraft et al. 1991). The practical solution was to bin the one-second count rate spectra in time, wavelength, or both, as needed to escape the regime of very low integrated counts. We defined our variance to be proportional to the count rate, so that the errors in the one-second count rate spectra gave the correct errors in the binned spectra when added in quadrature.

The reference solar spectra we derived were also used as a consistency check on our model of the solar spectrum at sub-pixel resolution. For the solar spectrum between 66.8 and 148.7 nm, we used high-resolution spectra obtained by the SUMER instrument on SOHO, with a median resolution of 0.0043 nm (Curdt et al., 2001). The SUMER atlas covered 66.8 to 148.7 nm and contained reference spectra for coronal hole, sunspot and quiet Sun areas. Wilhelm (2009) confirmed that the SUMER reference spectra are accurate "typical" snapshots of solar emissions. We combined the three SUMER reference spectra, with ratios determined geometrically from SDO/AIA (Curdt, personal communication), using channel 304 images for coronal hole areas and channel 195 images for sunspot areas, with the complement defining the quiet-Sun areas. This gave linear coefficients of 7% coronal hole, 8% sunspot, and 85% quiet Sun. The resulting high-resolution spectrum compared well with the LISIRD (Pankratz et al 2015) measurement of the composite Ly-α flux ($4.50\text{x}10^{11}$ photon $\text{cm}^{-2}$ $\text{s}^{-1}$ on 2015 July 14). Moreover, the soft X-ray flux as measured by GOES 15 in the long wavelength channel, indicative for emission from solar plage areas, was (0.5 - 0.8) $10^{-6}$ W $\text{m}^{-2}$ on 2015 July 14, in agreement with an 8% sunspot/active region. For wavelengths outside the SUMER range, we used data from TIMED/SEE (Woods et al. 2005) taken on 2015 July 15, during which TIMED observed the same face of the Sun that New Horizons observed during the solar occultation. This data spanned the full range of the Alice bandpass at a resolution of 0.54 nm (0.18145 nm per pixel), with data gaps near Ly-α at 114.2-118.9 and 123.1-128.9 nm. To convert from the model Sun ($S$, in photon $\text{cm}^{-2}$ $\text{s}^{-1}$ $\text{nm}^{-1}$), we multiplied by the effective aperture ($A_{\text{eff}}$, in $\text{cm}^2$, which quantified the end-to-end efficiency of the instrument) and the pixel resolution (δ, in nm $\text{pixel}^{-1}$, which varied from 0.177 to 0.183 nm $\text{pixel}^{-1}$). The resulting unocculted solar rate ($R$(λ), in counts $\text{pixel}^{-1}$ $\text{s}^{-1}$), is expressed as

$$R(\lambda) = S(\lambda) A_{eff}(\lambda) \delta(\lambda) \tag{1}$$

The nominal effective area was defined in Stern et al. (2008). The nominal effective area was adjusted for the small (0.396 nm) wavelength shift that arises from the Sun having been offset from the center line of the slit. This shift was insignificant over most of the wavelength range. The exception was near the photocathode gap, which was fixed in detector coordinates. That is, as the image of the Sun was shifted, the locations in wavelength space of the photocathode transition were also shifted. To account for this, we separated the effective area into a slowly varying factor and a factor due to the photocathode gap (interpolating the nominal effective area over 113.333 to 129.148 nm to derive the slowly varying factor), allowing us to shift the photocathode gap independently.

We began with the effective area from Stern et al. (2008), which was derived using Spica and other hot stars, and then derived a correction factor, $a$, using direct observations of the Sun, even though other calibrations were available closer-in-time to the Pluto flyby. Since one purpose was to demonstrate rough agreement with the solar flux model, using an effective area based on stars rather than the Sun avoids circular reasoning. Also, Spica and the other hot stars have less extreme wavelength-to-wavelength spectral variation than the Sun.

We also modeled the instrumental line-spread function. Because the solar spectrum has many emission lines that are several orders of magnitude brighter than the continuum, our goal was to define the line-spread function at levels 100 to 1000 times less than the peak. To define the line-spread function (LSF) and its uncertainty, we used two pre-launch observations of Ar and Ne lamp lines from a point source observed through the SOCC channel (Stern et al. 2008, their Fig 15), and Alice observations of the unocculted Sun. For the far wing behavior, we used pre-launch measurements of a monochromatic source at 121.6 nm measured through the airglow aperture. This measurement has high signal-to-noise ratio (SNR) out to 15 nm from the line center (data taken from Stern et al. 2008, their Fig 20), which shows a far-wing behavior of LSF proportional to $(\Delta\lambda)^{-1.2}$, where $\Delta\lambda$ is the distance from line center. We found a compromise LSF, based on both modeled lamps and the unocculted solar spectra (Fig 5). For the lamps, we identified 9 Ar lines that affect the spectrum from 80-100 nm, and 41 Ne lines, for which we scaled, shifted, and summed the LSF. The kernel for use with the solar spectrum was additionally modified to account for the angular size of the Sun in the spectral direction (the solar diameter was 0.0162°, as seen from New Horizons during the solar occultation of Pluto). Given the pixel scale and the pixel resolution of ~0.18 nm/pixel, this required an additional convolution by a semicircular kernel with full diameter of 0.149 nm. The adopted LSF was a piece-wise continuous function that was (i) Gaussian in the core, where $\Delta\lambda$ was less than 0.25 nm, with full-width half-max of 0.33 nm, (ii) proportional to $(\Delta\lambda)^{-2}$ from 0.25 to 10 nm, and (iii) proportional to $(\Delta\lambda)^{-1.2}$ for the far wings, where $\Delta\lambda$ was larger than 10 nm. The observed reference solar spectra at pixel $i$ and time $t$, $D_{ref}$, is given by

$$D_{ref}(i(\lambda), t) = a(i(\lambda), t)[k(\lambda) \ * R(\lambda)] \tag{2}$$

where $D_{ref,}$ is the reference (e.g., unocculted) solar spectrum for a given pixel index, $i$, and time, $t$. $k(\lambda) * R(\lambda)$ is the solar rate convolved with the line-spread function. $a(i, t)$ is a correction factor to the effective area that takes into account such effects as the repeller grid.

There was generally excellent agreement between the TIMED/SEE spectrum and the appropriately smoothed SOHO/SUMER spectrum in the region of overlap, and we show only the SOHO/SUMER spectrum in the region of overlap in Fig. 5. The effect of the grid shadows was evident as small decreases in flux that mimicked absorptions, for example at 81 and 84 nm (Fig 5, bottom). The Alice data was lower than the modeled TIMED/SEE spectrum by up to a factor of 0.4 at the extremes on the spectral range, i.e., between 67 and 75 nm and between 165 to 188 nm. Our analysis thus assumed a smaller effective area below 75 and above 165 nm; this adjusted effective area was used to generate Fig 5.

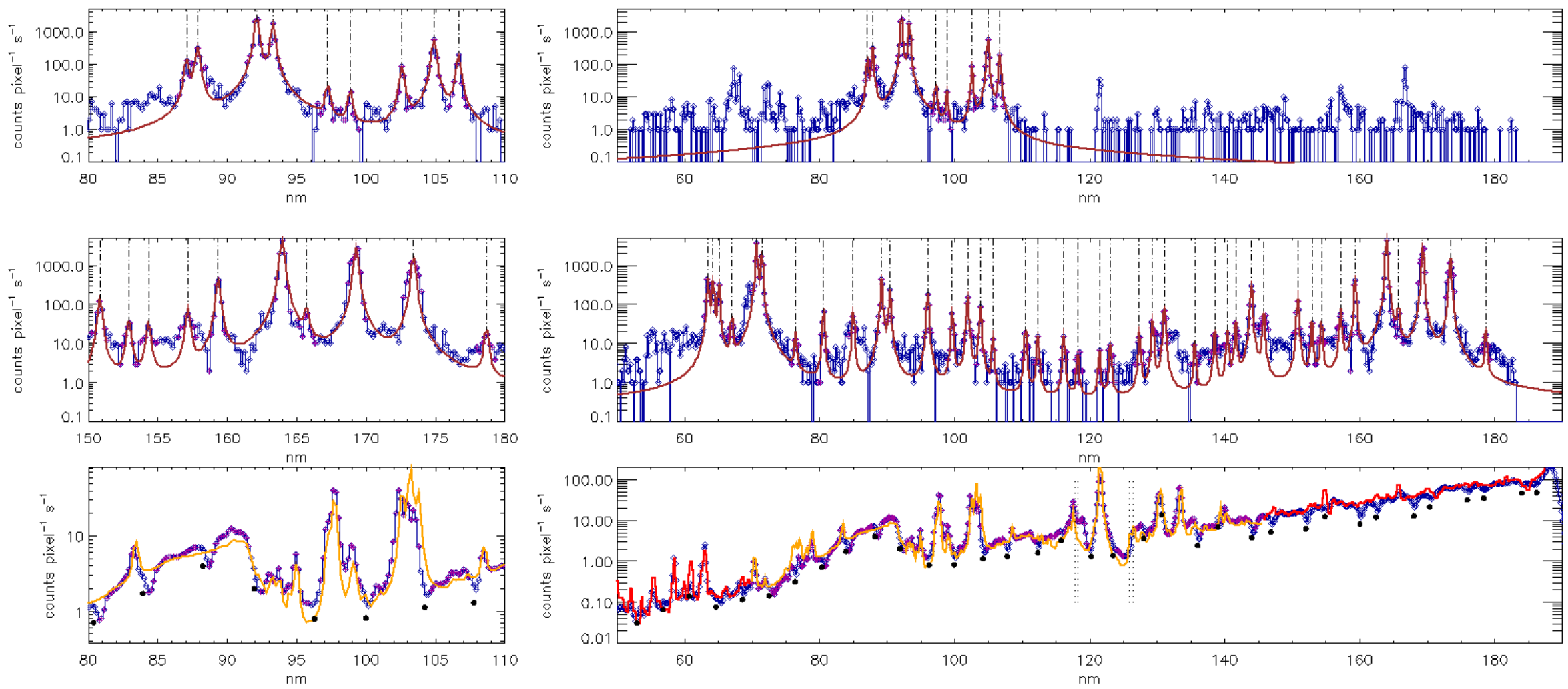

**FIGURE 5** Top: Ar lamp spectra over a 30-nm window (left) and the full Alice range (right), showing the effectiveness of our adopted line-spread function (LSF). Blue shows the pre-launch lab data. Red shows a selection of lines (dashed) convolved with the constructed line-spread function (see text). Middle: same, for Ne. Bottom: Solar spectrum in counts per pixel per second from the unocculted Alice data (blue), and modeled from SOHO/SUMER (yellow) and TIMED/SEE (red). Black dots show the wavelengths where grid shadows are predicted to decrease the solar flux. These align well with the observed localized decreases in the observed unocculted solar flux. The dashed lines show the wavelengths of the transition into the photocathode gap, which were shifted as described in the text. The shifted wavelengths of the transitions are 117.803 to 118.235 nm and 125.844 to 126.349 nm.

# 3. Cross Sections

For the ultraviolet solar occultation by Pluto observed by New Horizons, the refraction of Pluto's atmosphere can be ignored (Hinson et al. 2017), making the geometry of the occultation simple. The ray connecting the Sun and the New Horizons spacecraft has a minimum distance to the body center, called the tangent radius, $r'$, which can be defined the by the surface radius, $r_S$, and the height of the tangent point above the surface (the tangent height), $h$, by $r' = r_S + h$. At a distance along the ray, $x$, defined as 0 at the tangent point, the relationship between the radius in the atmosphere along the ray, $r$, and the tangent radius is simply $r^2 = r'^2 + x^2$. We can define the radius in the atmosphere as the sum of the surface radius and the altitude, $z$, by $r = r_S + z$.

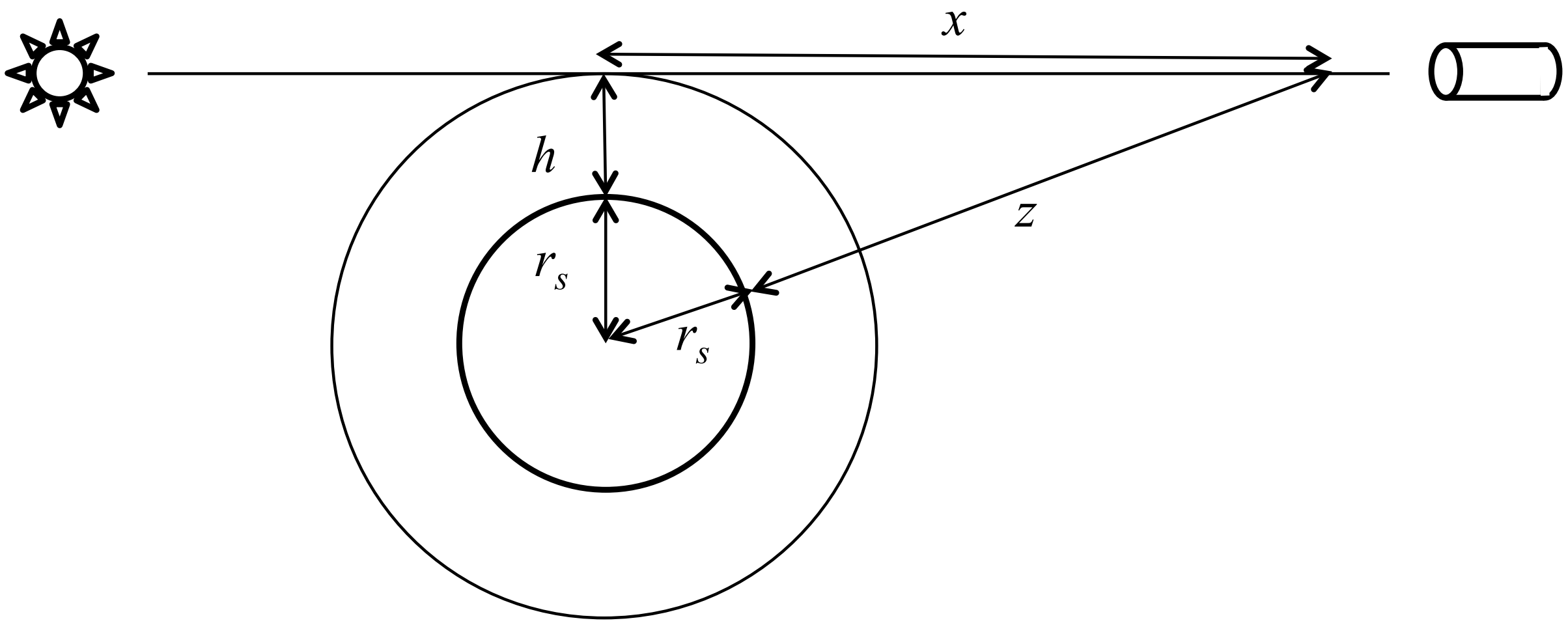

**FIGURE 6.** The geometry of the UV solar occultation. Light from the Sun (left) passes to the detector (right) on essentially a straight line. The minimum distance between the ray and the body center is the tangent radius, $r'$, which can be expressed as $r' = r_S + h$, where $r_S$ is the surface radius and $h$ is the tangent height. The solar flux is affected by the atmosphere along the entire line-of-sight, at different locations along the ray passing through the atmosphere, $x$, with $x = 0$ at the tangent point.

The incident UV solar flux is diminished by the absorption of an occulting atmosphere (e.g., Smith and Hunten 1990). The line-of-sight transmission, $Tr$, is a function of the line-of-sight optical depth, τ,

$$Tr(r',\lambda) = \exp[-\tau(r',\lambda)] \tag{3}$$

which is itself as a function of tangent radius, $r'$, and wavelength, λ:

$$\tau\left(r',\lambda\right) = \int_{-\infty}^{\infty}\left(\sum_{s} n_s(x)\sigma_s(x,\lambda)\right)dx \tag{4}$$

where $n_s$ is the local number density of species $s$, and $\sigma_s$ is the cross-section of species $s$. If the cross sections can be approximated as constant along the line-of-sight, Eq. (4) simplifies to

$$\tau\left(r',\lambda\right) = \sum_{s} N_s \sigma_s(\lambda) \tag{5}$$

where $N_s$ is the line-of-sight abundance of species $s$, defined by

$$N_s\left(r'\right) = \int_{-\infty}^{\infty} n_s(x)\,dx \tag{6}$$

The transmission, *Tr*, is weighted by the solar rate, *R*, before being convolved by the line-spread function, $k$, (with * indicating the convolution integral) and multiplied by an adjustment factor to account for factors such as the repeller grid, $a$, in a manner similar to the unocculted flux (Eq. 2)

$$D(i,t) = a(i,t)[k(\lambda) * (R(\lambda)Tr(\lambda))] \tag{7}$$

In practice, the observed reference solar spectra and Eq. 2 are used to eliminate the adjustment factor, $a$, giving

$$D(i,t) = D_{ref}(i,t)\frac{k(\lambda)*(R(\lambda)Tr(\lambda))}{k(\lambda)*R(\lambda)} \tag{8}$$

If the kernel is sharp compared to the variation in the transmission or solar flux, then Eq. 8 reduces to the simple $D(i,t) = D_{ref}(i,t)\ Tr(\lambda)$, as expected.

Sections 4 and 5 describe the derivation of line-of-sight number densities in Pluto's atmosphere from the solar occultation data. Table 1 presents a list of species known to be in Pluto's atmosphere, either from prior analysis of New Horizons data (Stern et al. 2015, Gladstone et al. 2016) or ground-based observations (e.g., Lellouch et al. 2011, 2016). The cross sections of the known species are plotted in Fig 7.

**Table 1 Absorption cross sections and sources**

| Formula | Name | Source |
|---|---|---|
| $N_2$ | Nitrogen | Shaw et al. 1992 (318.15 K); Chan et al. 1993 (298 K); Heays 2011 (70-140 K) |
| $CH_4$ | Methane | Kameta et al. 2002 (298 K); Chen & Wu 2004 (150 K);<br>Lee et al. 2001 (295 K) |
| CO | Carbon Monoxide | Chan et al. 1993b (298 K);<br>Visser et al 2009 (40-120 K); Stark et al. 2014 (40-120 K) |
| $C_2H_6$ | Ethane | Kameta et al. 1996 (298 K); Chen & Wu 2004 (150 K);<br>Lee et al. 2001 (295 K) |
| $C_2H_2$ | Acetylene | Cooper et al. 1995 (298 K); Nakayama and Watanabe 1964 (295 K);<br>Wu et al. 2001 (150 K) |
| $C_2H_4$ | Ethylene | Cooper et al. 1995 (298 K); Wu et al. 2004 (140 K). |
| HCN | Hydrogen cyanide | Barfield et al. 1972 (see text); West 1975 (~298 K) |
| Hazes | | Cheng et al. 2017 |

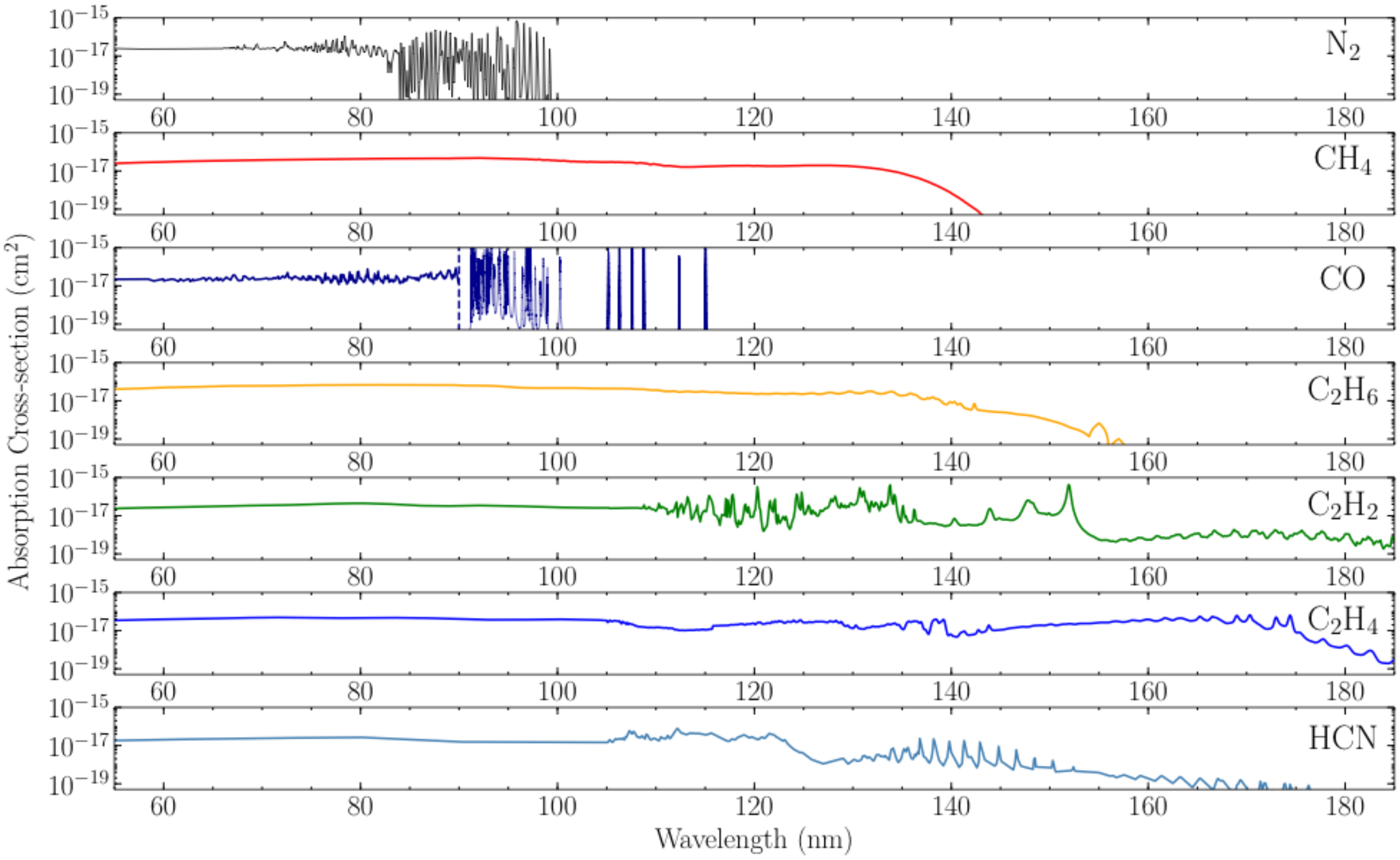

**FIGURE** 7: Absorption cross section vs. wavelength for the species known to be present in Pluto's atmosphere (Table I).

***Nitrogen ($N_2$)***. The ultraviolet cross section (Fig 7) of molecular nitrogen ($N_2$) is dominated by the ionization continuum shortward of the $B\ ^2\Sigma_u^+$ threshold at 66.123 nm (e.g., Huffman et al. 1963b). From 66.123 to ~100 nm, the spectrum becomes increasingly dominated by transitions to electronic states. Pre-dissociation complicates the determination of the individual line oscillator strengths. For this work, we used the measurements of Shaw et al. (1992) shortward of 78.6 nm, taken at 45 °C (318 K), which have a resolution of 0.02 nm. For the continuum, near 60 nm, the Shaw et al. absorption cross sections are within a few percent of measurements at 298 K (Chan et al. 1993a; Lee et al. 1973; Samson et al. 1977; Samson et al. 1987; Samson and Cairns 1964; Wight et al. 1976) or at 295 K (Huffman 1969; Wainfan 1955); one does not expect much variation with temperature at low pressures because the absorption is mostly due to ionization (Capalbo et al. 2015). In the wavelength range dominated by the electronic states, the Shaw et al. resolution is narrow compared with the width of solar lines (~0.1 nm), and narrow enough to separate one band from another (typical spacing ~0.1 - 0.5 nm), but not narrow enough to resolve the individual lines. At wavelengths longward of 84 nm, we used the semi-empirical coupled-channels calculations in Heays (2011) calculated at 0.005 nm resolution at 40 - 120 K (at 10 K resolution) and at 300 K. Even the semi-empirical coupled-channels calculations underestimated the cross sections shortward of ~84 nm, because the model can't keep up with the expanding number of Rydberg states, so we used Chan et al. 1993 between 79.6 and 84 nm. Rayleigh scattering by $N_2$ was not included; it is only $3.2 \times 10^{-23}$ $cm^{-2}$ at 100 nm and $6.6 \times 10^{-25}$ $cm^{-2}$ at 175 nm.

***Methane ($CH_4$).*** The UV absorption due to $CH_4$ (Fig 7), and all the alkanes, has both continuum absorption and broad Rydberg excitations that, unlike the narrow $N_2$ absorptions, are resolvable by most modern laboratory UV detectors (Chen & Wu 2004). The $CH_4$ cross section is characterized by a shoulder near 140 nm. In this work, we used the 298 K measurements of Kameta et al. (2002) from 55 - 120 nm, Chen and Wu (2004) at their coldest measurement of 150 K at 120-142 nm, and Lee et al. (2001; 295 K) longward of 142 nm. We scaled Kameta et al. (2002) and Lee et al. (2001) measurements to match that of Chen and Wu (2004) at 120 and 142 nm respectively, to avoid adding a discontinuity that fitting routines might treat as a spurious absorption edge. The effect of the warmer temperatures for the Kameta et al. and Lee et al. measurements were minor; Chen and Wu (2004) showed that the continuum shortward of 138 nm increases with decreasing temperature by only 2-5% per 100 K. Furthermore, we found that $C_2H_6$ absorption in Pluto's atmosphere dominated the $CH_4$ absorption longward of ~140 nm, so that even this small change in cross section with temperature on the shoulder had little effect on the total optical depth.

***Carbon Monoxide (CO).*** CO (Fig 7), like $N_2$, has narrow lines that are subject to saturation, or self-shielding (e.g., Visser et al. 2009), which can give misleading results for calculations that do not resolve the lines. For this analysis, we use the data of Chan et al. (1993b) shortward of 90 nm, taken at 298 K. Shortward of 90 nm, the lack of high-resolution or low-temperature spectra did not hinder our analysis of the Pluto solar occultation, since it is known from other measurements that the CO mixing ratio was less than 0.1% of $N_2$ (Lellouch et al. 2011, 2016). Longward of 90 nm, we use a combination of the Visser et al. (2009) cross sections with the data published in Stark et al. 2014, generated at 10 K intervals from 40 to 120 K, at ~2.7 x $10^{-5}$ nm resolution (Alan Heays, personal communication).

***Ethane ($C_2H_6$).*** Like $CH_4$ and the other alkanes, $C_2H_6$ (Fig 7) is characterized by continuum absorption and broad Rydberg excitations (Chen & Wu 2004). As described by Chen and Wu (2004), the shoulder for $C_2H_6$ is at a longer wavelength than for $CH_4$, a lighter alkane than $C_2H_6$. Longward of its absorption threshold, $C_2H_6$ has two diagnostic features at 155 and 157 nm. Similarly to $CH_4$, we used the data of Kameta et al. (1996) measured at 298 K shortward of 120 nm, the data from Chen and Wu (2004) at 150 K, their coldest temperature, for 120-150 nm, and Lee et al. (2001) at 295 K longward of 150 nm. We used the measurements of Chen and Wu (2004) taken at 370, 295, and 150 K to evaluate the expected effect of using 150 K for our extractions. In the range measured by Chen and Wu (2004), they showed that the broad Rydberg bands are slightly stronger at lower temperatures, so that e.g., the feature at the start of the shoulder at 142 nm would be 23% stronger at 70 K than at 295 K, based upon extrapolating from the measurements made at 150 K, 295 K, and 370 K. This particular feature is not diagnostic in the New Horizons Pluto solar occultation. The two features just shortward of 160 nm are longward of the Chen & Wu (2004) range. The broad shoulder between 140 and 150 nm is not greatly affected by temperature. At the end of Section 4, we discuss the effect that using hydrocarbon cross sections measured at 140-150 K had on the retrieved line-of-sight abundances.

***Acetylene ($C_2H_2$)***. The spectrum of $C_2H_2$ (Fig 7) has a complex and diffuse spectrum longward of 154 nm, and sharper and more well-defined features 140-154 nm (Wu et al. 2001). These sharper features are the more diagnostic ones for the Pluto solar occultation, including assorted features 110-135 nm, and more importantly, three bands at ~143.8, 148.0, and 152.0 nm. Cooper et al. (1995), measured at 298 K, extends from shortward of the Alice range, but is relatively coarse (3.8 nm spacing) near 100 nm. We transition from Cooper at el. to the Nakayama and Watanabe (1964) 295 K cross section at the short end of the Nakayama and Watanabe span, at 105.2 nm, and then to the Wu et al. (2001) measurements at 150 K longward of 117 nm.

***Ethylene ($C_2H_4$)***. In Alice's spectral range, the cross section of $C_2H_4$ (Fig 7) has an underlying continuum that peaks at 125 and 163 nm, exhibits a series of sharp bands near and shortward of 174.4 nm, weak bands near and shortward of 150.2 nm, and a complicated series of bands ~124-144 nm (Wu et al. 2004). In this work, we use the cross sections of Cooper et al. (1995) measured at 298 K shortward of 115.6 nm, and the cross sections from Wu et al. (2004) measured at 140 K longward of 115.6 nm.

***Hydrogen cyanide (HCN).*** We use the cross section of HCN (Fig 7) compiled in Huebner et al., 1992, which synthesizes the cross section from the atomic cross sections of H, C, and N shortward of 90.0 nm (Barfield et al. 1972), and uses West (1975) longward of 105.0 nm.

***Hazes.*** The composition of the hazes (cross section not plotted) is a current subject of study, as is their sizes and spherical vs. aggregate nature. The hazes are the dominant absorbers in the solar occultation longward of 175 nm, and contribute significantly to the absorption longward of ~150 nm. Using Mie scattering theory and the UV optical constants of tholin particles (Khare et al. 1984), we find that the haze cross section is nearly flat over 175-188 nm (variation less than 3%) for particles larger than 0.2 micron. As reported in Cheng et al. (2017), the scattering properties of the haze varies with altitude, being consistent with 0.5 μm spherical particles below 45 km altitude, transitioning to fractal aggregates at higher altitudes. Since it is not likely that the haze cross section was constant with altitude, we derived haze optical depth in place of line-of-sight abundance, and extinction coefficient in place of number density.

## 4. Line-of-sight abundances: hydrocarbons and haze

The retrieval of $N_2$, $CH_4$, $C_2H_6$, $C_2H_2$, $C_2H_4$, and haze line-of-sight abundances presented in this paper was performed individually at each altitude, in a method very similar to that used in Gladstone et al. (2016). We began by fitting for the hazes and hydrocarbons using only the wavelengths 100-180 nm, where the signal-to-noise of the occultation data was highest, and $N_2$ did not contribute. In a later step (Section 5), we included $N_2$ to analyze the wavelengths below 65 nm. As discussed below, HCN and CO, while known to be present in Pluto's atmosphere, were not detected in the Alice solar occultation by Pluto.

Retrievals were performed separately for ingress and egress. The retrieval started at a tangent height 2000 km altitude, above the first measureable absorption, and progressed downward toward the surface. To retrieve line-of-sight abundances from each spectrum, the spectrum at the next higher altitude step acted as the initial condition in a weighted Levenberg-Marquardt least-squares fit to minimize the weighted sum of squared residuals, or $\chi^2$ (Press et al. 2007). Fitting to individual one-second spectra gave unacceptably large errors and unstable solutions to the non-linear least-squares fit, so we averaged multiple spectra before retrieving line-of-sight abundances. Because of the 3.586 km $s^{-1}$ sky-plane velocity of the Sun, averaging in time corresponded to averaging in altitude. To investigate the impact of different averaging lengths, we performed one retrieval that averaged 23 seconds (82.5 km) for altitudes > 1000 km, 11 seconds (39.5 km) for 1000 to 800 km, and 5 seconds (17.9 km) below 800 km, scaling the errors per spectrum appropriately (Fig 8. left). A second retrieval averaged 23 seconds throughout the entire span (Fig 8, right). The 5-second averaging scale was matched to the angular size of the Sun as seen by New Horizons. At altitudes below 700 km, the 23-second averaging sampled less than one point per scale height. In the context of Eq. 8, we retrieved line-of-sight abundances that produced a transmission, $Tr$, that best matched the time-averaged data, $D$, when weighted by the solar rate, $R$, convolved by the line-spread function, $k$, normalized by the convolved solar rate, $k*R$, and multiplied by the time-averaged reference spectrum, $D_{ref}$ (Eq. 8). For each averaged spectrum, we fitted for the log of the line-of-sight abundance, because that quantity guaranteed a positive $N_s$, and because the surfaces of $\chi^2$ were more symmetric in ln($N$) than in $N$ itself.

Our retrievals (Fig 8) are plotted linearly with geopotential height $\xi$, defined by:

$$\xi = (r - r_s)\left(\frac{r_s}{r}\right) \tag{9}$$

where $r_s$ is the surface radius (e.g., Chamberlain & Hunten 1987). Plotting vs. geopotential height is visually useful because ln($p$) and ln($n$) are linear with $\xi$, and ln($N$) is nearly linear with $\xi$ (e.g., Young 2009), for constant temperature and hydrostatic equilibrium. We took $r_s$ to be 1190 km; this simplified comparison with the radio occultation (Hinson et al. 2017) who found $r_s$ = 1187.4 ± 3.6 km at ingress and 1192.4 ± 3.6 km at egress, and was similar to the value of 1188.3 ± 1.6 km from imaging data (Nimmo et al., 2017).

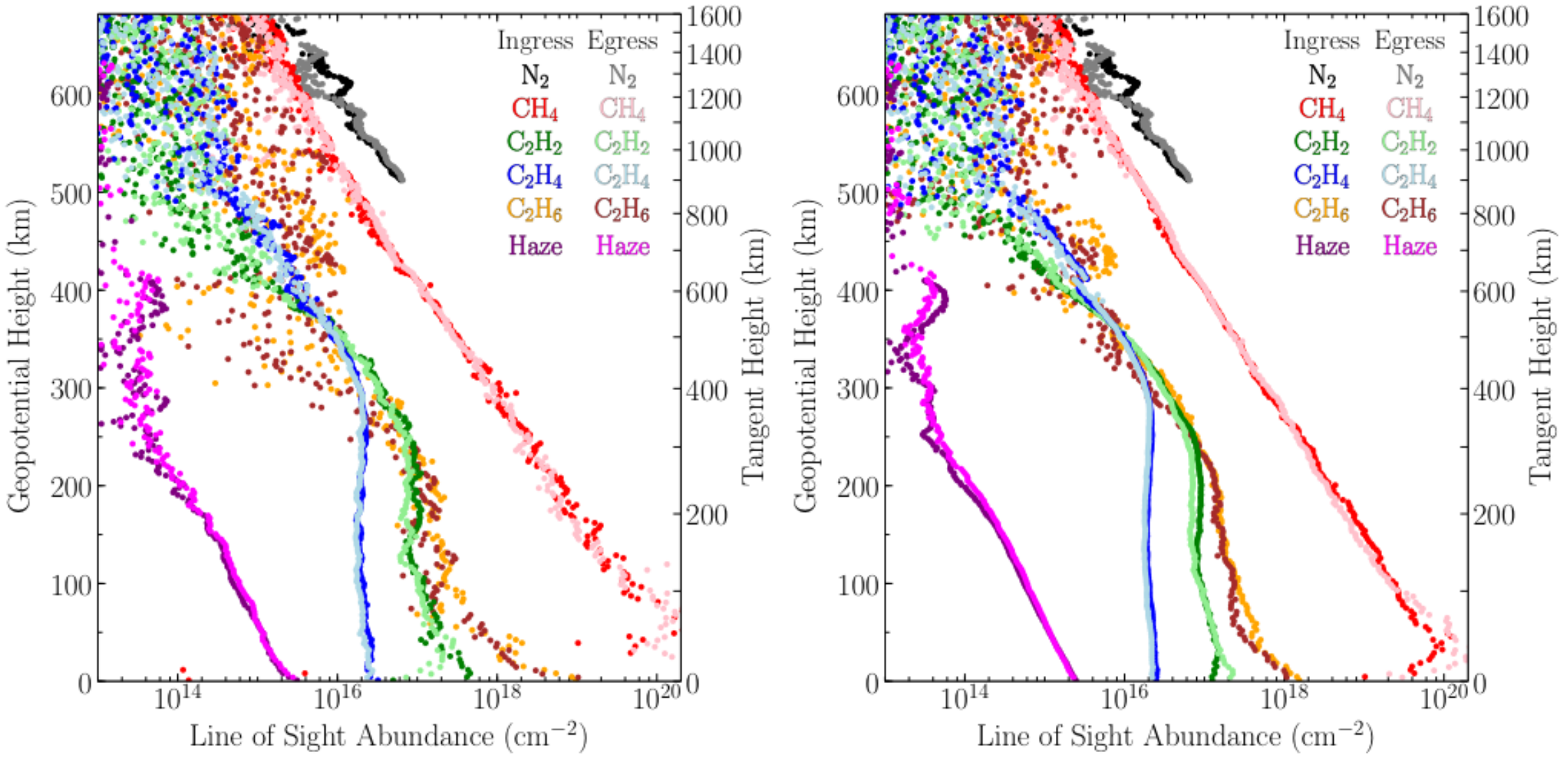

**FIGURE 8**. Line-of-sight abundance of $CH_4$ (red, ingress; pink, egress), $C_2H_6$ (brown, ingress; mustard, egress), $C_2H_2$ (dark green, ingress; light green, egress), $C_2H_4$ (dark blue, ingress; light blue, egress). For haze (dark purple, ingress; light purple, egress), we plot $10^{15}$ times the line-of-sight optical depth (that is, optical depth at the surface is slightly larger than 2). Minor ticks on the x-axis are plotted at 2, 4, 6, and 8 times the major tick values. (Left) Variable altitude smoothing, 23-second (82.5-km) smoothing above 1000 km, 11-second (39-km) smoothing from1000 to 800 km, and 5-second (18-km) below 800 km. (Right) Constant altitude smoothing of 23 seconds (82.5 km) throughout. For reference, the line-of-sight abundances for $N_2$ (gray, ingress; black egress) are also plotted, with an altitude smoothing of 23 seconds for both left and right (See Section 5).

The advantage of smoothing over longer intervals was improved signal-to-noise ratio (SNR), or decreased point-to-point scatter. This was particularly important for $C_2H_6$, which had large scatter at all altitudes in the left-hand panel of Fig 8. The improvement was quantified by looking at the formal errors (Press et al. 2007), in Fig 9, where we show errors in the fitted parameter, of $\sigma(\ln N)$ vs. altitude. For small errors, this is mathematically identical to the fractional error, $\sigma(\ln N) = \sigma(N)/N$. At larger errors, one can think of errors in $\ln N$ as asymmetric errors in $N$. That is, if $\sigma(\ln N) = 1$, then $N$ is measured to within a factor of 2.7, one-sigma. We chose $\sigma(\ln N) = 0.3$ as our criterion for a quality retrieval, equivalent to a fractional error of 30%. With the 23-second (82.5 km) smoothing, the fractional error on $C_2H_6$ stayed below 30% for altitudes below 400 km (Fig 9, lower). Because of the importance of $C_2H_6$ in Pluto's atmosphere, we focused on the results of the 23-second (82.5 km) smoothing for the remainder of the paper. The error analysis suggested that tangent heights over which the retrievals are valid spanned 80-1200 km for $CH_4$, 0 to 600 km for $C_2H_2$, 0 to 650 km for $C_2H_4$, 40-550 km for $C_2H_6$, and 0-350 for the haze. What appeared to be a second range of valid retrievals near 550 to 650 km altitude in the haze is likely to be an artifact of the reduction, as haze was a small contributor to the transmission at those altitudes (Fig. 11).

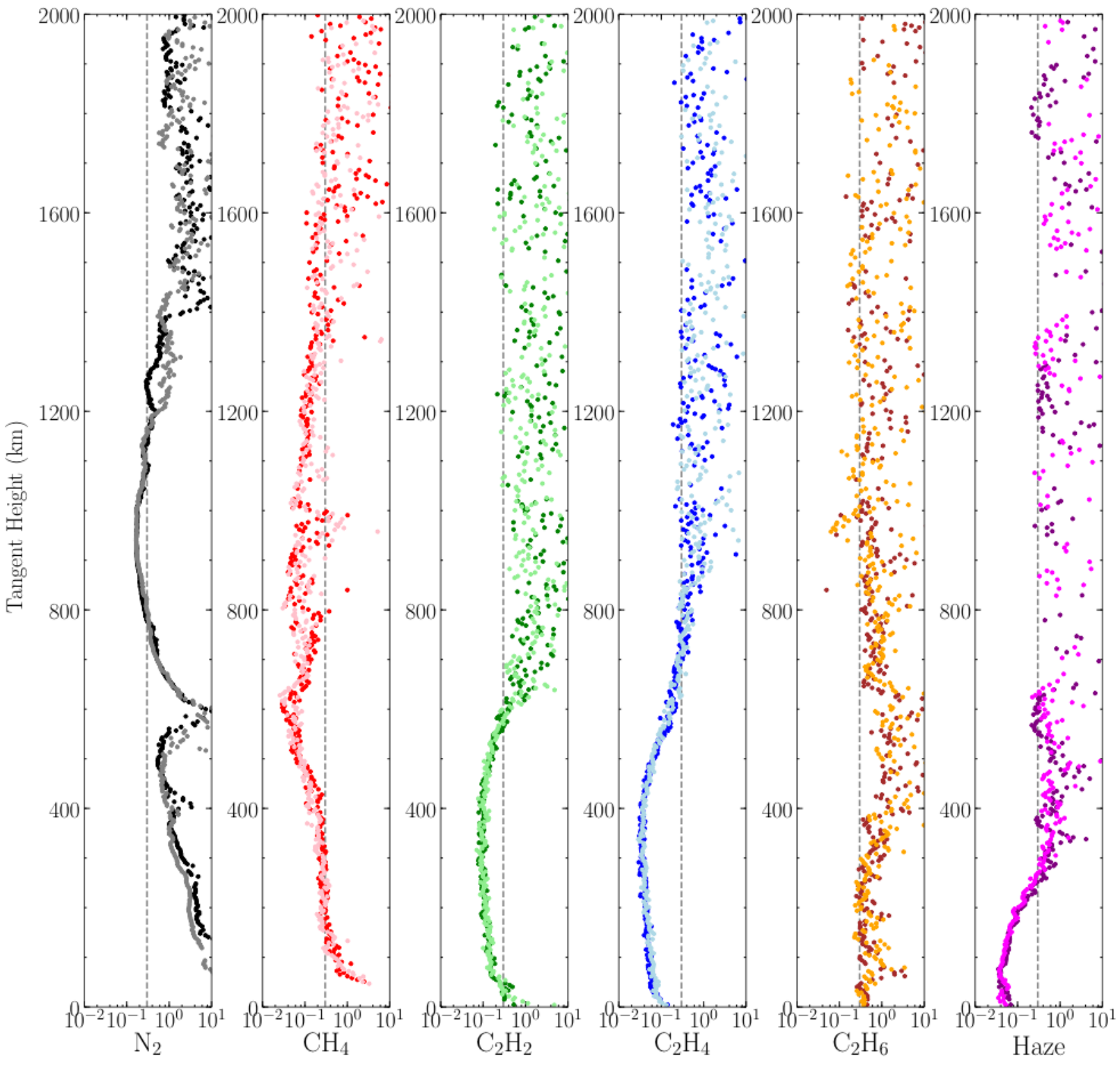

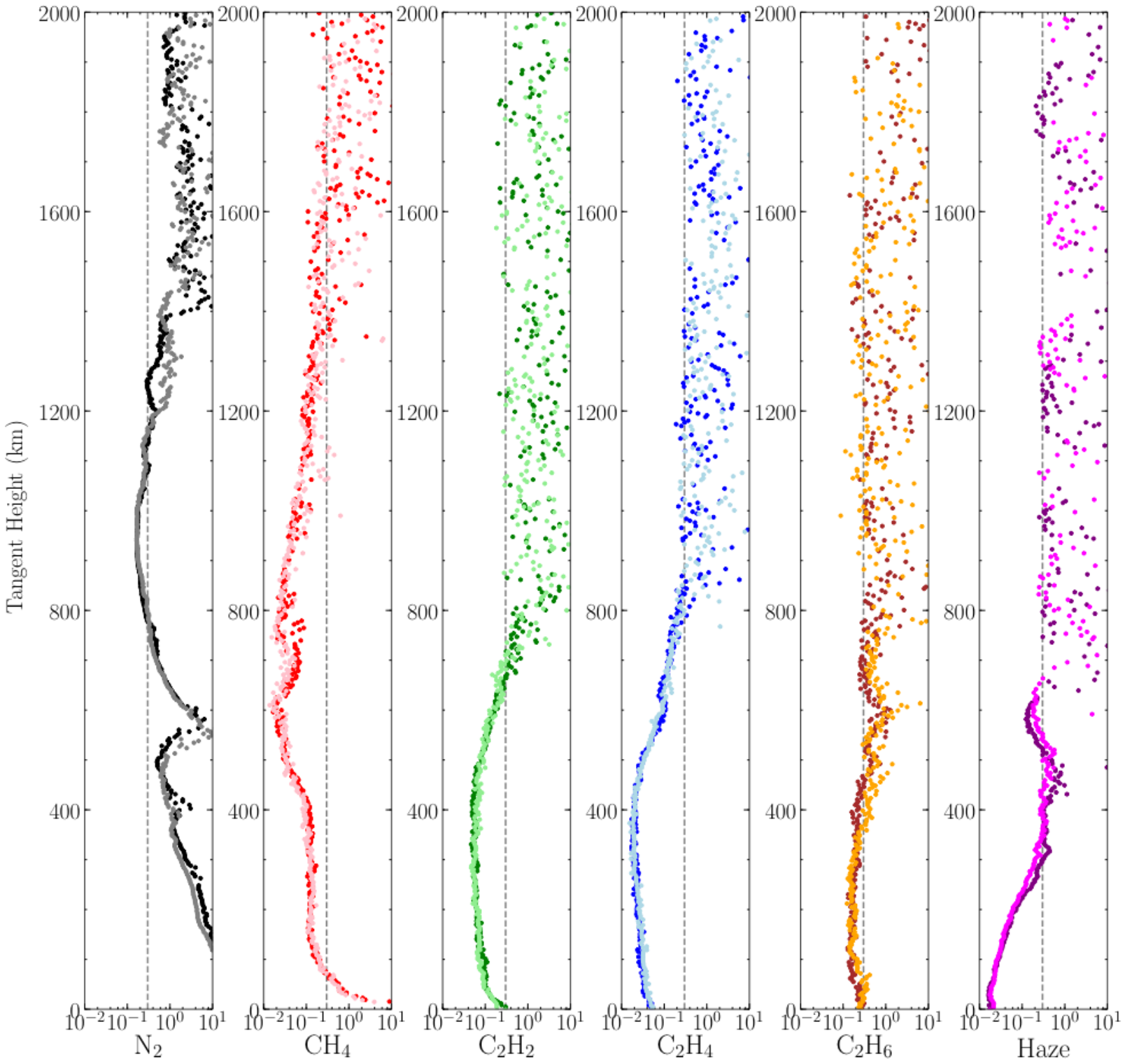

**FIGURE 9**: Fractional errors on the line-of-sight column densities (or, for haze, on line-of-sight optical depth), $\sigma_{\ln(N)} \approx \sigma_N/N$. $N_2$ (black, ingress; gray egress; See Section 5), $CH_4$ (red, ingress; pink, egress), $C_2H_6$ (brown, ingress; mustard egress), $C_2H_2$ (dark green, ingress; light green, egress), $C_2H_4$ (dark blue, ingress; light blue, egress), and haze (dark purple, ingress; light purple, egress). The 30% error criterion is shown as a dashed line. (Upper panel) Errors derived when variable altitude smoothing was used, as in Fig 8, left panel. (Lower panel) Errors derived when a constant altitude smoothing of 23 seconds (82.5 km) was used throughout. These are formal errors; there may be systematic errors of order 10-20% due to the use of cross sections warmer than Pluto's temperature.

The errors in the $C_2H_6$ retrieval were larger than those of $C_2H_2$ and $C_2H_4$. This was a result of the correlation, $c$, (Press et al. 2007) between the different species (Fig 10). $C_2H_6$ line-of-sight number densities were anti-correlated with those of $CH_4$, reaching $c \sim -0.8$ to $-0.85$ at tangent heights of 250 to 400 km, and a second peak of high anti-correlation at tangent heights of 650 to 800 km. $C_2H_4$ and haze were also anti-correlated, with $c \sim -0.8$ at tangent heights of 100 to 700 km.

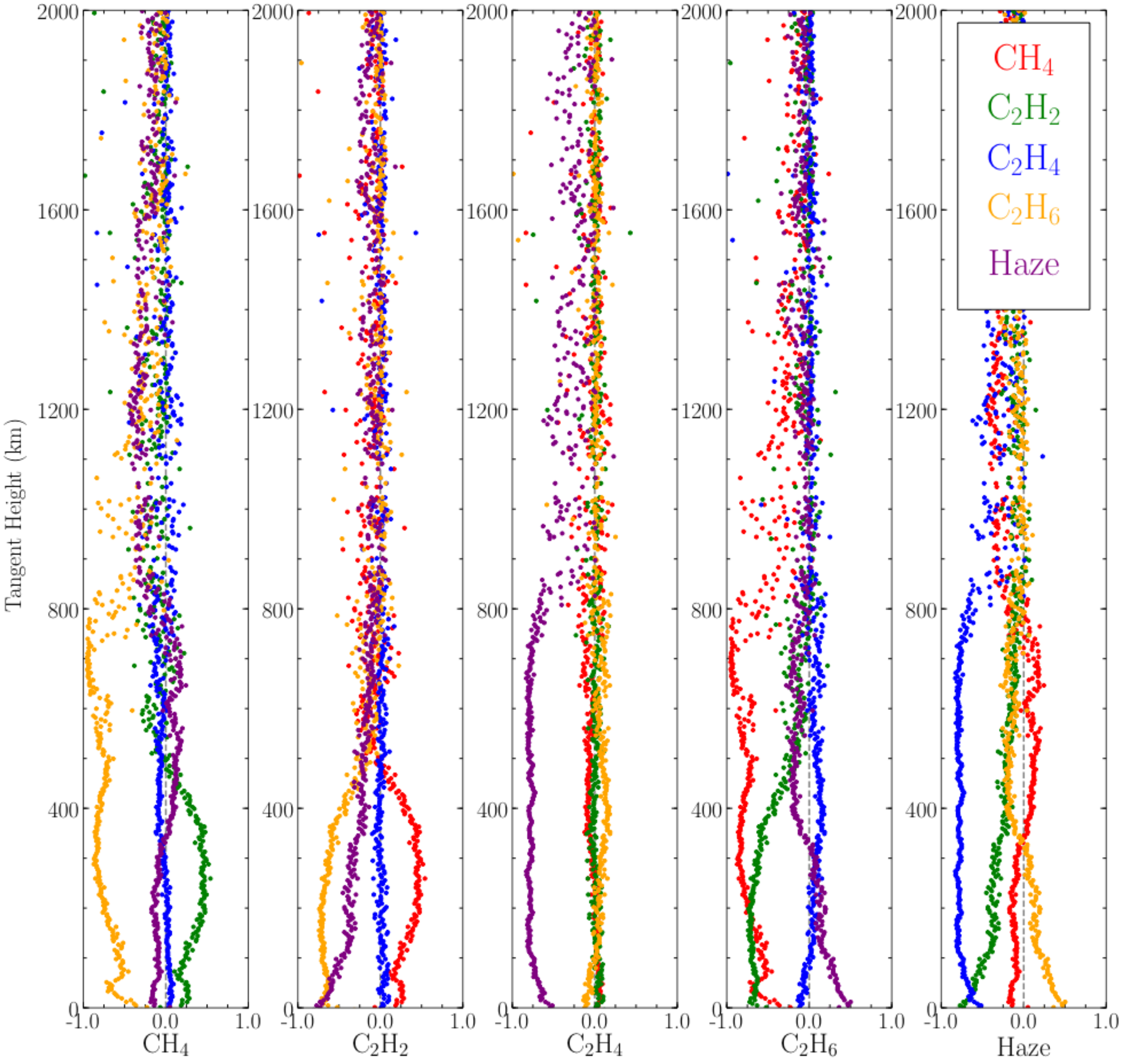

**FIGURE 10**: Pair-wise correlation coefficients, for $CH_4$ (red), $C_2H_6$ (mustard), $C_2H_2$ (dark green), $C_2H_4$ (dark blue), and haze (dark purple). $CH_4$ and $C_2H_6$ are anti-correlated near 250-400 km and above 650 km. Haze and $C_2H_4$ are anti-correlated at 100-700 km. Haze and $C_2H_2$ are anti-correlated just at the surface.

The quality of the fit, the altitudes of validity, and the issues of correlation are illustrated by plots of the transmission vs. tangent height at selected wavelengths (aka *lightcurves*, Fig 11), and by plots of transmission vs. wavelength at selected altitudes (aka *spectra*, Fig 12). $CH_4$ was the major absorber at 100-120 nm. In contrast, the wavelengths between 140 and 145 nm probed the slight offset in the shoulders of $CH_4$ and $C_2H_6$ between 250 and 400 km altitude; the similarity of their transmission spectra (Fig. 7) was the source of the $CH_4$-$C_2H_6$ anti-correlation. $C_2H_2$ had relatively weak absorption at 150 nm and strong absorption at 152 nm; a comparison of these pairs shows that the Pluto solar UV occultation is clearly sensitive to $C_2H_2$. Similarly, $C_2H_4$ had stronger absorption at 165 nm than at 172 nm, and was easily detected until the lowest altitudes, where haze became dominant.

Haze was nearly the sole source of extinction at wavelengths longer than 175 nm. The apparent higher peak in the haze at a tangent height of 600 km (Fig. 8) was not clearly evident in the 178 nm lightcurve (Fig 11). An investigation of the spectrum at 500 km (Fig 12) suggested that the haze measurement at 600 km might have been related to the details of the shape of the $C_2H_4$ absorption at 160 to 175 nm. As the $C_2H_4$ cross-sections that we used were measured at 140 K, we suggest caution in the inference of haze at >350 km altitude.

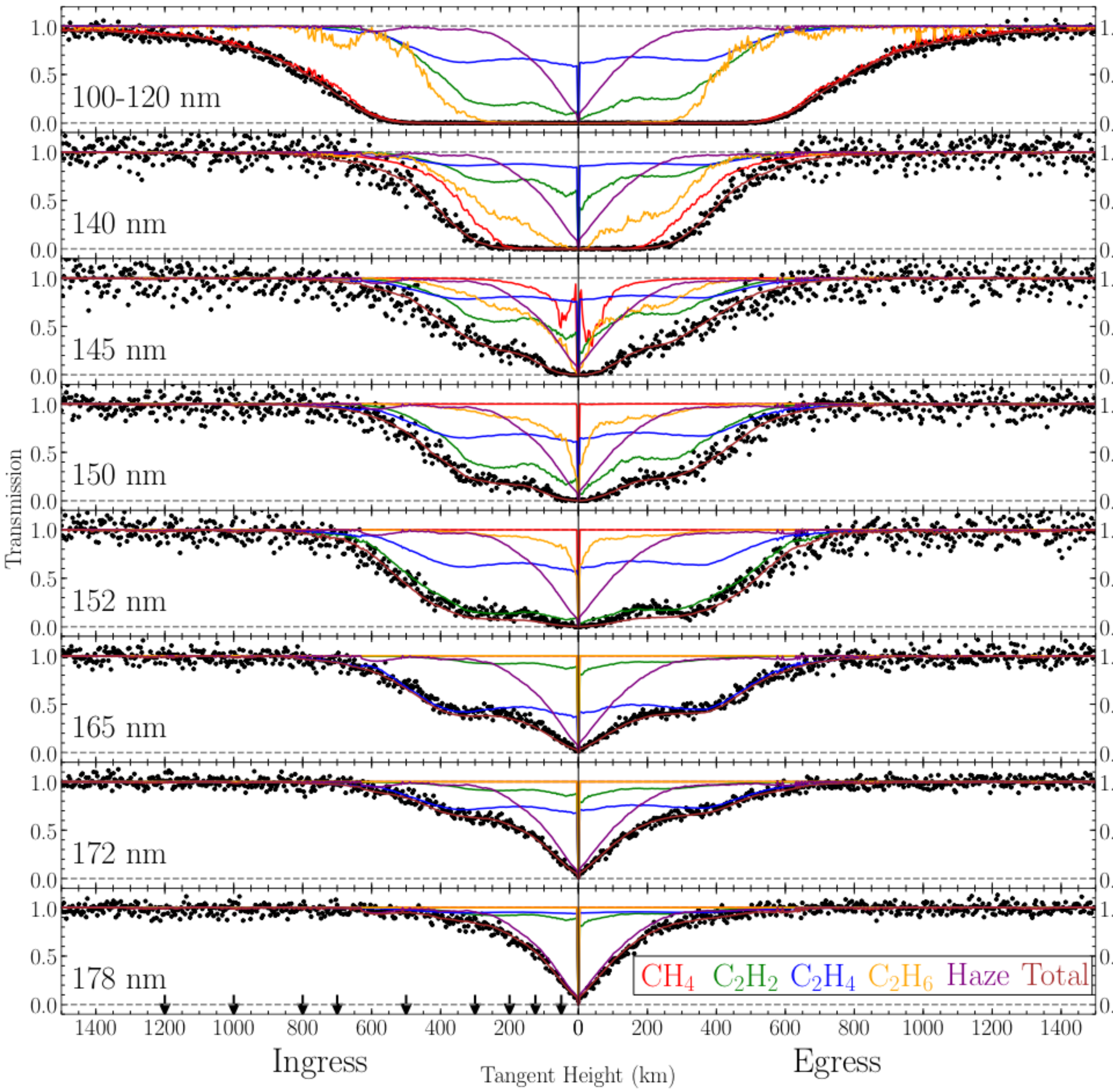

**FIGURE 11**: Models of the transmission lightcurves at selected wavelengths (smooth lines) compared with the observed lightcurves (dots). Both model and data are summed over 9 wavelength bins (1.6 nm) for the plot but were fit at full spectral resolution. Brown: total (all species). Red: $CH_4$. Mustard: $C_2H_6$. Green: $C_2H_2$. Blue: $C_2H_4$. Purple: Haze. Arrows indicate the altitudes shown in Figure 12.

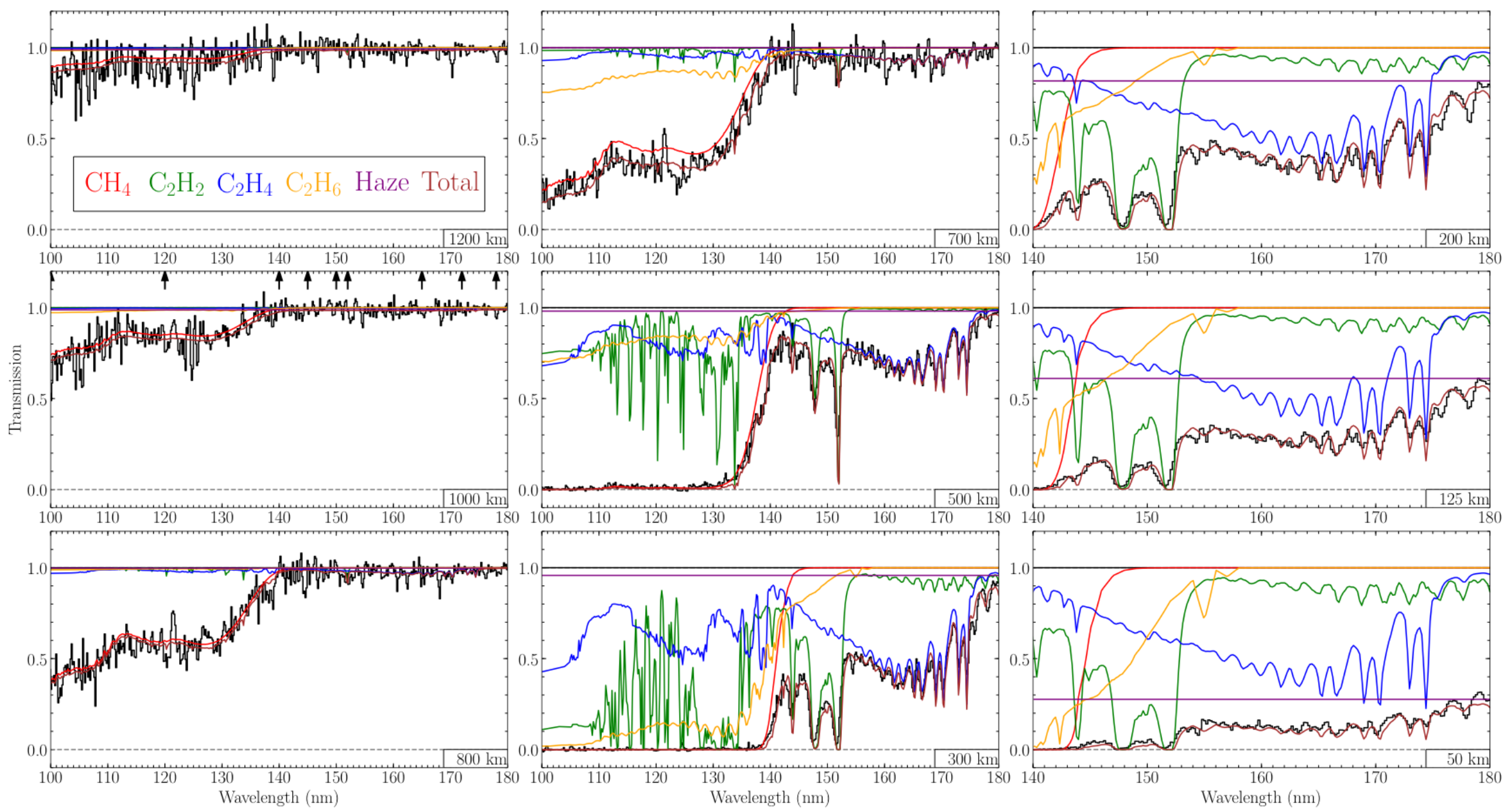

**FIGURE 12**: Comparison of transmission spectra averaged at 23 seconds at selected altitudes for ingress with model transmission based on the 23-second retrieval. Data are shown in black. Smooth lines show modeled transmission. Brown: total. Red: $CH_4$. Green: $C_2H_2$. Blue: $C_2H_4$. Mustard: $C_2H_6$. Purple: Haze. Arrows indicate the altitudes shown in Figure 11.

We found no indication of any HCN absorption above 200 km. Below 200 km, contribution by HCN absorption formally improved the fit, but there were no reliable diagnostic features of HCN present in the data. The CO continuum absorption was indistinguishable from that of $N_2$. We inspected spectra from 700 to 1000 km altitude for evidence of absorption by narrow CO features between 100 and 120 nm, with no convincing detection. Thus, we did not present retrievals of HCN or CO in this paper. This result was consistent with ALMA's detection of HCN with a mixing ratio of only $10^{-8}$ to $10^{-7}$ and CO of only $(5.15\pm 0.04) \times 10^{-4}$ (Lellouch et al. 2017).

These line-of-sight abundances were derived using $C_2H_2$, $C_2H_4$, and $C_2H_6$ cross-sections measured at 150 K, 140 K, and 150 K, respectively. As shown by Gladstone et al. (2016) and discussed in Section 7 of this paper, temperatures in Pluto's atmosphere were near 70 K above ~300 km altitude, or 70-80 K colder than the laboratory temperatures. To investigate the effect of cross-section temperature on the retrievals, we derived line-of-sight abundances using room-temperature cross sections, compared these to the retrievals that used ~150 K cross sections outlined in Section 3, and extrapolated the effect to the 70 K seen in Pluto's atmosphere. Because the temperature dependence of derived line-of-sight abundances may be non-linear with temperature, we only used these results to suggest the magnitude of the temperature effects. We found that the line-of-sight abundance of $C_2H_2$ decreased with decreasing temperature of the cross-section measurement, by 14% per 100 K near 300 km altitude. For $C_2H_4$, line-of-sight abundance also decreased at about 14% per 100 K, from the surface to ~600 km altitude. For $C_2H_6$, the line-of-sight abundance increased with decreasing temperature of the cross-section measurement, by 25% per 100 K, near 300 km altitude. Thus, the line-of-sight abundances for $C_2H_2$ and $C_2H_4$ presented here may be overestimated by ~10% at some altitudes, and the line-of-sight abundances for $C_2H_6$ may be underestimated by ~18%.

## 5. Line-of-sight abundances: Nitrogen

As described in Section 3, the $N_2$ cross section is dominated by continuum absorption shortward of 66.123 nm. At longer wavelengths, the retrieval of $N_2$ is complicated by the interaction of the very narrow ro-vibrational lines of $N_2$ electronic states, the solar spectrum, and the instrumental line-spread function. For this paper, we analyzed the $N_2$ continuum, and deferred analysis of the discrete ro-vibrational $N_2$ absorption spectral region to a later study.

The Alice bandpass that contained the $N_2$ continuum absorption spanned 52 to 66.123 nm (cf. Section 3). We concentrated on the 55-65 nm region (Fig 13). In this region, the $N_2$ and $CH_4$ absorption cross sections changed slowly with wavelength, and the solar flux was dominated by two strong solar lines, at 58.4 nm (He I) and 63.0 (O V), which are also seen as two horizontal turquoise lines near 60 nm at the top Fig 4A. Shortward of 55 nm, the solar flux was low and did not add to the signal-to-noise ratio. Longward of 65 nm, the solar flux included a weak line near 70 nm and increased substantially at 75-80 nm (e.g., the cyan colors in Fig 4A); however, for wavelengths greater than ~67 nm, the $N_2$ absorption cross section is significantly complicated by a forest of ro-vibrational transitions between discrete electronic states.

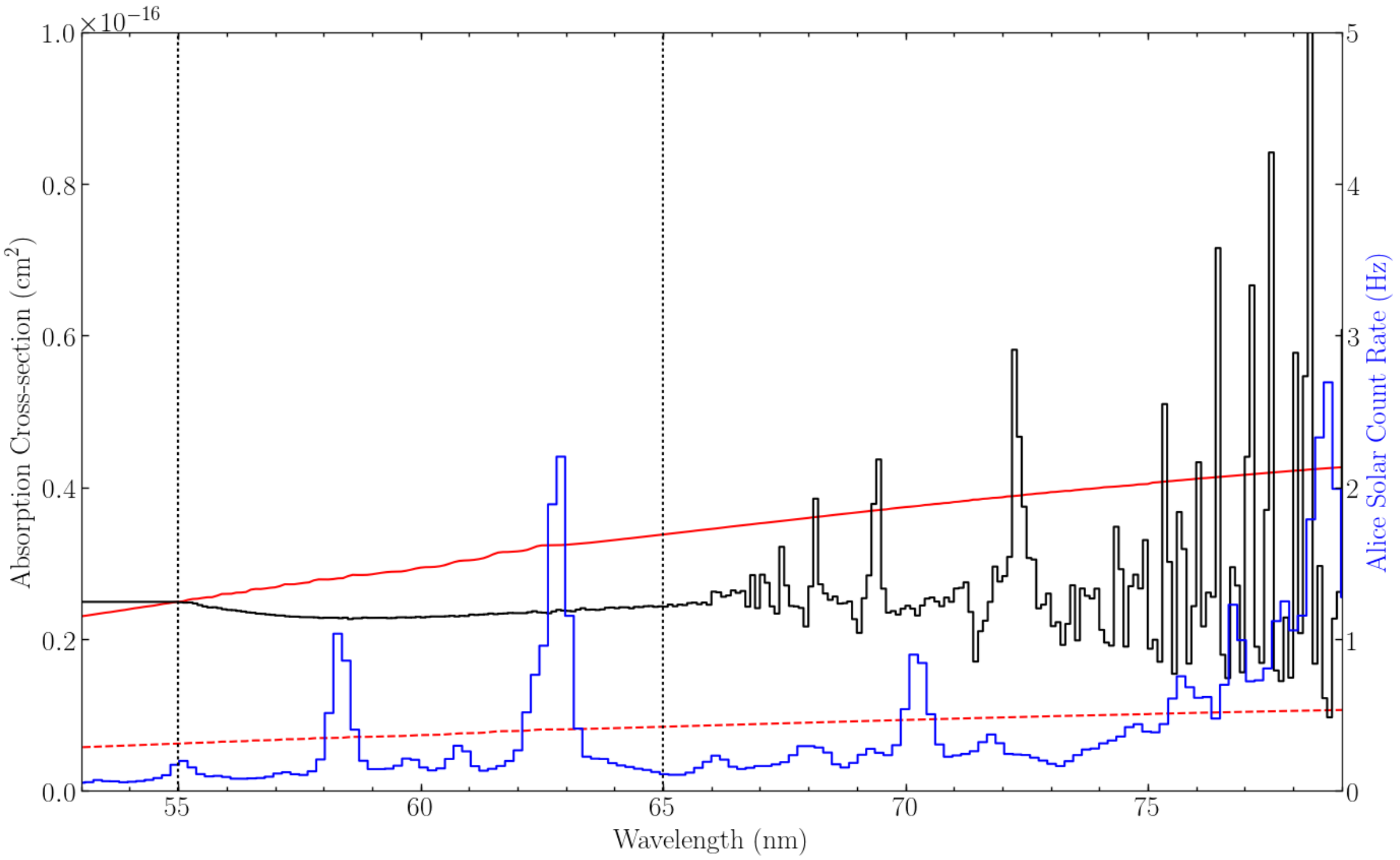

**FIGURE 13**. The observed unocculted solar spectrum (blue) in the 55-65 nm range is dominated by two strong solar lines, the He I 58.4 and O V 63.0 nm lines (Heroux & Higgins 1977). The $N_2$ absorption (black) is nearly constant over this region, as is the $CH_4$ absorption (red, solid). For reference, we also plot the $CH_4$ cross section divided by four (red, dashed), which is roughly scaled to the proportion of column $CH_4$ to $N_2$ at ~1000 km altitude.

The sum of the observed counts from 55-65 nm smoothed in altitude, $D$, and sum of the reference solar spectrum over the same wavelength range and altitude smoothing, $D_{ref}$, were related by the mean transmission, $<Tr>$, via $D = D_{ref} <Tr>$. At the altitudes of interest (above 800 km), the only known absorbers of consequence were $N_2$ and $CH_4$. The mean transmission can be expressed as:

$$\langle Tr \rangle = \frac{\int_{55}^{65} R(\lambda) Tr_{CH4}(\lambda) Tr_{N2}(\lambda) d\lambda}{\int_{55}^{65} R(\lambda) d\lambda(\lambda)} \quad (10)$$

Because the cross sections varied slowly compared with both the solar rate, $R(\lambda)$, and the smoothing kernel, we separated the mean transmission into a $CH_4$ factor and an $N_2$ factor.

$$\langle Tr \rangle \approx \langle Tr_{CH4} \rangle \langle Tr_{N2} \rangle = \left[ \frac{\int_{55}^{65} R(\lambda) Tr_{CH4}(\lambda) d\lambda}{\int_{55}^{65} R(\lambda) d\lambda(\lambda)} \right] \left[ \frac{\int_{55}^{65} R(\lambda) Tr_{N2}(\lambda) d\lambda}{\int_{55}^{65} R(\lambda) d\lambda(\lambda)} \right] \quad (11)$$

The $CH_4$ transmission accounted for some, but not all, of the absorption. The transmission due to $CH_4$ alone, $Tr_{CH4}$, was fixed at the values derived in Section 4. This reduced the problem to one of finding the single parameter, the line-of-sight abundance ($N_{N2}$), to match the average $N_2$ transmission, $<Tr_{N2}>$.

The retrieved $N_2$ line-of-sight abundances are plotted in Fig. 14, and plotted in context with the hydrocarbons and haze in Fig. 8. The fractional errors of $N_2$ are plotted with the other species in Fig. 9. When we applied a criterion of fractional errors < 30% to define the altitudes of valid retrievals, as we did for the hydrocarbons and haze, we found that the $N_2$ retrievals were valid from ~800 to ~1100 km tangent height (Fig 9). However, we felt that it was prudent to restrict the range. This is shown in the transmission vs. tangent height at 55-65 nm (aka *lightcurves*, Fig 15). At altitudes below 800, no information on $N_2$ was derived at these wavelengths. At altitudes between 800 and 900 km, some $N_2$ must have been present to account for the difference between the $CH_4$-only lightcurve and the observations; however, since the transmission was consistent with zero at these altitudes, we could only place lower limits on the $N_2$ line-of-sight column. The "roll over" in the $N_2$ line-of-sight abundance between 900 and 800 km tangent height was an indication of the difficulty in retrieving line-of-sight abundances at these altitudes. Between 900 to 1500 km tangent height, we derived $N_2$ line-of-sight column at each averaged point. Above 1500 km tangent height, we only placed upper limits on the $N_2$ line-of-sight column.

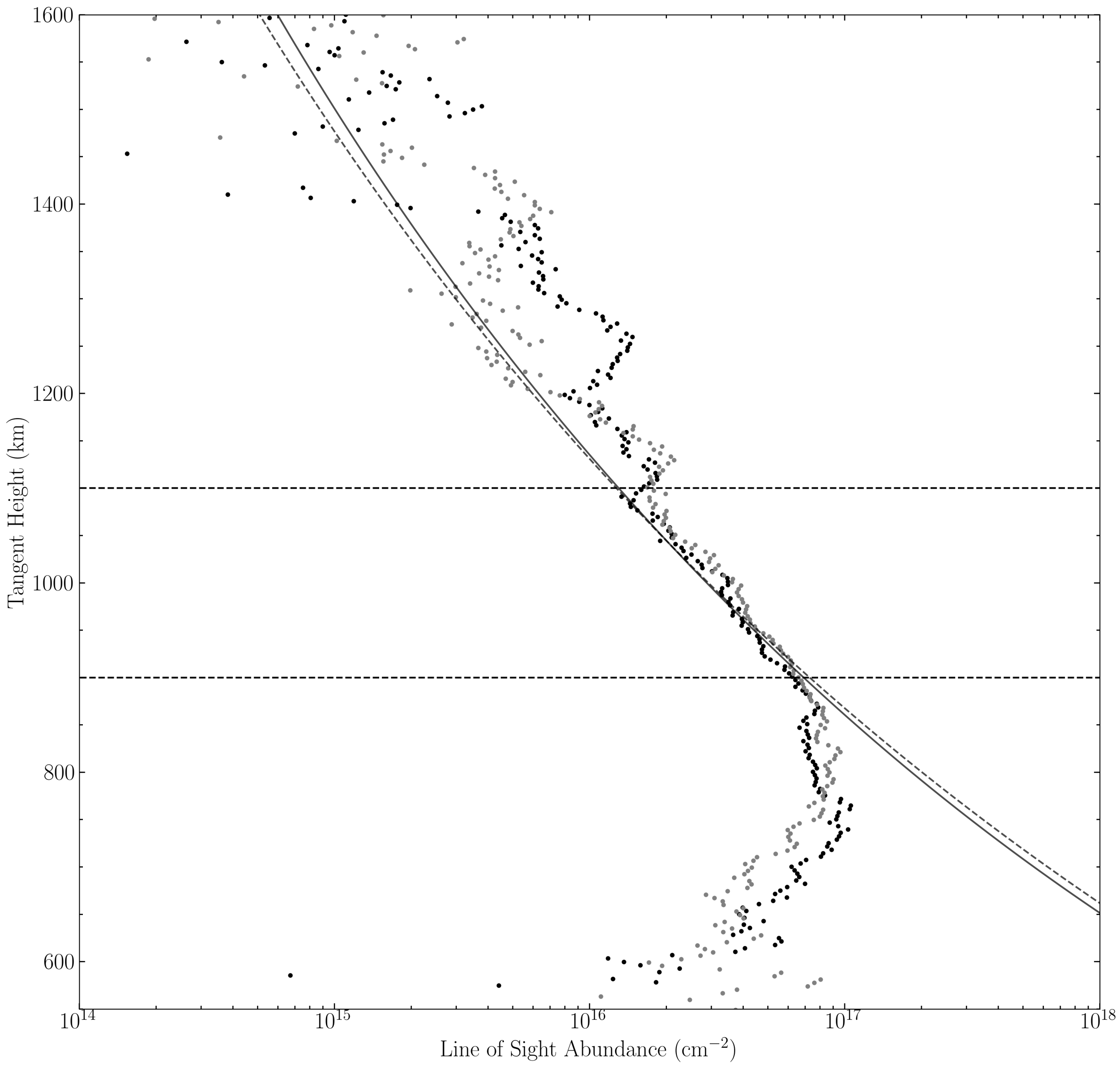

**FIGURE 14**. Line-of-sight abundance of $N_2$ (black, ingress; gray, egress). The valid altitudes of $N_2$ retrieval are 900-1100 km, between the horizontal dashed lines. Also plotted are two models of $N_2$ line-of-sight abundance constructed in Section 7, the nominal model profile (solid) and the flux-constrained model (dashed), which are nearly identical within the region of validity.

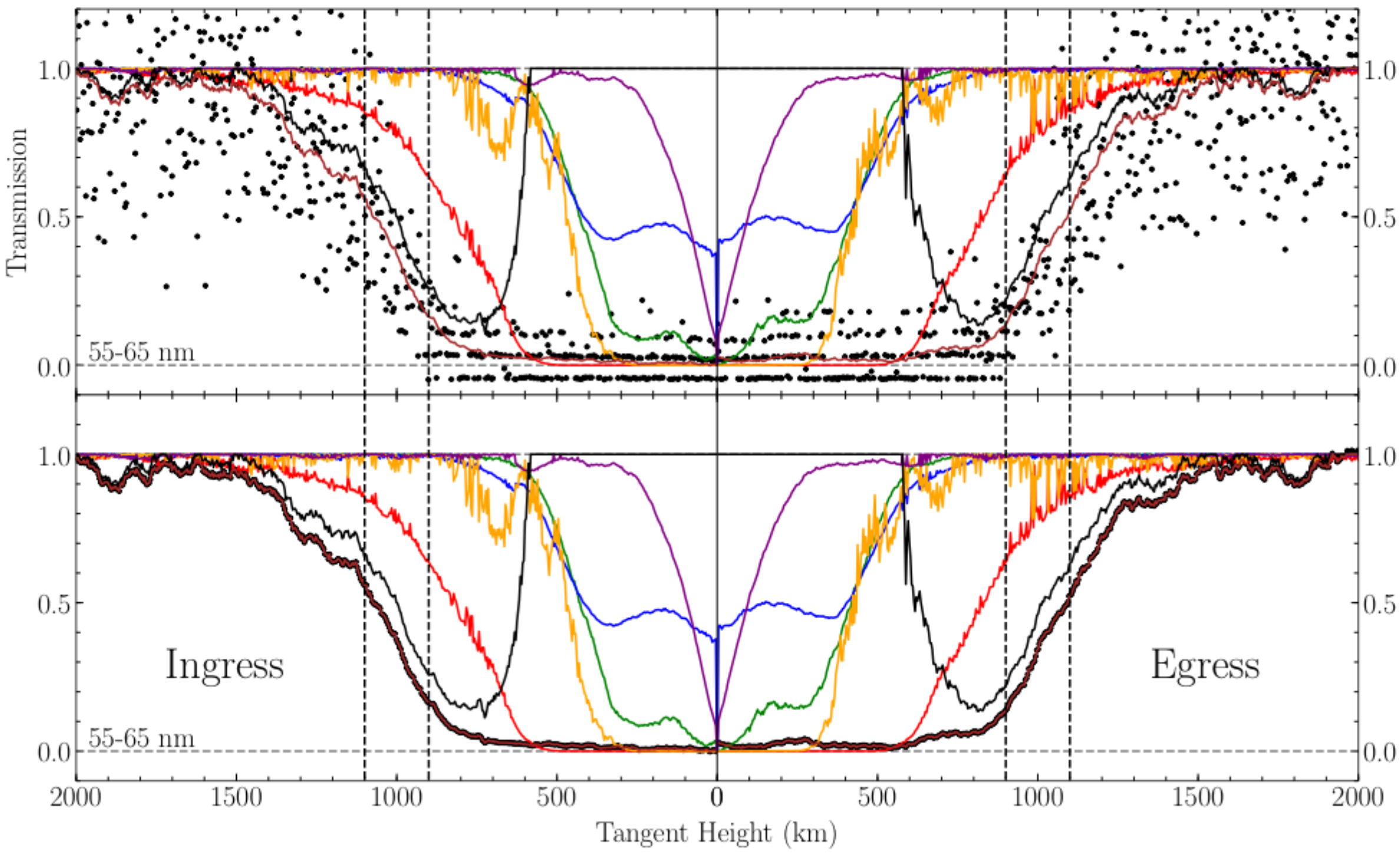

**FIGURE 15**: Models of the transmission lightcurves at 55-65 nm (smooth lines) compared with the observed spectra (dots). Brown: total (all species). Black: $N_2$. Red: $CH_4$. Mustard: $C_2H_6$. Green: $C_2H_2$. Blue: $C_2H_4$. Purple: Haze. The upper plot shows the transmission at 1-second resolution. The digitization is obvious at the lower baseline, and was one of the motivators for our choosing the range of validity for the $N_2$ retrieval. The lower plot shows the same model, with data averaged at 23 seconds. In both panels, the minimum in the $N_2$ transmission at ~800 km on ingress and egress is another indication of the altitude limits of the valid $N_2$ retrievals (900-1100 km), between the vertical dashed lines.

## 6. Local Number Density

The line-of-sight abundance, *N*, is the integral of the local number density, *n*, along the line-of-sight (Fig 6 and Eq. 6). Under certain assumptions, one can invert this relationship to derive *n* given *N*. By far, the most common assumption is that of local spherical symmetry (that is, within the region where the ray path intersects Pluto's atmosphere). For Pluto's upper atmosphere, spherical symmetry appeared to be a very good assumption. In the Alice Pluto solar occultation data itself, there was close agreement between the ingress and egress lightcurves, and nearly identical profiles for ingress and egress for the line-of-sight abundances of all species considered here (c.f. Fig. 8). Similarly, the REX Pluto Earth radio occultation (Hinson et al. 2017) showed similar density profiles for ingress and egress above 30 km altitude. For these two New Horizons datasets, the ingress and egress latitudes differed by only ~30°, raising the possibility that the similarity is simply due to ingress and egress probing similar latitudes. Much wider latitude ranges are typically probed by ground-based stellar occultations of Pluto's atmosphere, including one just two weeks prior to the New Horizons flyby (Sicardy et al. 2016). Analyses of the main occultation drop in ground-based stellar occultations (probing ~10 km to 400 km altitude) do not show evidence for statistically significant ellipticity (Person 2001) or only very rare cases of statistically significant dawn/dusk or summer/winter differences (Zangari 2013). These observations perforce only probed dawn/dusk asymmetries, and not noon/midnight asymmetries. However, there are robust theoretical reasons to expect very small horizontal variations in temperature over all the body (at a given pressure level) in the part of the atmosphere not influenced by topography because of the very long radiative timescale of the atmosphere. This is predicted by 3D Global Circulation Models even when methane is not horizontally well mixed (Toigo et al. 2015, Forget et al. 2017).

In contrast with its upper atmosphere, Pluto's lower atmosphere did show evidence for spatial variability. The REX Pluto Earth radio occultation below 30 km showed density, temperature, and pressure differences interpreted as being associated with planetary boundary layer effects and underlying surface temperatures (Hinson et al. 2017). Central flashes observed in ground-based stellar occultations (Olkin et al. 2015) implied Pluto's lower atmosphere was not spherically symmetric. The haze imaging by the LORRI instrument at New Horizons (Cheng et al. 2017) showed a latitudinal variation in the haze brightness in the lowest 200 km of Pluto's atmosphere. We proceed with the assumption of spherical symmetry, with the caveat that this might lead to inaccuracies in derived number densities in the lowest 30 km.

Given a spherically symmetric atmosphere, the classic method for deriving local number density from line-of-sight abundance is the Abel transform (Roble & Hays 1972).

$$n(r) = -\frac{1}{\pi}\int_r^\infty \frac{\left[dN(r')/dr'\right]}{\sqrt{r'^2 - r^2}}dr' \tag{12}$$

The net effect of the derivative and the integral in Eq. 12 is essentially a "half derivative" (e.g., Young 2009), which amplifies the noise in the profile. There are various methods for dealing with this noise, for example by imposing functional forms (Roble & Hays 1972) or by imposing a smoothness constraint using a Tikhonov regularization (Quémerais et al. 2006). Because we already smoothed our data over 23-second (82.5-km) intervals, and imposed a quality constraint of < 30% errors to define the valid altitudes, we used the unmodified Abel transform technique. For the haze analysis, the relationship between line-of-sight optical depth, τ, and local extinction coefficient, ε, was mathematically equivalent to the relationship between *N* and *n*.

The Abel transform formally includes an integral to infinity. Above the highest valid data point, we needed to specify the atmosphere. We did this by defining an altitude region over which we fit a functional form *N*(*r*), and then extrapolated the function to a radius where the contributions to the integral in Eq. 12 were negligible (in practice, this was taken to be 2000 km altitude). The higher altitude of the fitting region, $h_1$, was near the top of the region of valid retrievals of the line-of-sight abundance, as determined by the errors in the lower panel of Fig. 9. We fit to altitudes sampled every 23 seconds (~82.5 km) to avoid fitting to correlated points, so the lower altitude of the fitting region, $h_0$, was a multiple of 82.5 km below $h_1$. This step required some judgment, to balance using a larger region to better constrain the fits, while using a smaller region to avoid non-exponential changes in the profile. For example for $C_2H_6$, we kept our region above the roll-over at ~270 km altitude (Fig 8), choosing also to include a point above the 30% cut-off used to define the altitudes of valid line-of-sight abundance retrieval.

**Table 2 Altitudes of retrieval and extrapolation**

| Species | | Altitudes of Valid Line-of-sight Abundance Retrieval (km) | Altitudes of Fitting Region, $h_0$ - $h_1$ (km) | $N_0$, molecule/cm$^2$ | $H_0$, km |
|---|---|---|---|---|---|
| $N_2$ | Ingress | 900-1100 | 900-1066 | (6.37±0.98) x $10^{16}$ | 113.7±23.9 |
| | Egress | 900-1100 | 900-1065 | (6.79±1.09) x $10^{16}$ | 117.8± 25.8 |
| $CH_4$ | Ingress | 80-1200 | 869-1200 | (17.81±0.53) x $10^{15}$ | 146.5±6.5 |
| | Egress | 80-1200 | 868-1200 | (18.15±0.62) x $10^{15}$ | 154.4±9.4 |
| $C_2H_6$ | Ingress | 40-550 | 270-600 | (1.29±0.28) x $10^{17}$ | 74.8±9.6 |
| | Egress | 40-550 | 270-600 | (1.23±0.17) x $10^{17}$ | 64.0±6.2 |
| $C_2H_2$ | Ingress | 0-600 | 435-600 | (2.28±0.13) x $10^{16}$ | 63.1±7.9 |
| | Egress | 0-600 | 434-600 | (2.20±0.20) x $10^{16}$ | 64.7±5.1 |
| $C_2H_4$ | Ingress | 0-650 | 485-650 | (10.95±0.28) x $10^{15}$ | 91.6±7.2 |
| | Egress | 0-650 | 484-650 | (10.43±0.31) x $10^{15}$ | 94.1±9.1 |
| Hazes | Ingress | 0-350 | 184-350 | (2.38±0.14) x $10^{14}$ | 69.2±10.9 |
| | Egress | 0-350 | 183-350 | (2.60±0.14) x $10^{14}$ | 67.7±6.3 |

Within the extrapolation altitudes, we fit a simple function that assumed that each species has a constant ratio of temperature, *T*, to molecular weight $\mu$. We fit for two parameters at the lower extrapolation altitude: the line-of-sight abundance, $N_0$, and the scale height $H_0$. The scale height increases proportionally with the square of the radius, $r = r_s + z$, due to variable gravity. The number density, *n*, is an exponential in geopotential (Eq. 13). In order to define the scale height and number density at $r_0$, we define a new geopotential referenced to $r_0 = r_s + h_0$, $\xi' = (r - r_0)\, r_0/r$ (or, for $N(r')$, substitute $r'$ for *r*). Because the scale height was not small compared with the radius on Pluto, the line-of-sight column, *N*, was nearly but not quite an exponential in geopotential (Eq. 13; Young 2009).

$$\left.\begin{aligned} H &= H_0 \frac{r^2}{r_0^2} \\ n &= n_0 e^{-\frac{\xi'}{H_0}} \\ N &= N_0 e^{-\frac{\xi'}{H_0}} \left(\frac{r}{r_0}\right)^{3/2} \frac{1+\frac{9H}{8r}}{1+\frac{9H_0}{8r_0}} \end{aligned}\right\} \text{,for constant } \frac{\mathrm{T}}{\mu} \tag{13}$$

The resulting fits to $N_0$ and $H_0$ are tabulated in Table 2. The points in the extrapolated altitudes and the resulting functional fit are plotted in Fig. 16.

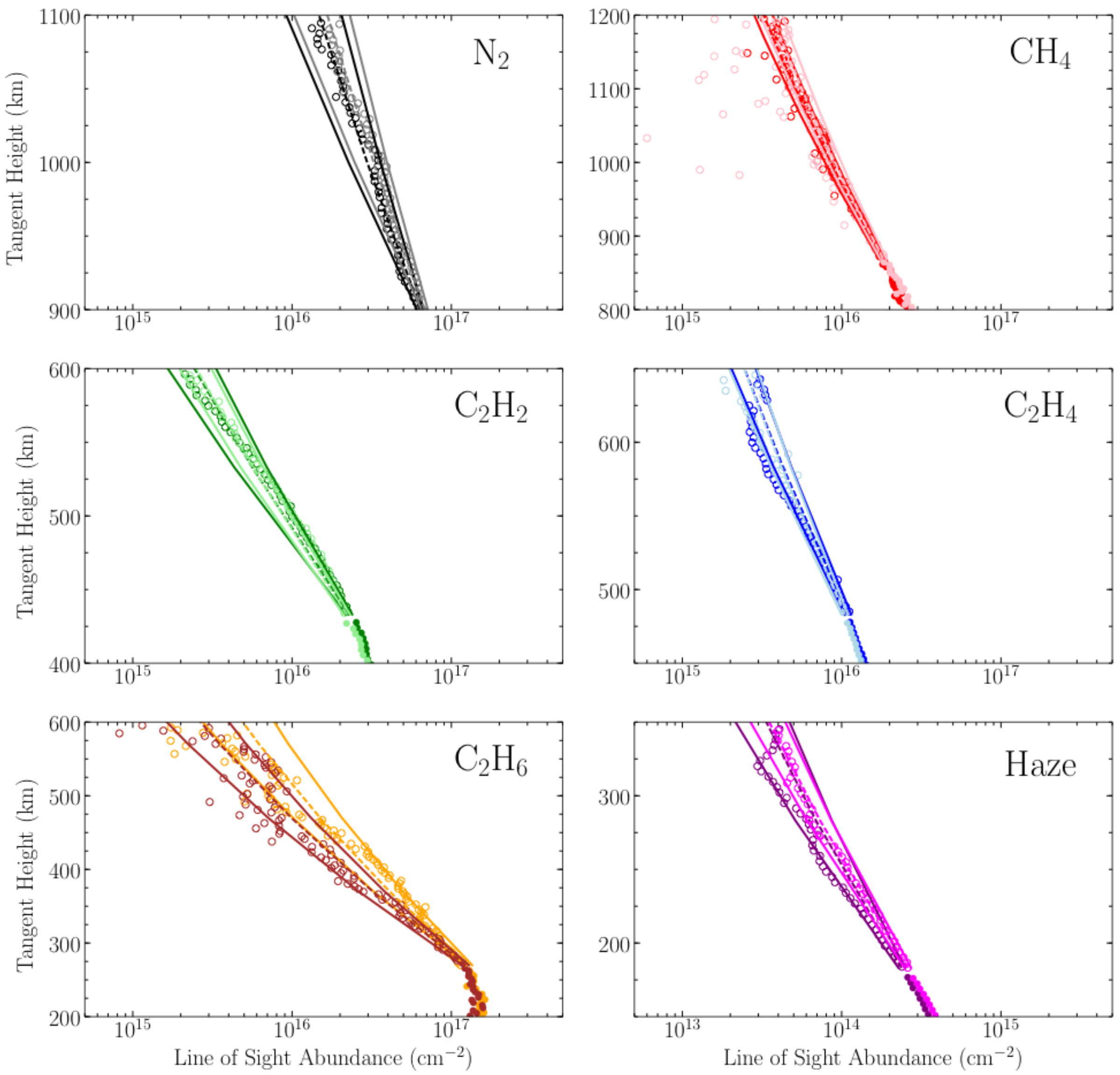

**FIGURE 16**: Fits of a simple function (constant $T/\mu$, see text) to data within the fitting altitude range. Minor ticks on the x-axis are plotted at 2, 4, 6, and 8 times the major tick values. Open dots are points in the fitting altitudes. Filled dots are in the retrieval altitudes. Dashed lines indicate the fitted function. Solid "funnels" around the dashed lines indicate the error on the fitted function. $N_2$ (black, ingress; gray egress), $CH_4$ (red, ingress; pink, egress), $C_2H_6$ (brown, ingress; mustard, egress), $C_2H_2$ (dark green, ingress; light green, egress), $C_2H_4$ (dark blue, ingress; light blue, egress), and haze (dark purple, ingress; light purple, egress).

We used the functional form established from the fitting altitude range to calculate the densities up to 2000 km for use with the Abel transform (Eq. 12). Fig. 17 shows the line-of-sight abundances (top) and the derived densities (bottom).

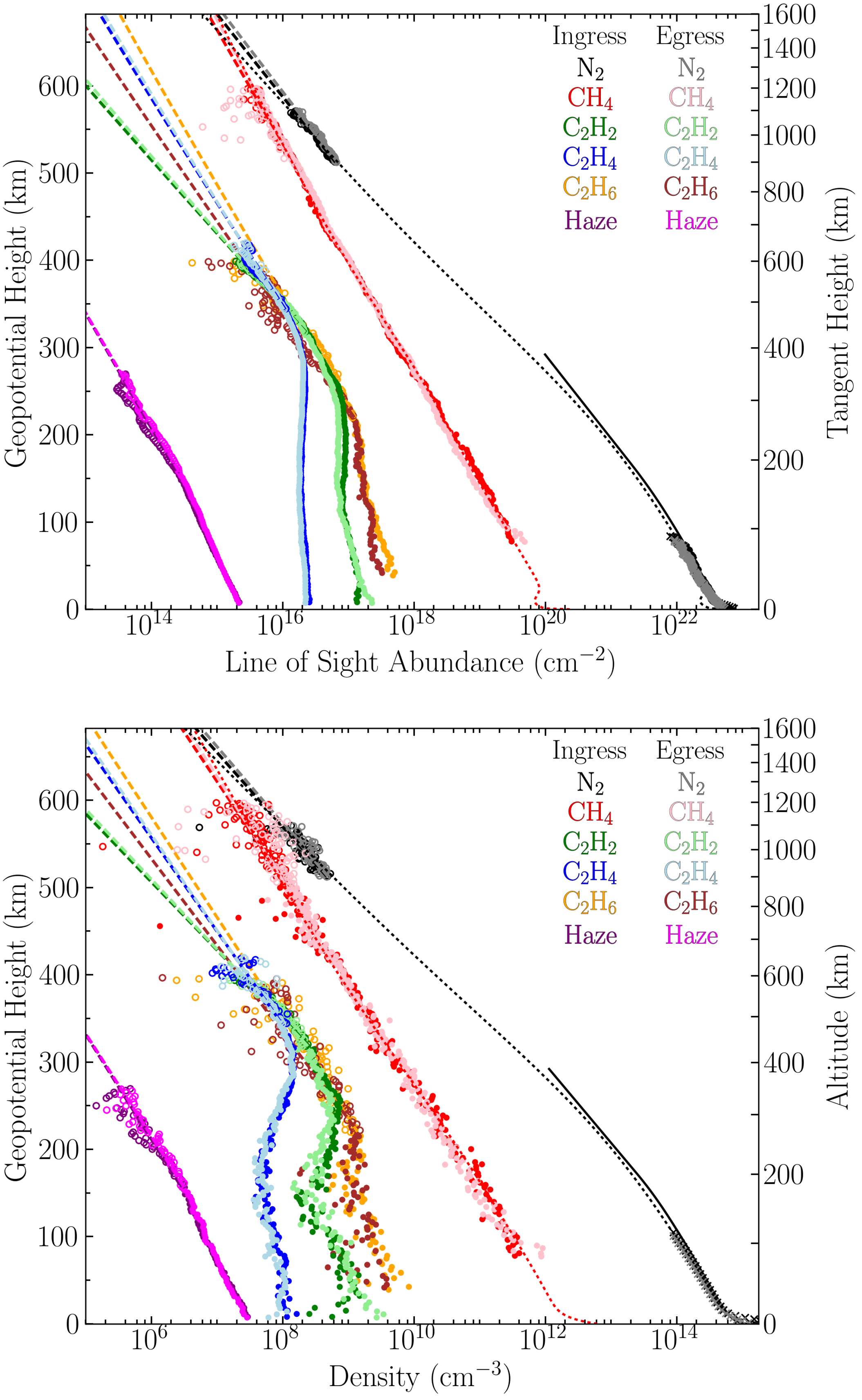

**FIGURE 17**: Line-of-sight abundance (top) and local density (bottom) plotted linearly with geopotential height on the left axis and tangent height or altitude on the right axis. Minor ticks on the x-axis are plotted at 2, 4, 6, and 8 times the major tick values. Open circles represent the points used to derive a functional approximation to the line-of-sight abundance. Dashed lines represent the fitted function, which is linear vs. geopotential for density, and nearly linear for line-of-sight abundance. $N_2$ (black, ingress; gray egress), $CH_4$ (red, ingress; pink, egress), $C_2H_6$ (brown, ingress; mustard, egress), $C_2H_2$ (dark green, ingress; light green, egress), $C_2H_4$ (dark blue, ingress; light blue, egress), and haze (dark purple, ingress; light purple, egress). Haze values are plotted as the optical depth times $10^{15}$ $cm^{-2}$ (top) or the extinction coefficient multiplied by $10^{15}$ $cm^{-2}$ (bottom). Also plotted are the nominal model for $CH_4$ and $N_2$ (dotted lines; See Section 7), and measurements of $N_2$ from the REX radio occultation below 110 km altitude (black crosses, ingress; gray crosses, egress) and the 2015 ground-based stellar occultation (solid black line).

Errors on the density were calculated by propagation of errors on the line-of-sight abundances (Fig. 18; Quémerais et al. 2006). As expected, the errors on the density were larger than those on the line-of-sight abundance. Whereas the altitudes on the line-of-sight abundance were selected to hold fractional errors to $<$ 30%, the fractional errors on the densities for many of the species were near one. This means they are only measured to within a factor of ~2.7, or exp(1). See Section 4 for a discussion on the interpretation of the fractional error.

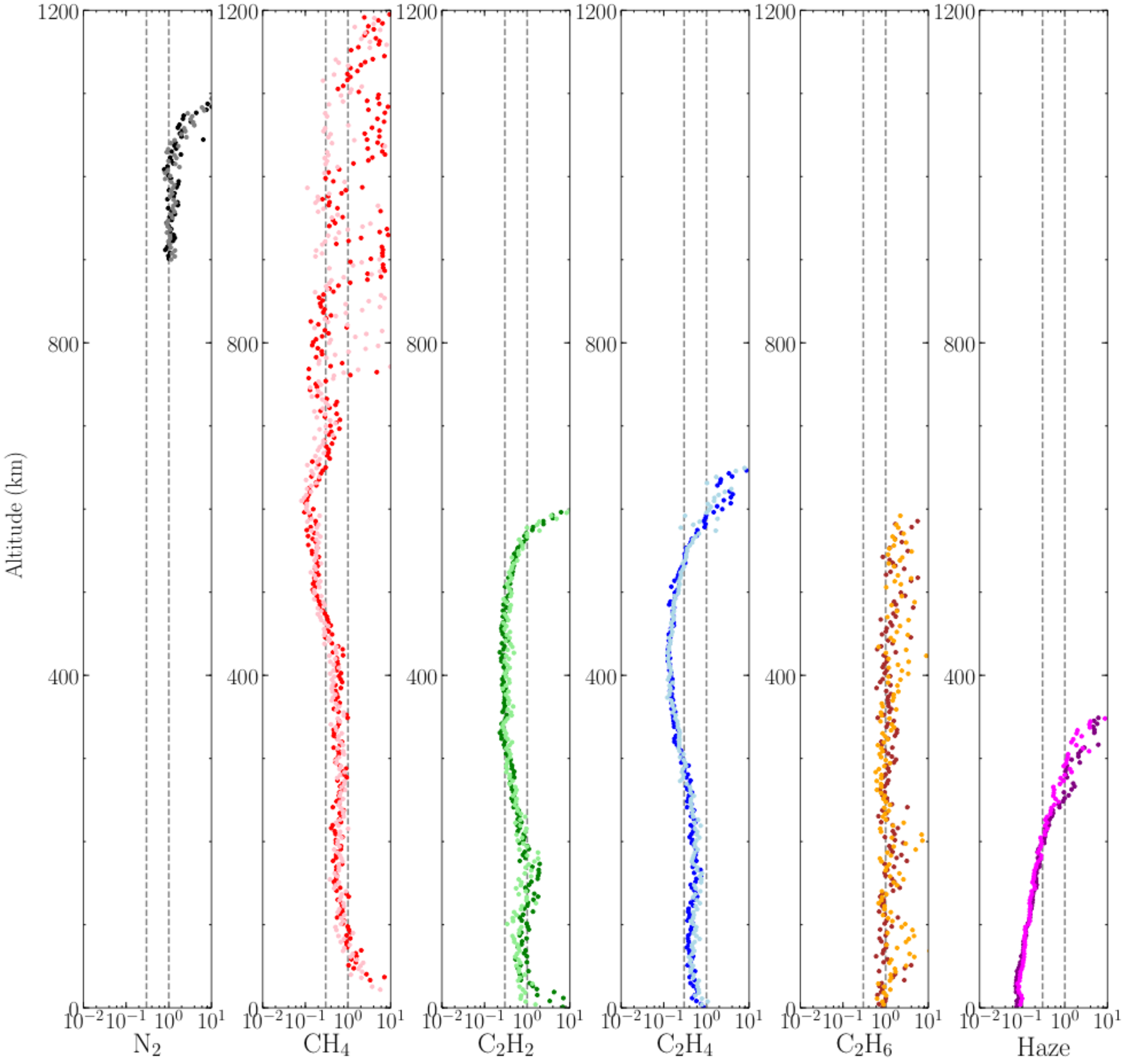

**FIGURE 18**: Fractional errors in local number density (or, for haze, in extinction coefficient), $\sigma_{\ln(n)} \approx \sigma_n/n$. $N_2$ (black, ingress; gray egress; See Section 5), $CH_4$ (red, ingress; pink, egress), $C_2H_6$ (brown, ingress; mustard egress), $C_2H_2$ (dark green, ingress; light green, egress), $C_2H_4$ (dark blue, ingress; light blue, egress), and haze (dark purple, ingress; light purple, egress). The $\sigma_{\ln(n)}$ = 0.3 and 1.0 errors are shown as dashed lines.

## 7. Temperatures and mixing ratios

If the density of $N_2$ were directly measured from Pluto's surface to 1200 km altitude, then pressures could be derived from hydrostatic equilibrium, temperatures could be directly measured from the scale height, and mixing ratios could be derived by simple ratios of densities. However, as shown in Fig. 17, this was not possible without some modeling to interpolate between the $N_2$ density measured by the Alice solar occultation and that measured by the REX radio occultation.

The altitude of unit slant-path optical depth for $N_2$ at 50-100 nm was much closer to that for $CH_4$ in the Alice Pluto solar occultation than expected pre-encounter (Stern et al. 2015, Gladstone et al. 2016). This meant that the $CH_4$ had much more influence on the $N_2$ retrieval than expected. In particular, the $CH_4$ continuum absorption became significant at 600-800 km altitude at the wavelengths of the $N_2$ continuum. The $N_2$ line-of-sight abundance was measured at 900-1100 km altitude, spanning about two scale heights. This range was large enough to attempt to directly measure the scale height, assuming a constant temperature (Eq. 13; Table 2). From this, we concluded that, if Pluto's upper atmosphere reaches isothermal temperatures at ~1000 km, then temperatures as measured solely by the Alice data were 76±16 K at ingress, and 79±17 K at egress (Fig 19). These errors were only weakly constraining, and this analysis made no attempt to connect the upper atmosphere at 1000 km altitude to the lower atmosphere. Fortunately, we could do better.

Other means were needed to understand Pluto's pressure and temperature profile. For this we relied on the simultaneous REX radio observation that probed nearer Pluto's surface (Hinson et al. 2017), and simple physical models of the $N_2$ and $CH_4$ density profiles which included (i) a hydrostatic model for $N_2$, and a simple physical model for the $CH_4$ density profile that included mixing and diffusive separation and (ii) a numerical calculation with mixing, diffusive separation, escape and simple photochemistry. We also compared results with the fortuitous ground-based stellar occultation that was widely observed just two weeks prior to the New Horizons flyby (e.g., Sicardy et al. 2016).

The analysis proceeded as in Gladstone et al. (2016), with updated values of $N_2$ line-of-sight abundance at 900-1100 km altitude (Section 5), $CH_4$ line-of-sight abundance at 80-1200 km altitude (Section 4), and $N_2$ line-of-sight abundance from the surface to 110 km derived from the New Horizons REX occultation (Hinson et al. 2017).

In the first step, we constructed a temperature profile in order to interpolate between the Alice and REX $N_2$ line-of-sight abundances, as in Gladstone et al. (2016). This started with the REX inferred temperature profiles where they are the same for both ingress and egress (~30 km altitude) and extrapolated the temperature up to the Alice $N_2$ line-of-sight abundances beginning at 900 km altitude in a trial and error iterative process that yielded a best fit to the Alice data. A constructed, analytic form of the temperature profile was used to derive an $N_2$ pressure profile, assuming hydrostatic equilibrium; the pressure profile was normalized at 32.4 km altitude ($r$ = 1222.4 km) where the REX ingress and egress pressures were both 5.14 μbar. The pressure and temperature profiles were combined to produce a local $N_2$ number density profile, and then integrated to produce a line-of-sight abundance profile for $N_2$, which was compared with both the Alice and REX data.

The constructed, analytic temperature profile was defined using sums of the log of ratios of hyperbolic cosines (Eq. 12, Lindzen and Hong 1974), a form mainly chosen because the temperature structure is infinitely differentiable with no discontinuities. The temperature profile depends on the surface temperature, $T_s$; the altitudes, $Z$, and widths, $\delta$, that define the shape of the terms within the sum; and 10 additional parameters, $C$ (Table 3). These 11 free parameters ($T_s$ and $C_0$.. $C_9$) are constrained by the REX occultation at altitudes below ~100 km, and by the Alice $N_2$ measurement near 900 km, and assumed to trend to isothermal above ~900 km. By trial and error, the inferred temperature profile illustrated in Fig. 19 and the parameters given in Table 3 yielded the best fit to the Alice data, while anchored by the REX temperature profile. We refer to this as the nominal temperature profile, with a trend toward isothermal temperatures of $T_{inf}$ = 67.8 K at high altitudes. A second profile with $T_{inf}$ = 65 K, constrained by conservation of $CH_4$ flux, is presented later.

The nominal model of $N_2$ line-of-sight abundances and number densities are plotted, along with the measured values from the Alice solar occultation, in Fig. 17. This joint analysis suggested that the upper boundary condition for the REX temperature profile of 95.5 K (Hinson et al., 2017) at an altitude of 112.4 km ($r$ = 1302.4 km) adopted from Sicardy et al. (2016) was too high and a value of 93.9 K was more consistent with the trend to a minimum temperature of 62.4 K at an altitude of 470 km than the asymptotic temperature of 80 K used by Sicardy et al. (2016).

$$T(z) = T_s + C_0\frac{h}{2} + \frac{1}{2}\sum_{i=0}^{8}\delta_i\left(C_{i+1} - C_i\right)\ln\left[\frac{\cosh\left(\frac{z - Z_i}{\delta_i}\right)}{\cosh\left(\frac{Z_i}{\delta_i}\right)}\right] \tag{14}$$

**Table 3: Terms for Nominal Constructed Analytic Temperature Structure**

| index, *i* | *Z*, km | δ, km | *C*, K/km |
|---|---|---|---|
| 0 | 8.51 | 10 | 7.850325 |
| 1 | 15 | 40 | -0.548625 |
| 2 | 100 | 70 | -0.099750 |
| 3 | 200 | 125 | -0.199500 |
| 4 | 280 | 115 | -0.169575 |
| 5 | 440 | 100 | -0.019950 |
| 6 | 800 | 135 | 0.019950 |
| 7 | 1200 | 200 | 0.000000 |
| 8 | 1500 | 440 | 0.000000 |
| 9 | 7400 | 100 | -0.009975 |

$T_s$ = 38.9025 K

The constructed analytic temperature structure trends to an isothermal value of 67.8 K. Attempts to vary this while still matching the observed Alice and REX line-of-sight abundance suggests that this value is good to ~3 K. Thus, the joint analysis is far more constraining than the analysis of the Alice data alone.

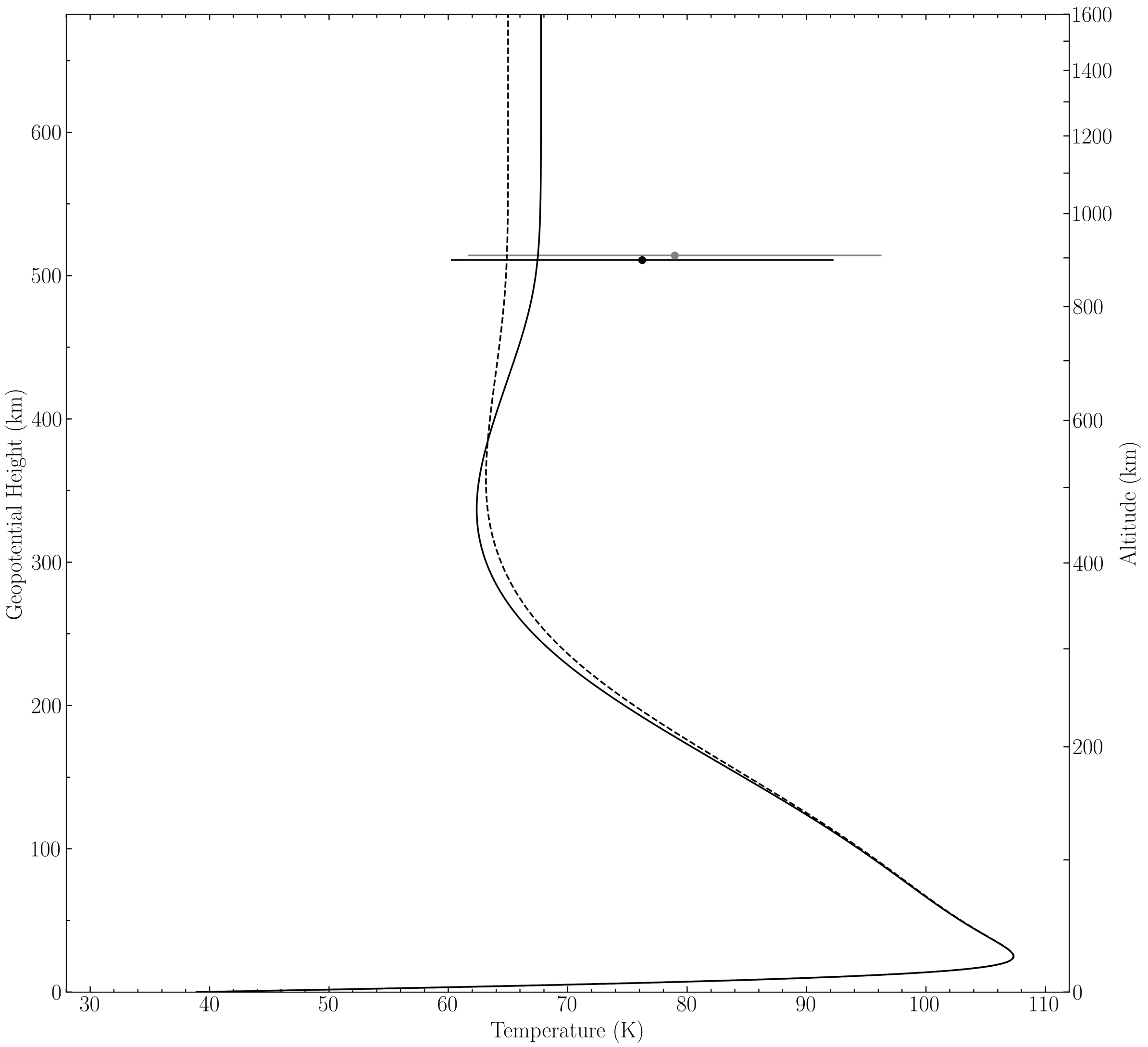

**Figure 19**. Nominal (solid) and flux-constrained (dashed) analytic temperature profiles consistent with $N_2$ line-of-sight abundances from the New Horizons Alice solar occultation and the REX radio occultation. Also plotted are temperatures derived only from the $N_2$ line-of-sight abundances within the $N_2$ fitting region of 900-1066 km altitude (dots; black, ingress; gray egress).

The second step of the analysis was to estimate the $CH_4$ mixing ratio. Again following Gladstone et al. (2016), we specified a simple form for the volume mixing ratio of $CH_4$, $q_{\mathrm{CH4}}$ (Eq 15).

$$q_{CH4}(h) = \frac{n_{CH4}(h)}{n_{total}(h)} = q_0 \left[ 1 + \exp\left( \frac{r - r_h}{H_c \frac{r}{r_h}} \right) \right]^{1-\gamma} \tag{15}$$

where $q_0$ is a mixing ratio at $r << r_h$, $\gamma = 16/28$ is the ratio of the molecular weights of $N_2$ and $CH_4$, $r$ is the radius from the center of Pluto, $r_h$ is the radius marking a transition from well-mixed at lower altitudes to diffusive equilibrium at higher ones, and $H_c$ is a scale height of this transition. In this work, we took slightly different parameter values than in Gladstone et al. (2016): $q_0 = 0.003$, $r_h = 1420$ km, and $H_c = 50$ km. Note that $H_c$ and $r_h$ should not be interpreted as a physical scale height and a homopause level, as they were only used here to define a smooth form for the mixing ratio; a physical description was invoked in the next step.

For the escape, we used the enhanced Jean's escape (Zhu et al. 2014), as in Gladstone et al. (2016), using the temperature and $N_2$ density profiles found in the first step and the $CH_4$ density profile derived in the second step. For the nominal temperature profile, the resulting $CH_4$ exobase was located at 1695 km altitude (2885 km radius), about 600 km above the levels probed by the Alice solar occultation, with $T = 67.78$ K, with $N_2$ and $CH_4$ densities of 1.87 $10^6$ and 2.71 $10^6$ molecule $cm^{-3}$, respectively. The enhanced Jeans escape rates were 7.2 x $10^{22}$ $N_2$ $s^{-1}$ and 6.4 x $10^{25}$ $CH_4$ $s^{-1}$, with the enhancement factor in Fig. 1 of Zhu et al. (2014). These steps yielded the direct interpretation and synthesis of the Alice and REX data sets without solving continuity and transport equations that require photochemistry, eddy and molecular diffusion coefficients, and supersede the earlier escape rates given in Gladstone et al. (2016) of 1x$10^{23}$ $s^{-1}$ for $N_2$ and 5 x $10^{25}$ $s^{-1}$ for $CH_4$.

For the third step, we wished to provide a physical basis for the $N_2$ and $CH_4$ density profiles in order to understand parameters such as $CH_4$ surface mixing ratio, $q_0$, eddy diffusion coefficient, $K_{zz}$, and the flux resupplying $CH_4$ at the surface, $\phi_{CH4}$. For this, it was necessary to demonstrate that the density profiles were solutions to their respective continuity and diffusive transport equations for a spherically symmetric atmosphere. This step was an iterative process, which began with the $q_0$ from Eq. 15, and ended by comparing the modeled $CH_4$ line-of-sight abundance with the Alice observations. At its essence, the 1-D transport code modeled the combined effect on the $CH_4$ profile of eddy diffusion (which drives $CH_4$ to be evenly mixed with $N_2$) and molecular diffusion (which drives $CH_4$ to increase with altitude relative to the heavier $N_2$).

The 1-D transport model of Strobel et al. (2009) was adopted for this task. This transport model (i) is valid even when the secondary species has a non-negligible mixing ratio, (ii) includes spherical geometry—needed in an extended atmosphere such as Pluto's, where the scale height is not small compared with the radius, (iii) includes an escape rate from the top of the atmosphere, and (iv) includes simplified estimates of photochemical destruction of $CH_4$. We used the temperature profile from step 1; the internal energy (heat) equation was not solved simultaneously with the transport.

The method that Strobel (2009) used to solve the equations numerically was a fourth order Runge–Kutta algorithm with all boundary conditions specified at the surface: $q_0$, $K_{ZZ}$, and $\phi_{CH4}$. The upward flux at the surface, $\phi_{CH4}$, was balanced by the sum of photochemical loss and escape of $CH_4$. We included a simple photochemical scheme in the numerical model. Photochemistry is discussed further in Section 8 and in Wong et al. (2017). The calculation the $CH_4$ profile depended only weakly on the details of the photochemical model, as was also found by Wong et al. (2017). Our model found two peaks of the photochemical destruction of $CH_4$. In this, we defined the $CH_4$ dissociation rate, $J_{CH4}$, and an $N_2$ ionization rate $I_{N2}$. $J_{CH4}$ is the effective dissociation rate for zero optical depth with an assumed absorption cross section for dissociation of $\sigma(CH_4)$ = 1.9 x $10^{-17}$ $cm^2$. $J_{CH4}$ is essentially due to solar Lyman alpha. $I_{N2}$ is the effective ionization rate of $N_2$ for zero optical depth with $\sigma(N_2)$ = 1.2 x $10^{-17}$ $cm^2$. The effective rates with adopted cross section yielded integrated rates of dissociation and ionization consistent with Pluto's extended atmosphere presenting a cross section area inside the respective unit optical depths circles. We let the $CH_4$ dissociation rate be $J_{CH4}$ = 2 x $10^{-9}$ $s^{-1}$, the $N_2$ ionization rate be $I_{N2}$ = 2 x $10^{-10}$ $s^{-1}$, with radiation incident at a zenith angle of 60°, and solar rates that were reduced by a factor of 2 for diurnal average. The main peak was near an altitude of ~460 km, based on absorption of Ly-α photons from the sun and the interplanetary medium. A second peak was at an altitude of ~720 km, and described the effect of $N_2$ being ionized to $N_2^+$, eventually leading to the destruction of two $CH_4$ molecules. The profiles of destruction rate had a form similar to that of a Chapman function (with, however, a variable scale height). This simple model yielded integrated $CH_4$ dissociation rates directly by Ly-α photolysis, 2.2 x $10^{25}$ $s^{-1}$, and indirectly by $N_2$ ionization, 4.4 x $10^{24}$ $s^{-1}$.

With the nominal temperature profile, we selected $q_0$ = 0.3% for the surface $CH_4$ mixing ratio (Table 4, run C); $q_0$ = 0.28-0.32% (run B, D) were also acceptable values when used with the nominal temperature profile. Our improved analysis of the Alice occultation data and extension of the $CH_4$ line-of-sight abundances to z ~ 80 km implied a lower $CH_4$ mixing ratio than the $q_0$ = 0.6-0.84% obtained by Gladstone et al. (2016) or the 0.4% surface mixing ratio adopted by Wong et al. (2017). The molecular diffusion coefficient of $CH_4$-$N_2$ mixture at the surface was 550 $cm^2$ $s^{-1}$. With a $CH_4$ mixing ratio at the surface of ~0.3%, a constant eddy diffusion coefficient of $10^3$ $cm^2$ $s^{-1}$ was preferred; this gave a derived homopause level (where the eddy diffusion coefficient equals the molecular diffusion coefficient) near the surface, at ~2 km altitude (Run C). An eddy diffusion coefficient equal to 2 x $10^3$ $cm^2$ $s^{-1}$ was marginally acceptable (Run D), with the homopause at 5 km altitude, probably the thickness of the planetary boundary layer (Hinson et al. 2017). With $q_0$ = 0.28%, the solution is a surface homopause (Run B). Based on the ratio of $CH_4$ to $N_2$ line-of-sight column densities we rejected $q_0$ = 0.2% for all values eddy diffusion (Run A), and also $q_0$ = 0.4%, even with eddy coefficients as large as 5 x $10^4$ $cm^2$ $s^{-1}$, with a homopause at ~160 km altitude (Run E). Thus, the homopause must have been below the altitude where the Alice UV solar occultation directly measures the $CH_4$ absorption.

**Table 4: Transport Model Results**

| Run | $T_{inf}$, K | $q_0$ | Homopause altitude, km | $K_{zz}$, cm$^2$ s$^{-1}$ | *Notes* |
|---|---|---|---|---|---|
| A | 67.8 | 0.20% | 0 | $<5.5 \times 10^2$ | ruled out by data |
| B | 67.8 | 0.28% | 0 | $5.5 \times 10^2$ | acceptable solution |
| **C** | **67.8** | **0.30%** | **2** | $\mathbf{1 \times 10^3}$ | **nominal solution** |
| D | 67.8 | 0.32% | 5 | $2 \times 10^3$ | acceptable solution |
| E | 67.8 | 0.40% | 150 | $5 \times 10^4$ | ruled out by data |
| F | 65.0 | 0.30% | 2 | $1 \times 10^3$ | acceptable solution |
| **G** | **65.0** | **0.35%** | **12** | $\mathbf{4 \times 10^3}$ | **flux-constrained solution** |

The upward flux, $\phi_{CH4}$, depended weakly on the eddy diffusion coefficient. For the nominal temperature profile and $q_0 = 0.3\%$, we found that solutions with $K_{ZZ} = 100$, 1000 (the preferred value), and 2000 cm$^2$ s$^{-1}$ gave values of $\phi_{CH4}$ of $10.7 \times 10^{25}$, $10.3 \times 10^{25}$, and $10.1 \times 10^{25}$ $CH_4$ s$^{-1}$, respectively (Table 2, runs B, C, D). If we set the $CH_4$ dissociation rates to zero, but kept the flux boundary condition for $CH_4$ the same, so that $CH_4$ had an escape rate of $10.3 \times 10^{25}$ $CH_4$ s$^{-1}$ (e.g., equal to the upward flux at the surface), then the $N_2$ density remained the same and the $CH_4$ density decreased by a mere 5% at an altitude of 1710 km ($r$ = 2900 km); this reinforced the conclusion that the exact partitioning between chemistry and escape had a second-order effect on the resultant $CH_4$ density profile in the 1D transport model.

The $N_2$ and $CH_4$ line-of-sight abundances and densities as determined from the nominal 1-D transport model are plotted in Fig 17. The integrated $N_2$ and $CH_4$ column densities from the transport models were $4.2 \times 10^{21}$ $N_2$ cm$^{-2}$ and $1.4 \times 10^{19}$ $CH_2$ cm$^{-2}$.

As $N_2$ was in hydrostatic, gravitational diffusive equilibrium throughout the atmosphere as the major species, the $N_2$ escape rate determined from the nominal 1-D transport model was the same as that determined at the end of step 2: $7.2 \times 10^{22}$ $N_2$ s$^{-1}$. For $CH_4$, when we changed from Eq. (15) to numerical solutions to the $CH_4$ transport equation, which extrapolated its density based on the physics of molecular diffusion, the exobase was raised from 1695 km altitude (2885 km radius) to 1735 km altitude (2925 km radius), and the $CH_4$ escape rate increased from $6.4 \times 10^{25}$ $CH_4$ s$^{-1}$ to $7.7 \times 10^{25}$ $CH_4$ s$^{-1}$. We found that the $CH_4$ density and escape rate at the exobase were insensitive to various input parameters such as $K_{zz}$ and $q_0$, when calculated with the nominal temperature profile in Fig. 19.

We investigated a second temperature profile, motivated by the consideration of the balance of upward flux from $CH_4$ at the surface with photochemical destruction and escape. The solutions to the nominal temperature profile required that only one third of the $CH_4$ dissociations led to irreversible removal. We increased the net dissociation rate by lowering $T_{inf}$ to 65 K, which lowered the escape rates to $3.2 \times 10^{22}$ $N_2$ s$^{-1}$ and 4.5-4.8 x $10^{25}$ $CH_4$ s$^{-1}$. With this lower escape rate, we found two acceptable solutions. One (Run F) had a $CH_4$ profile in the lower atmosphere that was similar to our nominal run (Run C): $\phi_{CH4} = 10.7 \times 10^{25}$, $q_0 = 0.3\%$, $K_{ZZ} = 10^3$, and a homopause at 2 km altitude. This run had 70% irreversible removal of $CH_4$. Another (Run G) increased the $CH_4$ density in the lower atmosphere, and fit the observations less well below 200 km, with $\phi_{CH4}$ = 11.9 x $10^{25}$ $CH_4$ $s^{-1}$, $q_0$ = 0.35%, $K_{ZZ}$ = 4 x $10^3$, and a homopause at 12 km altitude. Run G allowed a solution with 94% irreversible removal of $CH_4$ at the expense a poorer fit to the $CH_4$ line-of-sight abundance profile below 200 km. We therefore adopted Run G as a second alternate solution, which we term "flux-constrained," as the motivation for the lower $T_{inf}$ and the $CH_4$ profile were to balance of upward $CH_4$ flux (11.9 x $10^{25}$ $CH_4$ $s^{-1}$) with escape (~4.6 x $10^{25}$ $CH_4$ $s^{-1}$) and photochemical destruction (~ 7.3 x $10^{25}$ $CH_4$ $s^{-1}$).

Because $N_2$ and $CH_4$ dominated the atmosphere at all altitudes probed by the Alice solar occultation of Pluto, the total number density was well approximated as the sum of the $N_2$ modeled in the first step, and the $CH_4$ modeled in the third step: $n_{\text{total}} = n_{\text{CH4}} + n_{\text{N2}}$. Dividing measured densities by this modeled total density gave profiles of mixing ratio for all measured species (Fig. 20).

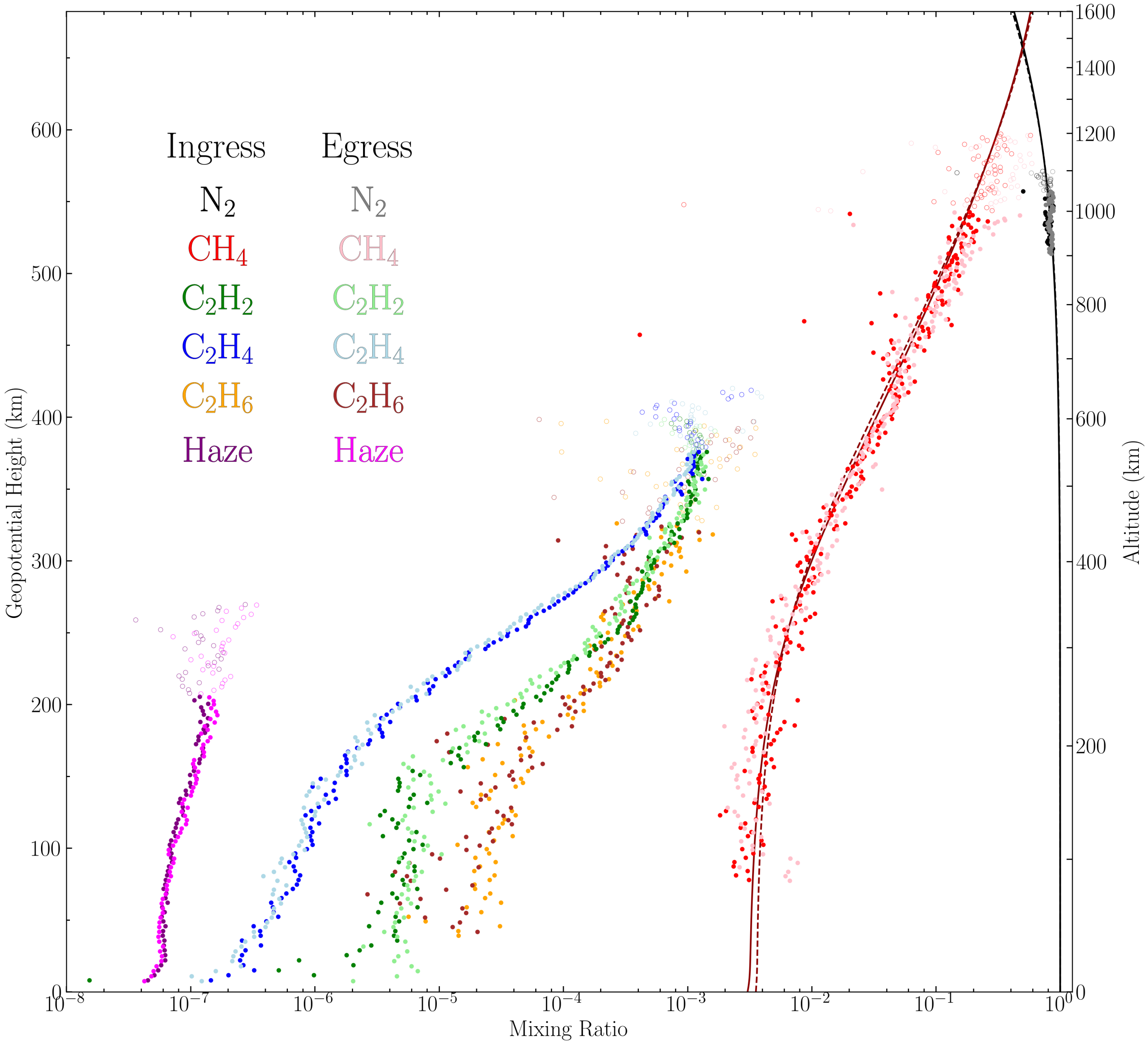

**Figure 20.** Plot of mixing ratio ($n/n_{\text{total}}$) vs. geopotential height on the left axis and altitude on the right axis. For haze, the value plotted is ε*$10^{15}$ $cm^2/n_{\text{total}}$, where ε is the extinction coefficient. Minor ticks on the x-axis are plotted at 2, 4, 6, and 8 times the major tick values. Solid lines are based on the nominal model, for which we interpolated between Alice and REX observation for $N_2$ (black), and appled a 1-D transport model to the Alice observations of $CH_4$ (red). The dashed line is based on the flux-constrained model for $CH_4$, for which we additionally adjusted the temperature profile and $CH_4$ transport to be more consistent with photochemical destruction of $CH_4$, at the expense of a poorer match to observations below 200 km altitude. Dots are observed densities divided the modeled total density: $N_2$ (black, ingress; gray egress), $CH_4$ (red, ingress; pink, egress), $C_2H_6$ (brown, ingress; mustard egress), $C_2H_2$ (dark green, ingress; light green, egress), $C_2H_4$ (dark blue, ingress; light blue, egress), and haze (dark purple, ingress; light purple, egress). This figure plots open circles for all species above where the $\sigma_{\ln(n)} > 1$. From Fig 18, this is estimated this to be at the following altitudes: $N_2$: 1050 km, $CH_4$: 1000 km, $C_2H_2$: 550 km, $C_2H_4$: 550 km, $C_2H_6$: 450 km, and haze: 250 km.

# 8. Discussion and conclusions

## *8.1 Summary of observations*

The present work supersedes Gladstone et al. (2016) by including new and more rigorous reductions of the Alice solar UV occultation, with improved analysis and error propagation. Additionally, the temperature and mixing ratio analysis presented here incorporated newer analysis of the REX radio occultation (Hinson et al., 2017), which used the complete REX dataset — something that had not yet been downlinked for the Gladstone et al. (2016) report. From this work, we found the following observational results:

(i) $N_2$ absorption was measured from 900 to 1100 km altitude by the Alice solar occultation with a line-of-sight abundance at 900 km altitude of $(6.37\pm0.98) \times 10^{16}$ and $(6.79\pm1.09) \times 10^{16}$ molecule cm$^{-2}$ for ingress and egress respectively (Figs 17). The temperature at 1000 km altitude was only weakly constrained by the Alice measurements alone (76 ± 16 K and 79±17 K for ingress and egress, respectively). We constructed an analytic temperature structure (Fig 19) using a joint analysis with the REX occultation, which also implied cold temperatures in Pluto's upper atmosphere, ~65-68 K. This temperature profile, along with some key processes, is plotted in Fig 21.

(ii) $CH_4$ was measured from 80 to 1200 km altitude (Figs 17, 20). Diffusive separation was clearly evident, with $CH_4$ varying from ~0.3%-0.35% at 80 km altitude, to ~3% at 550 km, and 18% at 1000 km. A numerical 1-D transport model implied Pluto had a very stable atmosphere with a small eddy diffusion coefficient, ~$(1\text{-}4) \times 10^3$, with a homopause near 2-12 km (Fig 21A).

(iii) Light hydrocarbons ($C_2H_2$, $C_2H_4$, $C_2H_6$) were detected throughout Pluto's middle atmosphere, from at or near the surface to ~600 km altitude. While their mixing ratio was near 0.1% at 550 km altitude for all three species, at 100 km it dropped to $2 \times 10^{-5}$, $5 \times 10^{-6}$, and $6 \times 10^{-7}$ for $C_2H_6$, $C_2H_2$, and $C_2H_4$ respectively. This is direct evidence for chemistry or condensation; as the molecular weights of these species are similar to that for $N_2$, they would have been well mixed in the absence of destruction or production.

(iv) The haze extinction coefficient, ε, was nearly proportional to the $N_2$ density from 26 to 100 km altitude, with $\varepsilon \approx 6 \times 10^7\, n_{N2}$ (with ε in cm$^{-1}$ and $n_{N2}$ in cm$^{-3}$). This was evidence that the haze may have had concentrations proportional to $N_2$ density, or (more likely) showed the competing effects of smaller particle sizes but higher number density at the altitudes of greater haze production.

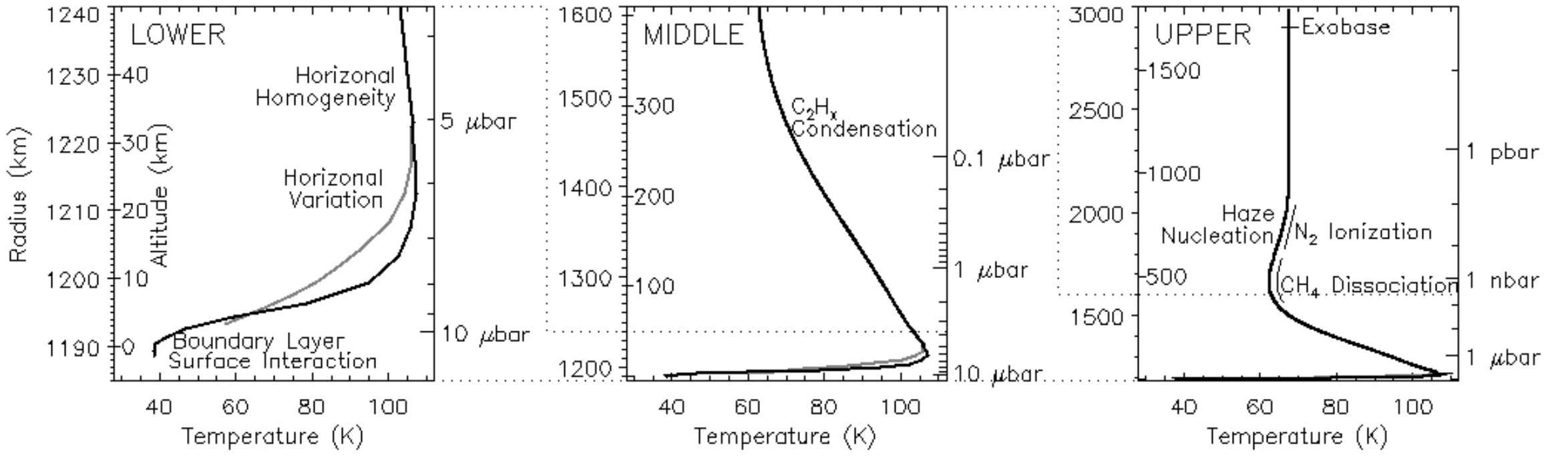

**Figure 21.** Summary of the temperature structure in Pluto's atmosphere from the joint analysis of the Alice solar occultation (this work) and the REX radio occultation (REX ingress in black, REX egress in gray). Radii and approximate altitude (referenced to a surface radius of 1190 km) are both plotted. Major processes are indicated.

### *8.2 Lower Atmosphere (~0 to 30 km altitude)*

The temperatures and bulk density in Pluto's lower atmosphere (Fig 21, left panel) were measured with the REX occultation by Hinson et al. (2017), who described horizontal variation between ingress vs. egress temperatures below 30 km and horizontal homogeneity above 30 km. Our analysis of the Alice solar occultation also showed horizontal homogeneity between ingress and egress, but could not shed light on this variation below 30 km. The absorption due to $N_2$, $CH_4$, and $C_2H_6$ in the lower atmosphere was not well constrained by the Alice solar occultation. $C_2H_2$, $C_2H_4$ and haze do have measurable absorption to the surface, but our analysis averaged data by 82.5 km, making it difficult to thoroughly explore the lowest 30 km. Future analysis might be designed to particularly search for ingress vs. egress differences at these altitudes in the Alice solar occultation.

In Pluto's lower atmosphere, the improved $CH_4$ retrieval presented here dramatically changed our picture of the vertical mixing in Pluto's atmosphere vs. Gladstone et al. (2016). We now see that Pluto's atmosphere was extremely stable. Observations were consistent with a very small eddy diffusion coefficient, $K_{zz}$, (~(0.5-4) x $10^3$ cm$^2$ s$^{-1}$), a very low homopause altitude, $z_h$, (~0-12 km, at the surface or within a planetary boundary layer), and a surface mixing ratio, $q_0(CH_4)$ of ~0.28-0.35%. Attempts to fit the New Horizon measurements with various combinations of input parameters ruled out $K_{zz} > 10^4$ cm$^2$ s$^{-1}$, $z_h >$ ~60 km, $q_0$ ($CH_4$) < 0.2% or $q_0$ ($CH_4$) > 0.4% (Table 4). Throughout much of Pluto's atmosphere, only molecular diffusion applies, not eddy diffusion.

The REX observations were consistent with $N_2$ in vapor-pressure equilibrium with the $N_2$ ice, as expected. The physics of the interaction of solid $CH_4$ with the atmosphere is much less well understood. Our analysis of the Alice observations give $CH_4$ mixing ratios similar to the results in the lowest few scale heights from ground-based near-IR observations 2008 - 2014 (~0.44%, Lellouch et al. 2015; ~0.5%, Cook et al. submitted). These measurements of atmospheric $CH_4$ abundance were orders of magnitude more than expected from Raoult's law for $CH_4$ gas over $CH_4$ in solid solution with $N_2$-rich ice (Owen et al., 1993). The source of the high gaseous $CH_4$ abundance might well be the $CH_4$-rich ice seen on Pluto (Grundy et al. 2016, Stansberry et al., 1996, Forget et al. 2017).

For our nominal solution of $q_0(CH_4)$ = ~0.3% and $K_{zz} = 10^3$ cm$^2$ s$^{-1}$, we found upward fluxes at the surface of 10.3 x $10^{25}$ $CH_4$ s$^{-1}$. This can be compared to the Hunten limiting flux, which is the flux at which a minor species is well-mixed throughout the entire atmosphere (Hunten 1973, Strobel 2012) and could be considered the maximum allowable flux. The Hunten limiting flux on Pluto was 11.5 x $10^{25}$ $CH_4$ s$^{-1}$ for the nominal solution, and is proportional to $q_0$. A flux of 10.3 x $10^{25}$ $CH_4$ s$^{-1}$ implies a global average of 5.8 x $10^8$ $CH_4$ cm$^{-2}$ s$^{-1}$, or only 4.9 x $10^{-7}$ g cm$^{-2}$ yr$^{-1}$. This was not an unreasonably large value for net global sublimation, since local sublimation rates over warm $CH_4$ patches can easily be 5 to 7 orders of magnitude larger than this (Stansberry et al. 1996).

In the earlier report of Gladstone et al. (2016), we did not have measurements of $CH_4$ line-of-sight abundances below 200 km. In that paper, we presented a model with a higher surface $CH_4$ mixing ratio of 0.6%-0.84% and a much higher eddy diffusion coefficient profile with $7.5\times10^5$ cm$^2$ s$^{-1}$ at the surface and asymptotically reaching $3\times10^6$ cm$^2$ s$^{-1}$ at z = 210 km, which yielded a homopause at z = 390 km. The low value of $K_{zz}$ derived here was very similar to that derived in Wong et al. (2017), using a simpler diffusion model that neglected both $CH_4$ escape and $CH_4$ dissociation by ionospheric chemistry. The small $K_{zz}$ is likely to be tied to the extremely large vertical thermal gradients in Pluto's lower atmosphere (Sicardy et al. 2016, Hinson et al. 2017), which were a consequence of $CH_4$ heating being balanced by thermal conduction (e.g., Zhu et al. 2014). Large thermal gradients lead to large values of the Brunt–Väisälä (buoyancy) frequency, a measure of atmospheric stability. It appears as though the large thermal gradient (large Brunt–Väisälä frequency) acted as a "lid," strongly suppressing vertical mixing between the lower and upper atmosphere. The suppression vertical mixing enabled the existence of local variation at the very lowest altitudes (Hinson et al. 2017), even while the upper atmosphere was horizontally well mixed. Thus, horizontal transport in the planetary boundary layer and lower atmosphere may play the most important role in redistribution of mass associated with sublimation and condensation of volatiles.

*8.3 Middle Atmosphere (~30 to 400 km altitude)*

Pluto's middle atmosphere (Fig 21, middle panel) is characterized by an approximate balance between $CH_4$ heating and $H_2O$, CO, and $C_2H_2$ cooling (Zhu et al. 2014, Strobel and Zhu 2017) and also by probable condensation of the three "$C_2H_x$ hydrocarbons," $C_2H_2$, $C_2H_4$, and $C_2H_6$ (Wong et al. 2017). The $CH_4$ profile was not much affected by the abundances of the $C_2H_x$ hydrocarbons (Wong et al. 2017). The $CH_4$ mixing ratio was observed to increase with altitude in the middle atmosphere to ~1% at 400 km altitude, due to diffusive equilibrium.

The line-of-sight abundances of $C_2H_2$, $C_2H_4$, and $C_2H_6$ were nearly constant with altitude from the surface to ~300 or 400 km altitude (Fig 8; Fig 17, upper panel), which led to a broad range of altitudes where the UV transmission from 150 to 170 nm also varied slowly (e.g., the cyan plateau in Fig 4B; Fig 11, panels labeled 150 nm, 152 nm, 165 nm and 172 nm). When converted to density (Fig 17B), these three "$C_2H_x$ hydrocarbons" had local maxima in the density near 410 km altitude for $C_2H_4$, 320 km for $C_2H_2$, and an inflection point or the suggestion of a local maximum at 260 km altitude for $C_2H_6$. The local maxima for $C_2H_4$ and $C_2H_2$ were seen also in Gladstone et al. (2016); in this work these maxima were much more pronounced. $C_2H_4$ and $C_2H_2$ also had clear density minima near 200 km and 170 km altitude, respectively. $C_2H_6$ was the hardest of the three $C_2H_x$ hydrocarbons to measure, and even this species appeared to have an inflection point or a density minimum near 170 to 200 km. The chemical models of Wong et al. (2017) also showed clear minima and maxima in density for $C_2H_4$ and $C_2H_2$. In Wong et al. (2017), the altitude of the $C_2H_4$ maximum is above that of the $C_2H_2$ maximum, as seen in the Alice solar occultation, although the observed altitude of both maxima is above what is modeled. The modeled minima for both $C_2H_4$ and $C_2H_2$ are near 200 km, with the dominant loss mechanism being condensation. The $C_2H_6$ chemistry may be more uncertain, as the chemical models do not fully reproduce the inflection points (and possible minima and maxima) seen in our analysis of the Alice data. In Wong et al. (2017), the dominant means of removal of $C_2H_x$ is by condensation via heterogeneous nucleation on hydrocarbon/nitrile aerosols between 200 and 400 km. The evidence for the removal of these species is also seen in plots of the mixing ratios (Fig. 20). The steep decrease in mixing ratio between 400 and 200 km for all three species was indicative of removal at these altitudes.

Condensation of the $C_2H_x$ hydrocarbons between 400 and 200 km altitude changes the size and sphericity of Pluto's haze particles. We do not measure the haze abundance or density directly, as its UV cross section is uncertain and likely to vary with altitude (Cheng et al. 2017). Rather, we measure the line-of-sight optical depth from the UV transmission, and, through inversion, calculate the extinction coefficient. If the cross section of haze particles varies slowly in altitude, then this suggests that the haze concentration is roughly constant with either $CH_4$ or $N_2$ density. However, the haze modeling of Gao et al. (2017) suggested that haze radii are larger near the surface than at 250 km. This implies that the actual haze concentration, once normalized for the increase in cross section, was smaller at the surface than at 250 km, near the modeled altitudes of the $C_2H_x$ hydrocarbon condensation. Gao et al. (2017) compared their haze model with an earlier retrieval of haze extinction from the Alice solar occultation. It may be that the change in the haze radius and the change in the haze density are roughly equivalent competing effects, leading to a nearly constant value of extinction coefficient divided by bulk number density.

*8.4 Upper Atmosphere (~400 to 2000 km altitude)*

We confirmed low temperatures in Pluto's upper atmosphere (Fig 21, right panel) from the analysis of the $N_2$ transmission, as first reported in Gladstone et al. (2016), using a joint analysis of the $N_2$ line-of-sight abundances derived from the REX radio occultation (Hinson et al. 2017) and the Alice UV solar occultation. The Alice solar occultation measured $N_2$ at 900-1100 km altitude, and the REX radio occultation measured $N_2$ at about 0 to 100 km altitude (referenced to a surface radius of 1190 km), leaving a 800-km measurement gap between the radio and solar occultations. To span the gap isothermally would imply a temperature of 64.3 K, which was much colder than measured by REX or from ground-based stellar occultations (Sicardy et al. 2016); the temperature profile that we show here spanned the gap while avoiding discontinuities in the temperature, pressure, and number density profiles and their derivatives.

The modeled temperature minimum near 470 km altitude (Fig 21 C) was within the measurement gap. It was needed to decrease the $N_2$ fast enough to satisfy the observed $N_2$ line-of-sight abundances at both the REX measurements and the Alice measurements above 900 km altitude. There probably were other equally valid solutions that bridge this gap; however, the $N_2$ density and pressure are integrals of the temperature, and differences in the details of the temperature profile lead to only small differences in the $N_2$ density. Furthermore, the temperature minimum near 470 km was compatible with the broad temperature minimum that spanned 300-500 km altitude that Lellouch et al. (2017) inferred from ALMA observations of CO and HCN rotational line emissions in Pluto's atmosphere. A temperature minimum implies net cooling — as, indeed, does a positive curvature in the temperature profile ($(1/r^2)\ \partial/\partial r\ (r^2\ \kappa\ \partial T/\partial r) \approx \kappa\ \partial^2 T/\partial r^2 > 0$, where $\kappa$ is the thermal conductivity, e.g., Zhu et al. 2014). As discussed in Gladstone et al. (2016), Forget et al. (2017) and Strobel & Zhu (2017), cooling by CO, $C_2H_2$, and HCN are not sufficient, and there remains a mystery coolant in the upper atmosphere, for which Strobel & Zhu (2017) argue the case that it is $H_2O$.

$CH_4$ is photochemically destroyed in the upper atmosphere by solar Ly-α, interplanetary Ly-α, and ionization of $N_2$ followed by ion chemistry that dissociates two $CH_4$ for every $N_2$ ionized. The resulting formation of nitriles eventually leads to very large negatively charged macromolecules and aerosols that coagulate at lower altitudes into Pluto's detectable hazes (Cheng et al. 2017, Lavvas et al. 2010).

To compute the total photons absorbed by Pluto's atmosphere we assumed it presented a "bullseye" target defined by the radius of the optical depth $\tau = 1$ circle. The total dissociation loss rate of $CH_4$, $L_{CH4}$, was then

$$L_{CH4} = F_{L\alpha}\left(1+0.2\right)\pi\left(r_{CH4}\right)^2 + \pi I_0 4\pi\left(r_{CH4}\right)^2 + F_{N2}\pi\left(r_{N2}\right)^2 \qquad (16)$$

where $r_{CH4} = 1875$ km and $r_{N2} = 2100$ km are the $\tau = 1$ radii for $CH_4$ at Ly-a and for $N_2$, respectively. $F_{L\alpha}$ is the solar Ly-α flux, the factor 0.2 represents $CH_4$ dissociation by solar flux other than Ly-α, $\pi I_0$ is the downward interplanetary Ly-α flux, assumed isotropic over the spherical atmosphere, and $F_{N2}$ is the solar $N_2$ ionizing flux times 2 (because each ionized $N_2$ dissociates two $CH_4$). At the time of the New Horizons flyby of Pluto, the solar Ly-α flux at Pluto was $F_{L\alpha} \sim 3.5 \times 10^8$ photons cm$^{-2}$ s$^{-1}$, the interplanetary Ly-α average intensity was $4\pi I_0 \sim 145\ R$ (Gladstone et al. 2015), and the solar $N_2$ ionizing flux times 2 (because each ionized $N_2$ dissociates two $CH_4$) was $F_{N2} \sim 1 \times 10^8$ photons cm$^{-2}$ s$^{-1}$. The terms were in the ratio of 0.61:0.21:0.18, respectively, and the total was $7.7 \times 10^{25}$ $CH_4$ s$^{-1}$.

This photolysis rate led to an apparent conflict between the net flux at the surface ($10.3 \times 10^{25}$ s$^{-1}$) and the sum of the total photolysis and $CH_4$ escape ($7.7 \times 10^{25} + 7.7 \times 10^{25} = 15.4 \times 10^{25}$ s$^{-1}$). In fact, the sum of photolysis and escape exceeds the Hunten limiting flux of $11.5 \times 10^{25}$ $CH_4$ s$^{-1}$. The difference cannot have been due to a net increase in $CH_4$ column abundance with time. Observations of $CH_4$ from ground-based near-IR absorption (Lellouch et al. 2015, Cook et al., submitted) limit the change in the column of gaseous $CH_4$ to less than 20% yr$^{-1}$, or less than ~$1 \times 10^{11}$ $CH_4$ cm$^{-2}$ yr$^{-1}$. Thus, we needed to lower the photolysis rates or the escape rates, or both.

If one third of the $CH_4$ dissociations led to irreversible removal, then the net $CH_4$ dissociations would be to 2.6 x $10^{25}$ $s^{-1}$. This is what was assumed in the simple photochemical model in Section 7 and our nominal solution.

To determine the fraction of irreversible $CH_4$ dissociations, one needs a full photochemical model for Pluto's atmosphere. The simplicity of our photochemical model did not allow us to calculate the fraction of irreversible photodissociation events, but Wong et al. (2017) calculated the fraction of irreversible photodissociation events to be greater than 0.99 (Yuk Yung, private communication). The calculation of the $CH_4$ photodissociation rate appropriate to the Alice occultation is further complicated by long time constants. For solar Ly-α radiation, the optically thin photodissociation time constant was about 5 Earth years and at optical depth = 1, it is ~ 13 Earth years. Thus one would have to average the incident fluxes over the last solar cycle. At Pluto's orbital distance, interplanetary Ly-α is also important in $CH_4$ photodissociation with equivalent long time constants and would likewise have to be averaged over the past 13 years or more. Pluto is also receding from the Sun, which introduces another time variable in the magnitude of UV radiation incident on the atmosphere. Another time constant relevant to $CH_4$ is its residence time of three Pluto years in the atmosphere based on the current total photodissociation and escape rate ~ 10 x $10^{25}$ $CH_4$ $s^{-1}$ and the observed column of 1.4 x $10^{19}$ $CH_4$ $cm^{-2}$.

The second way to resolve the apparent conflict between the total loss rate and upward flux from the surface was to revisit the escape rate. Because $CH_4$ was the major constituent at the exobase, its density distribution determined the location of the exobase. Thus the best way to lower its escape rate was to lower the temperature at the exobase. To test how much we could lower the $CH_4$ escape rate and still fit the New Horizons occultation data, we did a numerical calculation with $q_0$ = 0.3%, $K_{zz}$ = $10^3$ $cm^2$ $s^{-1}$, and adjusted the temperature profile such that the upper atmospheric isothermal temperature decreased from 67.8 to 65 K . This increased the minimum temperature in the $r$ = 1600-1700 km region by ~0.5 K. For photochemistry, we set $J_{CH4}$ = 4 x $10^{-9}$ $s^{-1}$, and $I_{N2}$ = 5 x $10^{-10}$ $s^{-1}$, and obtained integrated $CH_4$ dissociation rates directly by Ly-α photolysis of 4.4 x $10^{25}$ $s^{-1}$, and indirectly by $N_2$ ionization of 1.1 x $10^{25}$ $s^{-1}$. With these changes, the escape rates were 3.2 x $10^{22}$ $N_2$ $s^{-1}$ and 4.8 x $10^{25}$ $CH_4$ $s^{-1}$. Even this lower escape rate still requires that only 70% of the $CH_4$ dissociations led to irreversible removal.

We searched for solutions with even higher total dissociation rates. If we kept the surface conditions of $CH_4$ were the same as nominal ($q_0$ = 0.3%, $K_{zz}$ = $10^3$ $cm^2$ $s^{-1}$), we did not find a solution where 100% of the $CH_4$ dissociations led to irreversible removal. This would have required that the escape rate was the difference between the vertical flux at the surface and the maximum dissociation loss rate, or (10.3 x $10^{25}$ - 7.7 x $10^{25}$) = 2.6 x $10^{25}$ $s^{-1}$. To drive the $CH_4$ escape rate down this low required the isothermal upper atmospheric temperature to be 62 K, which had a demonstrably poor fit to the $N_2$ line-of-sight column density profile. However, we found a compromise solution with $q_0$ = 0.35% and $K_{zz}$ = 4 x $10^3$ $cm^2$ $s^{-1}$, and an isothermal upper atmospheric temperature of 65 K. This solution matched the $N_2$ line-of-sight column density profile, but had some disagreement with the measured $CH_4$ abundance below 200 km. This "flux-constrained" solution yielded an escape rate of 3.2 x $10^{22}$ $N_2$ $s^{-1}$ and 4.6 x $10^{25}$ $CH_4$ $s^{-1}$. The surface flux of 11.9 x $10^{25}$ $CH_4$ $s^{-1}$ was balanced by this escape rate and a total dissociation rate of 7.3 x $10^{25}$ $s^{-1}$, or 94% of the 7.7 x $10^{25}$ $s^{-1}$ maximum.

Because of these considerations our recommended escape rates are as follows: $N_2$ = (3-7) x $10^{22}$ $s^{-1}$, and $CH_4$ = (4-8) x $10^{25}$ $s^{-1}$.

We examined the consistency of these new atmospheric results to those obtained from the solar wind plasma interaction measured by the Solar Wind Around Pluto (SWAP) instrument on New Horizons (McComas et al. 2008). SWAP observations showed a fascinating interaction between the solar wind and Pluto that is intermediate between the cometary and unmagnetized planetary (e.g., Mars and Venus) cases (McComas et al. 2016). In a subsequent study, Zirnstein et al. (2016) used SWAP data to show that that the heavy ions detected were more likely $CH_4$+ than $N_2$+, substantiating a methane-based solar wind interaction, confirming the analysis of the Alice data that the primary escaping species was $CH_4$. McComas et al. (2016) showed that for ideal, fluid assumptions, the standoff distance of Pluto's bow shock at ~4.5 Rp was consistent with a neutral $CH_4$ production rate Q = 5 × $10^{25}$ $s^{-1}$, outflow speed ~100 m $s^{-1}$, and solar wind Mach number M of ~10 (see Figure 8 of their paper). While the SWAP observations provide only an indirect measure of Pluto's atmospheric escape and thus these values may have some errors (we estimate less than an order of magnitude), the results from these studies are consistent with the values derived in the current study.

## 9. Future Work

This analysis used $C_2H_2$, $C_2H_4$, and $C_2H_6$ cross-sections measured in the laboratory at 150 K, 140 K, and 150 K, which was warmer than the ~65-70 K in much of Pluto's atmosphere. We estimated that the current cross sections might lead to systematic errors in derived abundances of ~10-20%. Additionally, we suspected that the temperature-dependence of the shape of the $C_2H_4$ absorption at 160 to 175 nm might have affected the retrieval of haze abundance at 500-600 km. New laboratory measurements can improve the accuracy of the retrieval.

Other species were predicted to be present, from Pluto photochemical models (e.g., Summers et al. 1997; Krasnopolsky & Cruikshank 1999; Wong et al. 2017), by analogy with Titan and Triton photochemisty (e.g., Yung et al. 1984; Strobel et al. 1990), by cosmochemical arguments (e.g., $^{29}N_2$, Ar), from the influx of $H_2O$ at the top of Pluto's atmosphere from incoming Kuiper Belt material, which is on the order of 50 kg $day^{-1}$, or 2 x $10^{22}$ molecules $s^{-1}$ if this all $H_2O$ (Horanyi et al. 2016), and minor species that might accompany the $H_2O$ (Despois et al. 2005). We discussed here that no definitive signal of HCN or CO absorption was seen. Upper limits on these and others can be quantified, as was recently done by Kammer et al. (submitted) for $O_2$.

Because the fractional error of $C_2H_4$ and haze are small even at the surface, and because these not are very correlated or anti-correlated at the surface, it should be possible to look for differences between entry and exit for these components. Such an analysis would need a different approach than that used here, which smoothed observations over 23 seconds (82.5 km).

There are electronic bands of $N_2$ that give additional information between 80 and 100 nm, particularly at 97-98 nm where the solar spectrum has some strong lines. These can be used to extend the measurement of the $N_2$ line-of-sight abundance, perhaps to ~800 km tangent height. This analysis is computationally complex, but all the components are in place, including high-resolution laboratory spectroscopy at the temperatures seen in Pluto's atmosphere, a model of the solar flux as seen by New Horizons, and a model of the instrumental line-spread function.

Our understanding of the temperature structure and the CO, $CH_4$ and HCN profiles would be improved by a joint analysis of the New Horizons solar occultation with ground-based high-resolution IR observations (Lellouch et al. 2015, Cook et al. submitted), ground-based stellar occultations (e.g., Sicardy et al. 2016), and ground-based radio observations (Lellouch et al. 2017).

Finally, the temperatures, densities, and mixing ratios presented here can be used in a variety of future studies, such as chemistry, haze production, seasonal modeling, energetics, and escape. A wide variety of studies were based on an earlier analysis, including Wong et al. (2017), Gao et al. (2017), Cheng et al. (2017), and Strobel & Zhu (2017).

## Acknowledgement

This work was supported, in part, by funding from NASA's New Horizons mission to the Pluto system. The New Horizons Mission Design and Navigation teams enabled us to watch this glorious sunset and sunrise. Werner Curdt provided the high spectral-resolution solar models. We gratefully acknowledge the publicly available solar data and spectroscopic data: LISIRD Lyman-alpha data from http://lasp.colorado.edu/lisird/lya/, GOES15 soft X-ray flux from http://www.swpc.noaa.gov/; and the Titan spectroscopic database at http://www.lisa.univ-paris12.fr/GPCOS/SCOOPweb/. Thanks go out to Julie Moses for providing an electronic version of the $C_2H_2$ cross sections from Wu et al. 2001.

## References

Barfield, W. 1972. Fits to new calculations of photoionization cross sections for low-Z elements. Journal of Quantitative Spectroscopy and Radiative Transfer 12, 1409-1433.

Bolovinos, A., J. Philis, E. Pantos, P. Tsekeris, and G. Andritsopoulos 1981. The methylbenzenes vis-a-vis benzene. Comparison of their spectra in the Rydberg series region. J. Chem. Phys. 75, 4343-4349.

Bolovinos, A., J. Philis, E. Pantos, P. Tsekeris, and G. Andritsopoulos 1982. The methylbenzenes vis-a-vis benzene. Comparison of their spectra in the valence shell transition region. J. Mol. Spectrosc. 94, 55-68.

Broadfoot, A. L., and 21 colleagues 1989. Ultraviolet spectrometer observations of Neptune and Triton. Science 246, 1459-1466.

Burton, G.R., W.F. Chan, G. Cooper, and C.E. Brion 1992. Absolute oscillator strengths for photoabsorption (6-360 eV) and ionic photofragmentation (10-80 eV) of methanol. Chem. Phys. 167, 349-367.

Burton, G.R., W.F. Chan, G. Cooper, and C.E. Brion 1993. The electronic absorption spectrum of NH3 in the valence shell discrete and continuum regions. Absolute oscillator-strengths for photoabsorption (5-200 eV). Chem. Phys. 177, 217-231.

Capalbo, F. J., Y. Bénilan, R. V. Yelle, and T. T. Koskinen 2015. Titan's Upper Atmosphere from Cassini/UVIS Solar Occultations. The Astrophysical Journal 814, 86.

Chamberlain, J. W. & D. M. Hunten 1987. Theory of planetary atmospheres: an introduction to their physics and chemistry, 2nd revised and enlarged edition. Academic Press, Inc. Orlando, FL.

Chan, W. F., G. Cooper, R. N. S. Sodhi, and C.E. Brion 1993a. Absolute optical oscillator strengths for discrete and continuum photoabsorption of molecular nitrogen (11-200 eV)," Chem. Phys. 170, 81-97.

Chan, W.F., G. Cooper, C.E. Brion 1993b. Absolute optical oscillator strengths for discrete and continuum photoabsorption of carbon monoxide (7–200 eV) and transition moments for the $X\ ^1\Sigma^+ \rightarrow A\ ^1\Pi$ system. Chemical Physics, Volume 170, 123-138.

Chen, F.Z. and C.Y. Robert Wu 2004. Temperature-dependent photoabsorption cross sections in the VUV-UV region. I. Methane and ethane. J Quant. Spectroscopy and Radiative Transfer 85, 195-209.

Chen, F.Z., D.L. Judge, C.Y.R. Wu, and J. Caldwell 1999. Low and room temperature photoabsorption cross sections of $NH_3$ in the UV region. Planet. Space Sci. 47, 261-266.

Cheng, B.-M., H.-C. Lu, H.-K. Chen, M. Bahou, Y.-P. Lee, A.M. Mebel, L.C. Lee, M.-C. Liang, and Y.L. Yung 2006. Absorption cross sections of $NH_3$, $NH_2D$, $NHD_2$, and $ND_3$ in the spectral range 140-220 nm and implications for planetary isotopic fractionation. Astrophys. J. 647, 1535-1542.

Cheng, B.-M., H.-F. Chen, H.-C. Lu, H.-K. Chen, M. S. Alam, S.-L. Chou, and M.-Y. Lin 2011. Absorption Cross Section of Gaseous Acetylene at 85 K in the Wavelength Range 110-155 nm. The Astrophysical Journal Supplement Series 196, 3.

Cheng, A. F., and 13 colleagues 2017. Haze in Pluto's atmosphere. Icarus 290, 112-133.

Cook, G. R. and P. H. Metzger 1964. Photoionization and Absorption Cross Sections of $H_2$ and $D_2$ in the Vacuum Ultraviolet Region. Journal of the Optical Society of America (1917-1983) 54, 968

Cooper G., G.R. Burton, and C.E. Brion 1995. Absolute UV and soft X-ray photoabsorption of acetylene by high-resolution dipole (e,e) spectroscopy. J. Electron Spectrosc. Relat. Phenom. 73 (2), 139-148.

Curdt, W., P. Brekke, U. Feldman, K. Wilhelm, B. N. Dwivedi, U. Schühle, and P. Lemaire 2001. The SUMER spectral atlas of solar-disk features. Astronomy and Astrophysics 375, 591-613.

Despois, D., N. Biver, D. Bockelée-Morvan, and J. Crovisier 2005. Observations of Molecules in Comets. Astrochemistry: Recent Successes and Current Challenges 231, 469-478.

Eden, S., P. Limão-Vieira, P. Kendall, N.J. Mason, S.V. Hoffmann, and S.M. Spyrou 2003. High resolution photo-absorption studies of acrylonitrile, $C_2H_3CN$, and acetonitrile, $CH_3CN$. Eur. Phys. J. D 26, 201-210.

Fahr, A. and A. K. Nayak 1994. Temperature dependent ultraviolet absorption cross sections of 1,3-butadiene and butadiyne. Chemical Physics 189, 725-731.

Feng, R., G. Cooper, and C.E. Brion 2002. Dipole (e,e) spectroscopic studies of benzene: quantitative photoabsorption in the UV, VUV and soft x-ray regions. J. Electron Spectrosc. Related Phenom. 123, 199-209.

Forget, F., T. Bertrand, M. Vangvichith, J. Leconte, E. Millour, and E. Lellouch 2017. A post-new horizons global climate model of Pluto including the $N_2$, $CH_4$ and CO cycles. Icarus 287, 54-71.

Gao, P., and 13 colleagues 2017. Constraints on the microphysics of Pluto's photochemical haze from New Horizons observations. Icarus 287, 116-123.

Gladstone, G. R., W. R. Pryor, and S. A. Stern 2015. Ly $\alpha$ @Pluto. Icarus 246, 279-284.

Gladstone, G. R., and 159 colleagues 2016. The atmosphere of Pluto as observed by New Horizons. Science 351, aad8866.

Greathouse, T. K., and 12 colleagues 2010. New Horizons Alice ultraviolet observations of a stellar occultation by Jupiter’s atmosphere. Icarus 208, 293-305.

Guo, Y. and R. W. Farquhar 2008. New Horizons Mission Design. Space Science Reviews 140, 49-74.

Heays, A.N., 2011. Photoabsorption and photodissociation in molecular nitrogen. PhD Thesis, The Australian National University.

Heays, A.N. et al., 2011. High-resolution Fourier-transform extreme ultraviolet photoabsorption spectroscopy of $^{14}N^{15}N$. J. Chem. Phys. 135, 244301-1– 244301-11.

Herbert, F., B. R. Sandel, R. V. Yelle, J. B. Holberg, A. L. Broadfoot, D. E. Shemansky, S. K. Atreya, and P. N. Romani 1987. The upper atmosphere of Uranus - EUV occultations observed by Voyager 2. Journal of Geophysical Research 92, 15093-15109.

Herbert, F. and B. R. Sandel 1991. $CH_4$ and haze in Triton's lower atmosphere. Journal of Geophysical Research Supplement 96.

Heroux, L. and J. E. Higgins 1977. Summary of full-disk solar fluxes between 250 and 1940 A. Journal of Geophysical Research 82, 3307-3310.

Hinson, D. P., I. R. Linscott, L. A. Young, G. L. Tyler, S. A. Stern, R. A. Beyer, M. K. Bird, K. Ennico, G. R. Gladstone, C. B. Olkin, M. Pätzold, P. M. Schenk, D. F. Strobel, M. E. Summers, H. A. Weaver, W. W. Woods, the New Horizons ATM Theme Team, the New Horizons Science Team 2017. Radio occultation measurements of Pluto’s neutral atmosphere with New Horizons. Icarus, submitted.

Horanyi, M., A. Poppe, and Z. Sternovsky 2016. Dust ablation in Pluto's atmosphere. EGU General Assembly Conference Abstracts 18, 3652

Huebner, W.F., Keady, J.J., Lyon, S.P., Solar Photo Rates for Planetary Atmospheres and Atmospheric Pollutants, Astrophys. Space Sci. 195, 1-294 (1992).

Huffman, R. E., Y. Tanaka, and J. C. Larrabee 1963a. Absorption Coefficients of Xenon and Argon in the 600-1025 Å; Wavelength Regions. Journal of Chemical Physics 39, 902-909.

Huffman, R. E., Y. Tanaka, and J. C. Larrabee 1963b. Absorption Coefficients of Nitrogen in the 1000-580 Å Wavelength Region. Journal of Chemical Physics 39, 910-925.

Huffman, R.E. 1969. Absorption cross-sections of atmospheric gases for use in aeronomy. Can. J. Chem. 47, 1823-1834.

Hunten, D. M. 1973. The escape of $H_2$ from Titan.. Journal of Atmospheric Sciences 30, 726-732.

Jessup, K. L., G. R. Gladstone, A. N. Heays, S. T. Gibson, B. R. Lewis, and G. Stark 2013. $^{14}N^{15}N$ detectability in Pluto’s atmosphere. Icarus 226, 1514-1526.

Kameta, K., S. Machida, M. Kitajama, M. Ukai, N. Kouchi, Y. Hatano, and K. Ito 1996. Photoabsorption, photoionization, and neutral-dissociation cross sections of $C_2H_6$ and $C_3H_8$ in the extreme-uv region. J. Electron Spectrosc. Related Phenom. 79, 391-393.

Kameta, K., N. Kouchi, M. Ukai, and Y. Hatano 2002. Photoabsorption, photoionization, and neutral-dissociation cross sections of simple hydrocarbons in the vacuum ultraviolet range. J. Electron Spectrosc. Related Phenom. 123, 225-238.

Kammer, J. A., D. E. Shemansky, X. Zhang, and Y. L. Yung 2013. Composition of Titan's upper atmosphere from Cassini UVIS EUV stellar occultations. Planetary and Space Science 88, 86-92.

Koskinen, T. T., R. V. Yelle, D. S. Snowden, P. Lavvas, B. R. Sandel, F. J. Capalbo, Y. Benilan, and R. A. West 2011. The mesosphere and lower thermosphere of Titan revealed by Cassini/UVIS stellar occultations. Icarus 216, 507-534.

Kraft, R. P., D. N. Burrows, and J. A. Nousek 1991. Determination of confidence limits for experiments with low numbers of counts. The Astrophysical Journal 374, 344-355.

Krasnopolsky, V. A. and D. P. Cruikshank 1999. Photochemistry of Pluto's atmosphere and ionosphere near perihelion. Journal of Geophysical Research 104, 21979-21996.

Krasnopolsky, V. A., B. R. Sandel, and F. Herbert 1992. Properties of haze in the atmosphere of Triton. Journal of Geophysical Research 97, 11.

Lavvas, P., Yelle, R.V., Griffith, C.A., 2010. Titan's vertical aerosol structure at the Huygens landing site: constraints on particle size, density, charge, and refractive in- dex. Icarus 210, 832–842.

Lee, L. C., R. W. Carlson, D. L. Judge, and M. Ogawa 1973. The absorption cross sections of $N_2$, $O_2$, CO, NO, $CO_2$, $N_2O$, $CH_4$, $C_2H_4$, $C_2H_6$, and $C_4H_{10}$ from 180 to 700 Å. J. Quant. Spectrosc. Radiat. Transfer 13, 1023-1031.

Lee, A.Y.T., Y.L. Yung, B.M. Cheng, M. Bahou, C.-Y. Chung, and Y.P. Lee 2001. Enhancement of deuterated ethane on Jupiter. Astrophys. J. 551, L93-L96.

Lellouch, E. C. de Bergh, B. Sicardy, H.U. Käufl, and A. Smette 2011. High resolution spectroscopy of Pluto's atmosphere: detection of the 2.3 $\mu$m CH4 bands and evidence for carbon monoxide. Astron. & Astrophys. 530, L4.

Lellouch, E., C. de Bergh, B. Sicardy, F. Forget, M. Vangvichith, and H.-U. Käufl 2015. Exploring the spatial, temporal, and vertical distribution of methane in Pluto's atmosphere. Icarus 246, 268-278.

Lellouch, E., and 18 colleagues 2017. Detection of CO and HCN in Pluto's atmosphere with ALMA. Icarus 286, 289-307.

Lindzen, R. S. and S.-S. Hong 1974. Effects of Mean Winds and Horizontal Temperature Gradients on Solar and Lunar Semidiurnal Tides in the Atmosphere.. Journal of Atmospheric Sciences 31, 1421-1446.

McCandliss, S. R. 2003. Molecular Hydrogen Optical Depth Templates for FUSE Data Analysis. Publications of the Astronomical Society of the Pacific 115, 651-661.

McComas, D., and 17 colleagues 2008. The Solar Wind Around Pluto (SWAP) Instrument Aboard New Horizons. Space Science Reviews 140, 261-313.

McComas, D. J., and 18 colleagues 2016. Pluto's interaction with the solar wind. Journal of Geophysical Research (Space Physics) 121, 4232-4246.

Marr, G. V. and J. B. West 1976. Absolute Photoionization Cross-Section Tables for Helium, Neon, Argon and Krypton in the VUV Spectral Regions. Atomic Data and Nuclear Data Tables 18, 497.

Morton, D. C. and L. Noreau 1994. A compilation of electronic transitions in the CO molecule and the interpretation of some puzzling interstellar absorption features. The Astrophysical Journal Supplement Series 95, 301-343.

Mota, R., and 10 colleagues 2005. Water VUV electronic state spectroscopy by synchrotron radiation. Chemical Physics Letters 416, 152-159.

Nakayama T. and K. Watanabe 1964. Absorption and photoionization coefficients of acetylene, propyne, and 1-butyne. J. Chem. Phys. 40, 558-561.

Nee, J.B., M. Suto, and L.C. Lee 1985. Photoexcitation processes of CH3OH: Rydberg states and photofragment fluorescence. Chem. Phys. 98, 147-155.

Nimmo, F., and 16 colleagues 2017. Mean radius and shape of Pluto and Charon from New Horizons images. Icarus 287, 12-29.

Okabe, H. 1981. Photochemistry of acetylene at 1470 Å. Journal of Chemical Physics 75, 2772-2778.

Olkin, C. B., L. A. Young, R. G. French, E. F. Young, M. W. Buie, R. R. Howell, J. Regester, C. R. Ruhland, T. Natusch, and D. J. Ramm 2014. Pluto's atmospheric structure from the July 2007 stellar occultation. Icarus 239, 15-22.

Parkinson, W.H. and K Yoshino 2003. Absorption cross-section measurements of water vapor in the wavelength region 181–199 nm. Chem Phys. 294, 31-35.

Pankratz, C. K., A. Wilson, M. A. Snow, D. M. Lindholm, T. N. Woods, T. Traver, and D. Woodraska 2015. Accessing Solar Irradiance Data via LISIRD, the Laboratory for Atmospheric and Space Physics Interactive Solar Irradiance Datacenter. AGU Fall Meeting Abstracts.

Person, Michael J. 2001. The Use of Stellar Occultations to Study the Figures and Atmospheres of Small Bodies in the Outer Solar System. Masters Thesis. MIT.

Press, William et al. 2007. Numerical Recipes: The Art of Scientific Computing, Third Edition. Cambridge University Press.

Quémerais, E., J.-L. Bertaux, O. Korablev, E. Dimarellis, C. Cot, B. R. Sandel, and D. Fussen 2006. Stellar occultations observed by SPICAM on Mars Express. Journal of Geophysical Research (Planets) 111, E09S04

Roble, R. G. and P. B. Hays 1972. A technique for recovering the vertical number density profile of atmospheric gases from planetary occultation data. Planetary and Space Science 20, 1727-1744.

Samson, J. A. R. and R. B. Cairns 1964. Absorption and Photoionization Cross Sections of $O_2$ and $N_2$ at Intense Solar Emission Lines. Journal of Geophysical Research 69, 4583-4590.

Samson, J.A.R., G. N. Haddad, and J. L. Gardner 1977. Total and partial photoionization cross sections of N2 from threshold to 100 Å. J. Phys. B: At. Mol. Opt. Phys. 10, 1749-1759.

Samson, J. A. R., T. Masuoka, P. N. Pareek, and G. C. Angel 1987. Total and dissociative photoionization cross sections of N2 from threshold to 107 eV. Journal of Chemical Physics 86, 6128-6132.

Shaw, D. A., D. M. P. Holland, M. A. MacDonald, A. Hopkirk, M. A. Hayes, and S. M. McSweeney 1992. A study of the absolute photoabsorption cross section and the photionization quantum efficiency of nitrogen from the ionization threshold to 485Å. Chemical Physics 166, 379-391.

Sicardy, B., and 67 colleagues 2016. Pluto's Atmosphere from the 2015 June 29 Ground-based Stellar Occultation at the Time of the New Horizons Flyby. The Astrophysical Journal 819, L38

Smith, G. R. and D. M. Hunten 1990. Study of planetary atmospheres by absorptive occultations. Reviews of Geophysics 28, 117-143.

Smith, P.L., K. Yoshino, W.H. Parkinson, K. Ito, and Stark. G. 1991. High-resolution, VUV (147-201 nm) photoabsorption cross sections for C2H2 at 195 and 295 K. J. Geophys. Res. E 96, 17529-17533.

Smith, G. R., D. F. Strobel, A. L. Broadfoot, B. R. Sandel, D. E. Shemansky, and J. B. Holberg 1982. Titan's upper atmosphere - Composition and temperature from the EUV solar occultation results. Journal of Geophysical Research 87, 1351-1359.

Stansberry, J. A., J. R. Spencer, B. Schmitt, A.-I. Benchkoura, R. V. Yelle, and J. I. Lunine 1996. A model for the overabundance of methane in the atmospheres of Pluto and Triton. Planetary and Space Science 44, 1051-1063.

Stark, G., K. Yoshino, P. L. Smith, K. Ito, and W. H. Parkinson 1991. High-resolution absorption cross sections of carbon monoxide bands at 295 K between 91.7 and 100.4 nanometers. The Astrophysical Journal 369, 574-580.

Stark, G., and 10 colleagues 2014. High-Resolution Oscillator Strength Measurements of the v' = 0,1 Bands of the B-X, C-X, and E-X Systems in Five Isotopologues of Carbon Monoxide. The Astrophysical Journal 788, 67

Stern, S. A., and 10 colleagues 2008. ALICE: The Ultraviolet Imaging Spectrograph Aboard the New Horizons Pluto-Kuiper Belt Mission. Space Science Reviews 140, 155-187.

Stern, S. A., and 150 colleagues 2015. The Pluto system: Initial results from its exploration by New Horizons. Science 350, aad1815.

Stevens, M. H., D. F. Strobel, M. E. Summers, and R. V. Yelle 1992. On the thermal structure of Triton's thermosphere. Geophysical Research Letters 19, 669-672.

Strobel D. F. 2009. Titan's hydrodynamically escaping atmosphere: Escape rates and the structure of the exobase region. Icarus 202, 632–641.

Strobel, D. F. 2012. Hydrogen and methane in Titan's atmosphere: chemistry, diffusion, escape, and the Hunten limiting flux principle. Canadian Journal of Physics 90, 795-805.

Strobel, D. F., M. E. Summers, F. Herbert, and B. R. Sandel 1990. The photochemistry of methane in the atmosphere of Triton. Geophysical Research Letters 17, 1729-1732.

Summers, M. E., D. F. Strobel, and G. R. Gladstone 1997. Chemical Models of Pluto's Atmosphere. Pluto and Charon 391.

Toigo, A. D., R. G. French, P. J. Gierasch, S. D. Guzewich, X. Zhu, and M. I. Richardson 2015. General circulation models of the dynamics of Pluto's volatile transport on the eve of the New Horizons encounter. Icarus 254, 306-323.

Tyler, G. L., I. R. Linscott, M. K. Bird, D. P. Hinson, D. F. Strobel, M. Pätzold, M. E. Summers, and K. Sivaramakrishnan 2008. The New Horizons Radio Science Experiment (REX). Space Science Reviews 140, 217-259.

Visser, R., van Dishoeck, E.F., Black, J.H., 2009. The photodissociation and chemistry of CO isotopologues: applications to interstellar clouds and circumstellar disks. Astron. Astrophys. 503, 323–343.

Wainfan, N., W. C. Walker, and G. L. Weissler 1955. Photoionization efficiencies and cross sections in $O_2$, $N_2$, $CO_2$, A, $H_2O$, $H_2$, and $CH_4$. Phys. Rev. 99, 542- 549.

West, G. A. 1975. Cyanide Radical Molecular Electronic and Vibrational Chemical Laser: Hydrogen Cyanide Polyatomic Chemical Laser. Thesis, University of Wisconsin, Madison, WI.

Wight, G. R., M. J. Van der Wiel, and C. E. Brion 1976. Dipole excitation, ionization and fragmentation of N2 and CO in the 10-60 eV region. J. Phys. B: Atom. Molec. Phys. 9, 675- 689.

Wilhelm, K. 2009, Solar energy spectrum. In: Trümper, J. (ed.) Landolt-Börnstein Database VI. Astronomy and Astrophysics, 4B the Solar System, Springer, Berlin, 10 – 20.

Woods, T., Tobiska, K., Rottman, G., Worden, J., "Improved solar Lyman alpha irradiance modeling from 1947 through 1999 based on UARS observations," Journal of Geophysical Research, Vol. 105, No. A12, December 1, 2000.

Woods, T. N., F. G. Eparvier, S. M. Bailey, P. C. Chamberlin, J. Lean, G. J. Rottman, S. C. Solomon, W. K. Tobiska, and D. L. Woodraska 2005. The Solar EUV Experiment (SEE): Mission overview and first results, J. Geophys. Res., 110, A01312.

Woods, T., "Lyman-Alpha Composite of Daily Solar Irradiance", version 3, released 1/23/15, Laboratory for Atmospheric and Space Physics (LASP), http://lasp.colorado.edu/lisird/lya, <the date and time you accessed the data>.

Wong, M. L, S. Fan, P. Gao, M Liang, Shia, Y. L. Yung, J. A. Kammer, M. E. Summers, G. R. Gladstone, L. A. Young, C. B. Olkin, K. Ennico, H. A. Weaver, S. A. Stern, and the New Horizons Science Team 2017, The photochemistry of Pluto's atmosphere as illuminated by New Horizons. Icarus, submitted.

Wu, C.Y.R., F.Z. Chen, and D.L. Judge 2001. Measurements of temperature-dependent absorption cross sections of $C_2H_2$ in the VUV-UV region. J. Geophys. Res. 106 E, 7629-7636.

Wu, C. Y. R., F. Z. Chen, and D. L. Judge 2004. Temperature-dependent photoabsorption cross sections in the VUV-UV region: Ethylene. Journal of Geophysical Research (Planets) 109, E07S15

Yoshino, K., J.R. Esmond, W.H. Parkinson, K. Ito, and T. Matsui 1996a. Absorption cross section measurements of water vapor in the wavelength region 120 to 188 nm. Chem Phys. 211, 387-391.

Yoshino, K., J. R. Esmond, Y. Sun, W. H. Parkinson, K. Ito, and T. Matsui 1996b. Absorption cross section measurements of carbon dioxide in the wavelength region 118.7-175.5 nm and the temperature dependence. Journal of Quantitative Spectroscopy and Radiative Transfer 55, 53-60.

Young, L. A., and 27 colleagues 2008. New Horizons: Anticipated Scientific Investigations at the Pluto System. Space Science Reviews 140, 93-127.

Young, L. A. 2009. Rapid Computation of Occultation Lightcurves Using Fourier Decomposition. The Astronomical Journal 137, 3398-3403.

Yung, Y. L., M. Allen, and J. P. Pinto 1984. Photochemistry of the atmosphere of Titan - Comparison between model and observations. The Astrophysical Journal Supplement Series 55, 465-506.

Zangari, Amanda M. 2013. Investigating spatial variation in the surface and atmosphere of Pluto through stellar occultations and PSF photometry. Ph. D. Thesis. MIT.

Zirnstein, E. J., and 10 colleagues 2016. Interplanetary Magnetic Field Sector from Solar Wind around Pluto (SWAP) Measurements of Heavy Ion Pickup near Pluto. The Astrophysical Journal 823, L30

Zhu, X., D. F. Strobel, and J. T. Erwin 2014. The density and thermal structure of Pluto's atmosphere and associated escape processes and rates. Icarus 228, 301-314.